\pdfoutput=1
\documentclass[
	reprint,
	amsmath,
	amssymb,
	aps,
	longbibliography,
	pra,
	floatfix,
	nofootinbib
]{revtex4-2}

\usepackage[utf8]{inputenc}

\usepackage{amsmath}
\usepackage{tensor}
\usepackage{multirow}

\usepackage[nomessages]{fp}	

\usepackage[hyperref]{xcolor}

\definecolor{goethe-blau}{cmyk}{1.0,0.2,0.0,0.4}
\definecolor{hellgrau}{cmyk}{0.04,0.04,0.05,0.02}
\definecolor{sandgrau}{cmyk}{0.12,0.09,0.13,0.0}
\definecolor{dunkelgrau}{cmyk}{0.25,0.25,0.30,0.75}
\definecolor{emo-rot}{cmyk}{0.04,1.0,0.8,0.07}
\definecolor{purple}{cmyk}{0.08,1.0,0.3,0.36}
\definecolor{senfgelb}{cmyk}{0.01,0.25,1.0,0.05}
\definecolor{gruen}{cmyk}{0.62,0.4,0.87,0.09}
\definecolor{magenta}{cmyk}{0.08,0.86,0.12,0.12}
\definecolor{orange}{cmyk}{0.0,0.7,1.0,0.04}
\definecolor{sonnengelb}{cmyk}{0.0,0.12,0.95,0.0}
\definecolor{helles-gruen}{cmyk}{0.4,0.17,0.81,0.07}
\definecolor{lichtblau}{cmyk}{0.8,0.0,0.06,0.04}

\usepackage[
colorlinks=true,
linkcolor=goethe-blau,
breaklinks,
pdfpagelabels,
pdfstartview=FitH,
bookmarksopen=true,
bookmarksopenlevel=2,
bookmarksnumbered=true,
plainpages=false,
hypertexnames=false,
citecolor=emo-rot,
urlcolor=purple,
pdfauthor={{S.~Jafarzade, A.~Koenigstein, F.~Giacosa}},
pdftitle={{Phenomenology of spin-3 tensor mesons}},
pdfsubject={{particle physics}}
]{hyperref}

\usepackage{bookmark}

\usepackage{upgreek}
\usepackage{cancel}

\usepackage{ wasysym }

\usepackage{graphicx}
\usepackage{dcolumn}
\usepackage{bm}

\allowdisplaybreaks

\begin{document}

\preprint{APS/123-QED}

\title{Phenomenology of \texorpdfstring{$J^{PC} = 3^{--}$}{JPC = 3--} tensor mesons}


\author{Shahriyar Jafarzade\textsuperscript{1}}
	\email{shahriyar.jzade@gmail.com}
	\affiliation{
		Institute of Physics, Jan Kochanowski University,
		\\
		ul.~Uniwersytecka 7, P-25-406 Kielce, Poland.
	}

\author{Adrian Koenigstein\textsuperscript{2}}
	\email{koenigstein@th.physik.uni-frankfurt.de}
	
	\affiliation{
		Institute for Theoretical Physics, Goethe-University,
		\\
		Max-von-Laue-Str.~1, D-60438 Frankfurt am Main, Germany.
	}
	
\author{Francesco Giacosa\textsuperscript{1,2}}
	\email{fgiacosa@ujk.edu.pl}
	\affiliation{
		Institute of Physics, Jan Kochanowski University,
		\\
		ul. Uniwersytecka 7, P-25-406 Kielce, Poland,
	}
	\affiliation{
		Institute for Theoretical Physics, Goethe-University,
		\\
		Max-von-Laue-Str.~1, D-60438 Frankfurt am Main, Germany.
	}

\date{\today}

\begin{abstract}
	We study the strong and radiative decays of the antiquark-quark ground state $J^{PC} = 3^{--}$ ($n^{2 S + 1} L_J = 1^3 D_3$) nonet \{$\rho_{3} (1690)$, $K_{3}^{\ast} (1780)$, $\phi_{3} (1850)$, $\omega_{3} (1670)$\} in the framework of an effective quantum field theory approach, based on the $SU_\mathrm{V}(3)$ flavor symmetry. The effective model is fitted to experimental data listed by the \textit{Particle Data Group}. We predict numerous experimentally unknown decay widths and branching ratios. An overall agreement of theory (fit and predictions) with experimental data confirms the $\bar{q} q$ nature of the states and qualitatively validates the effective approach. Naturally, experimental clarification as well as advanced theoretical description is needed for trustworthy quantitative predictions, which is observed from some of the decay channels. Besides conventional spin-$3$ mesons, theoretical predictions for ratios of strong and radiative decays of a hypothetical glueball state $G_3 (4200)$ with $J^{PC} = 3^{--}$ are also presented.
\end{abstract}

\maketitle

\tableofcontents

\newpage

\section{Introduction}
\label{sec:introduction}

	The spectroscopy and the phenomenological description of conventional mesons is important for many reasons. It allows to test quantum chromodynamics (QCD) in the nonperturbative regime. Furthermore, it aims to the correct and systematic understanding and assignment of experimentally measured states, resonances and mesons \cite{Zyla:2020zbs}. It is also fundamental for the search of mesons that go beyond the antiquark-quark ($\bar{q}q$) picture, such as glueballs, hybrids, and multiquark states. Namely, both in the light-quark and heavy-quark sectors, it is possible to search for ``exotics'' only, if the conventional $\bar{q} q$ mesons are fully under control \cite{Klempt:2007cp}.
	
	 Consequently, our work is just an incremental contribution to this extensive project in high-energy physics. In our work, we provide a (qualitative) analysis of the phenomenology of the spin-$3$ mesons $\rho_{3} (1690)$, $K_{3}^{\ast} (1780)$, $\phi_{3} (1850)$, $\omega_{3} (1670)$ with $J^{PC} = 3^{--}$ ($n^{2 S + 1} L_J = 1^3 D_3$) as well as the potential spin-$3$ tensor glueball $G_3 (4200)$ \cite{Morningstar:1999rf,Meyer:2004gx,Chen:2005mg}. This involves a fit of our effective model to experimental data as well as predictions for decay widths and ratios of the decays of the spin-$3$ mesons. Although we provide explicit values for the decay widths and branching ratios, these should be considered only as first estimates and mainly interpreted as qualitative results. Hence, this study may for example help in identifying dominant decay channels  \textit{etc.}. 

\subsection{Contextualization}

	In the low-energy sector of QCD, conventional mesons contain the light up ($u$), down ($d$), and strange ($s$) quarks, hence they can be grouped into nonets: $\bar{u} d$, $\bar{u} s$, \textit{etc.}. Nonets are classified by the quantum numbers $J^{PC}$, where $J$ is the absolute value of the total spin of the meson ($|L - S| \leq J \leq | L + S |$, where $L$ is the angular momentum and $S$ the spin of the system), $P$ is the sign-change under parity transformations, and $C$ the sign-change under charge conjugation. For $\bar{q} q$ states, besides $J^{PC}$, also the older spectroscopic notation $n^{2 S + 1} L_{J}$, with $L = S, P, D, \ldots$ and with $n = 1, 2, \ldots$, which is the radial quantum number, is used. In the following, we shall restrict to the radial ground state nonets with $n = 1$. For a $\bar{q} q$ system, where $P = ( - 1 )^{L + 1}$ and $C = ( - 1 )^{L + S}$, notice that $L$ and $S$ are, strictly speaking, not well defined quantum numbers in a relativistic setup, hence mixing of states with the same $J^{PC}$ but with different $L$ and $S$ is possible -- even if it is oftentimes a subordinate effect. However, this nomenclature for the states is still well upheld also in presence of mixing for practical purposes.
		
	In the low-energy regime of QCD the ``classical'' symmetries of QCD at high-energy scales still play a crucial role -- more precisely the spontaneous, anomalous or explicit breaking of these symmetries. Quickly recapitulated, the situation is as follows: gluons are ``democratic'': they interact equally strongly with right- and left-handed quarks implying that in the so-called chiral limit (\textit{i.e.}\ assuming negligibly small bare quark masses $m_u$, $m_d$, and $m_s$) the QCD Lagrangian is invariant under $U_\mathrm{L} (3) \times U_\mathrm{R} (3)$ transformations. However, this symmetry of the classical action is broken due to quantum effects, which is called the $U_\mathrm{A}(1)$-anomaly \cite{Adler:1969gk,Bell:1969ts,tHooft:1976rip,Jackiw:2008}. Still, the baryon number is conserved, which corresponds to the unaffected $U_\mathrm{V}(1)$ symmetry. The remaining $SU_\mathrm{L} (3) \times SU_\mathrm{R} (3)$ symmetry is spontaneously broken at low energies into $SU_\mathrm{V} (3)$ \cite{Peskin:1995ev,Weinberg:1996kr}. The $SU_\mathrm{V} (3)$ amounts to a rotation in the flavor space spanned by $u$, $d$ and $s$ and is consequently called \textit{flavor symmetry} or \textit{vector symmetry}. Hence, at the composite mesonic level, chiral partner-mesons, which are linked through a chiral transformation, are not degenerate anymore. After all, the residual realization of an approximate $SU_\mathrm{V} (3)$, which is only explicitly broken by the bare quark masses, is the reason, why nonets of $\bar{q} q$ states can be used to classify the low-energy spectrum of QCD \cite{GellMann:1961ky,Neeman:1961jhl,Greiner:1994}.
	
	Typical examples of very well-known mesonic nonets are the pseudoscalar states,
		\begin{align}
			\{ \pi, \, K, \, \eta^{\prime} (958), \, \eta \} \, ,	\vphantom{\bigg(\bigg)}	\label{eq:spin_0-+_nonet}
		\end{align}
	with $J^{PC} = 0^{-+}$ $(n^{2 S + 1} L_J = 1^{1} S_{0})$, the vector mesons
		\begin{align}
			\{ \rho(770), \, K^{\ast} (892), \, \phi (1020), \, \omega (782) \} \, ,	\vphantom{\bigg(\bigg)}	\label{eq:spin_1--_nonet}
		\end{align}		
	with $J^{PC} = 1^{--}$ $(n^{2 S + 1} L_J = 1^{3} S_1)$, the pseudovector mesons,
		\begin{align}
			\{ b_1 (1235), \, K_{1 , B}, \, h_1 (1415), \, h_1 (1170) \} \, ,	\vphantom{\bigg(\bigg)}	\label{eq:spin_1+-_nonet}
		\end{align}
	with $J^{PC} = 1^{+-}$ $(n^{2 S + 1} L_J = 1^{1} P_1)$, the axial-vector meson,
		\begin{align}
			\{ a_1 (1260), \, K_{1 , A}, \, f_1 (1420), \, f_1 (1285) \} \, ,	\vphantom{\bigg(\bigg)}	\label{eq:spin_1++_nonet}
		\end{align}
	with $J^{PC} = 1^{++}$ $(n^{2 S + 1} L_J =  1^{3} P_1)$, and the well-established tensor mesons,
		\begin{align}
			\{ a_{2} (1320), \, K_{2}^{\ast} (1430), \, f_{2}^{\prime} (1525), \, f_{2} (1270) \} \, ,	\vphantom{\bigg(\bigg)}	\label{eq:spin_2++_nonet}
		\end{align}
	with $J^{PC} = 2^{++}$ $(n^{2 S + 1} L_J =  1^{3} P_{2})$, compare with Ref.\ \cite{Godfrey:1985xj} or the \textit{quark model review} by the \textit{Particle Data Group} (PDG) \cite{Zyla:2020zbs}.
	
	The aim of this work is to study the $\bar{q} q$ nonet with quantum numbers $J^{PC}=3^{--}$, resulting from $L=2$ and $S=1$, therefore $n^{2 S + 1} L_J = 1^{3} D_{3}$ states, in a relativistic effective quantum field theory (QFT) model based on $SU_\mathrm{V}(3)$ flavor symmetry. The mesons belonging to this nonet, corresponding to the lightest states with the proper quantum numbers in the PDG \cite{Zyla:2020zbs}, are:
		\begin{align}
			\{ \rho_{3} (1690), \, K_{3}^{\ast} (1780), \, \phi_{3} (1850), \, \omega_{3} (1670) \} \, .	\label{eq:spin_3--_nonet}
		\end{align}
	The study of these states is very interesting for a series of reasons (for previous studies on related subjects see also Refs.\ \cite{Godfrey:1985xj,Wang:2016enc,Wang:2016hbl,Bergstrom:1990kf,Harnad:1972fe,Pang:2015eha,Caporale:2005me,Johnson:2020ilc} and Refs.\ therein):
	\begin{enumerate}
		\item	It is the only well-known mesonic ground-state nonet with $J > 2$, which is confirmed by several experiments \cite{Zyla:2020zbs}.
		
		\item	Many decay channels are known \cite{Zyla:2020zbs}, which allows for further theoretical and experimental tests of the assignment.
		
		\item	Predictions for not-yet measured strong and radiative decay rates are possible.
		
		\item	The spin-$3$ ground-state nonet \eqref{eq:spin_3--_nonet} is also measured in lattice QCD calculations \cite{Dudek:2013yja,Johnson:2020ilc}.
	\end{enumerate}
	This nonet is therefore tailor made for an effective QFT study of decays. In particular, we shall answer whether and to what extend the $\bar{q} q$ assignment works, we can test validity of flavor symmetry for various decay channels, make various ``postdictions'' and -- most interestingly -- predictions for many decay channels. In view of the ongoing experimental efforts in hadron physics at \textit{GlueX} \cite{Ghoul:2015ifw,Zihlmann:2010zz,Proceedings:2014joa} and \textit{CLAS12} \cite{Rizzo:2016idq} at \textit{Jefferson Lab}, at \textit{COMPASS} and \textit{LHCb} at \textit{CERN} \cite{Ryabchikov:2019rgx,Bediaga:2018lhg}, at \textit{BESIII} in Beijing \cite{Mezzadri:2015lrw,Marcello:2016gcn}, and at the future \textit{PANDA} experiment \cite{Lutz:2009ff} at the \textit{GSI} facility, we consider a revival of interest on such resonances valuable.
	
	As an additional application of our approach, we also study some decays of the $J^{PC} = 3^{--}$ glueball state. This can be done by a simple modification of the action of spin-$3$ mesons and by using the mass of approximately $4.2 \, \,$GeV found in lattice simulations \cite{Morningstar:1999rf,Meyer:2004gx,Chen:2005mg} (in the quenched approximation)  as an input. Due to the flavor-singlet nature of the glueball, only few decay channels are possible.

\subsection{Method}

	From a technical point of view, we construct $SU_\mathrm{V}(3)$-invariant effective actions/Lagrangians that involve the mesonic nonet \eqref{eq:spin_3--_nonet} as well as its various decay products consisting in the well-established $\bar{q} q$ nonets that were previously introduced. From a (Functional) Renormalization Group (FRG) perspective, the effective actions/Lagrangians can be interpreted as residual infrared (IR) effective actions with coupling constants that already involve all quantum effects from higher energy scales, the ultraviolet (UV). These coupling constants are determined via fits to experimental data, rather than by ab initio QCD-calculations. All calculations for the decays are therefore performed at tree-level. This approach was already implemented in earlier works for studies of various light mesonic nonets, such as tensor mesons \cite{Giacosa:2005bw}, axial-vector, and pseudovector mesons \cite{Divotgey:2013jba}, the scalar mesons \cite{Giacosa:2005zt,Giacosa:2005qr}, the pseudotensor mesons \cite{Koenigstein:2016tjw}, and the orbitally and radially excited vector mesons \cite{Piotrowska:2017rgt}.

\subsection{Organization and structure}
	
	We organize the paper as follows: in Chap.\ \ref{sec:themodel}, after a discussion of our low-energy effective model, we present antiquark-quark nonets, their transformation rules, and the corresponding Lagrangian of the model. It consists of seven interaction terms. In Chap.\ \ref{sec:results} we show the results for the decay widths and branching ratios of spin-$3$ mesons, while in Chap.\ \ref{sec:glueball} we list some branching ratios for an hypothetical heavy $3^ {--}$ glueball. A discussion of the validity of the employed effective interaction terms is reported in Chap.\ \ref{chap:discussion} and conclusions can be found in Chap.\ \ref{sec:conclusions}. Many technical aspects concerning the QFT treatment of $J = 3$ fields can be found in the numerous and detailed appendices.
	
\section{Effective model for \texorpdfstring{$J^{PC} = 3^{--}$}{JPC = 3--} mesons}
\label{sec:themodel}

	As mentioned in the introduction, in the chiral limit ($m_{u,d,s} = 0$) the $SU_\mathrm{L} (3) \times SU_\mathrm{R} (3) \times U_\mathrm{V} (1)$ symmetry of QCD is spontaneously broken into flavor symmetry $SU_\mathrm{V} (3)$ and baryon-number conservation -- the $U_\mathrm{V}(1)$ symmetry. One of the first successful models that describes this process using four-Fermi interactions is given by Refs.\ \cite{Nambu:1961tp,Nambu:1961fr}. At the level of confined light hadrons, this symmetry is evident: quark-antiquark mesons are clearly grouped into nonets, some of which were already listed in the introduction \cite{GellMann:1961ky,Neeman:1961jhl,Greiner:1994}. One possibility to describe low-energy QCD makes use of effective low-energy hadronic models. If the degrees of freedoms are only hadrons, then confinement and color-neutrality is automatically built in. Typically, these models are defined by a proper action that mimics the chiral symmetry of QCD and its spontaneous and explicit breaking.
	
	The chiral symmetry is for example the guiding principle of chiral perturbation theory (ChPT) \cite{Gasser:1983yg,Gasser:1984gg,Scherer:2002tk,Leutwyler:1993iq,Pich:1995bw,Bernard:2006gx,Schindler:2005ke,Jenkins:1995vb,Booth:1996hk,Terschlusen:2016kje,Cirigliano:2003yq}, in which chiral symmetry is nonlinearly realized, as well as for linear $\sigma$-models \cite{GellMann:1960np,Parganlija:2012fy,Ko:1994en,Carter:1995zi,Janowski:2014ppa,Koch:1997ei}, in which it is linearly realized. In effective models the nonets usually appear in pairs of chiral partners. The breaking of chiral symmetry generates a mass difference between them. This dynamics can also be studied at nonzero temperature and/or nonzero baryon or isospin chemical potentials. Within the last years, the Functional Renormalization Group \cite{Gies:2006wv,Polonyi:2001se,Wetterich:1992yh,Morris:1993qb,Dupuis:2020fhh,Pawlowski:2005xe}, turned out to be one particularly interesting framework to describe chiral symmetry breaking (and phase transitions) in these kind of models, see for example Refs.\ \cite{Grahl:2013pba,Eser:2015pka,Rennecke:2016tkm,Pawlowski:2014zaa,Eser:2018jqo,Divotgey:2019xea,Eser:2019pvd,Cichutek:2020bli,Otto:2019zjy,Otto:2020hoz,Jung:2019nnr,Heller:2015box}.
	
	Hadronic models in which only the residual flavor symmetry $SU_\mathrm{V} (3)$ is explicitly conserved have also been constructed for a variety of nonets in a series of publications by one of the authors and collaborators \cite{Giacosa:2005zt,Giacosa:2005qr,Piotrowska:2017rgt,Divotgey:2013jba,Giacosa:2005bw,Koenigstein:2016tjw}. These models can be interpreted as the effective emerging terms of chiral models after spontaneous symmetry breaking is worked out. When the effective action is written down, only flavor symmetry is retained and an expansion in dominant and subdominant terms in $1/N_\mathrm{c}$ is carried out. Moreover, terms that break explicitly flavor symmetry, either because of the underlying breaking due to nonzero and unequal quark masses ($m_u \approx 2 \, \,$MeV, $m_d \approx 5 \, \,$MeV and $m_s \approx 93 \, \,$MeV) or due to the chiral anomaly can be included.
	
	In this work, we study the decays of the $\bar{q} q$ ground-state $J^{PC} = 3^{--}$ nonet by constructing such a model. We shall consider only the dominant terms in the large-$N_\mathrm{c}$ expansion \cite{tHooft:1973alw,Witten:1979kh,Lebed:1998st} and neglect flavor symmetry breaking corrections, since the present level of data accuracy does not allow for their investigations.
	
\subsection{Particle content}

\subsubsection{Symmetries}

	Mesonic nonets that transform under the adjoint representation of the approximate flavor symmetry $SU_\mathrm{V} (3)$ are the main ingredients of the current work. Transformation rules under parity, charge conjugation and flavor transformations of the mesonic nonets of our model are summarized\footnote{Other mesonic nonets constructed in the same way can be found for instance in Refs.\ \cite{Parganlija:2012fy,Divotgey:2013jba,Giacosa:2016hrm,Koenigstein:2016tjw}.} in the Table \ref{tab:transformations}. On the other hand, the mesonic nonets are in direct correspondence to the physical states, which is listed in Table \ref{tab:resonances_states}.
	
	\begin{table}
		\centering
		\renewcommand{\arraystretch}{1.5}
		\caption{\label{tab:transformations} %
					Transformation properties of the pseudoscalar $P$ \eqref{eq:pseudoscalar_nonet}, the vector $V_1$ \eqref{eq:vector_nonet}, the pseudovector $B_1$ \eqref{eq:pvector_nonet}, the axial-vector $A_1$ \eqref{eq:avector_nonet}, the rank-$2$ tensor $A_2$ \eqref{eq:tensor_nonet}, and rank-$3$ tensor $W_3$ \eqref{eq:tensor3_nonet} nonets under parity transformations $P$, charge conjugation $C$, and $SU_\mathrm{V} (3)$ flavor transformations $U$. Notice the position of the Lorentz indices for parity transformations, since spatial and time-like indices do not transform identically. We use the Minkowski metric in the $(\eta_{\mu\nu}) = \mathrm{diag}( + 1, - 1, - 1, - 1 )$ convention.
		}
		\begin{ruledtabular}
			\begin{tabular}[c]{l c c c}
				\multicolumn{1}{c}{nonet}	&	parity									&	charge conjugation			&	flavor
				\\
				$J^{PC} = F_J^{\cdots}$		&	$P$									&	$C$						&	$SU_\mathrm{V} (3)$
				\\
				\colrule
				$0^{-+} = P$				&	$ - P ( t, - \vec{x} )$					&	$(P)^T$						&	$U P U^{\dagger}$
				\\
				$1^{--} = V_1^{\mu}$		&	$V_{1 , \mu} ( t, - \vec{x} )$			&	$ - ( V_1^{\mu} )^T$		&	$U V_1^{\mu} U^{\dagger}$
				\\
				$1^{+-} = B_1^{\mu}$		&	$ - B_{1 , \mu} ( t, - \vec{x} )$		&	$ - ( B_1^{\mu} )^T$		&	$U B_1^{\mu}U^{\dagger}$
				\\
				$1^{++} = A_1^{\mu}$		&	$ - A_{1 , \mu} ( t, - \vec{x} )$		&	$( A_1^{\mu} )^T$			&	$U A_1^{\mu} U^{\dagger}$
				\\
				$2^{++} = A_2^{\mu\nu}$		&	$A_{2 , \mu\nu} ( t, - \vec{x} )$		&	$( A_2^{\mu\nu} )^T$		&	$U A_2^{\mu\nu} U^{\dagger}$
				\\
				$3^{--} = W_3^{\mu\nu\rho}$	&	$W_{3 , \mu\nu\rho} ( t, - \vec{x} )$	&	$-( W_3^{\mu\nu\rho} )^T$	&	$U W_3^{\mu\nu\rho} U^{\dagger}$
			\end{tabular}
		\end{ruledtabular}
	\end{table}
	
	Each $\bar{q} q$ nonet can be assigned to a certain microscopic antiquark-quark current $N_{ij} \equiv ( \bar{q}_j \, \Gamma \, q_i ) / \sqrt{2}$, where $\Gamma$ is a combination of Dirac matrices and derivatives, which -- in the nonrelativistic limit -- reduces to the corresponding nonrelativistic configuration with the desired $L$ and $S$.

\subsubsection{Mesonic nonets in the model}

\paragraph*{Pseudoscalar mesons}
	
	The first nonet of Table \ref{tab:transformations} describes the matrix of pseudoscalar mesons $P$ with quantum numbers $L = S = 0$, leading to $J^{PC} = 0^{-+}$ $(n^{2 S + 1} L_J = 1^{1} S_{0})$. The elements are $P_{ij} \equiv ( \bar{q}_j \, \mathrm{i} \gamma^{5} \, q_i ) / \sqrt{2}$ and the mesons are \{$\pi$,	$K$, $\eta^{\prime} (958)$, $\eta$\}, where the first entry represents the isospin $I = 1$ triplet (the three pions), the second entry the two $I = 1/2$ isodoublets (the four kaons), and the last two entries the two isoscalar states (the etas). Pseudoscalars form the basis of almost all low-energy effective hadronic models/limits or theories of QCD, \textit{e.g.}\ ChPT \cite{Gasser:1983yg,Gasser:1984gg,Scherer:2002tk,Leutwyler:1993iq,Pich:1995bw,Bernard:2006gx,Schindler:2005ke,Jenkins:1995vb,Booth:1996hk,Terschlusen:2016kje} and $\sigma$-models \cite{GellMann:1960np,Ko:1994en,Carter:1995zi,Parganlija:2012fy,Koch:1997ei} (for an link between ChPT and $\sigma$-models, see Ref.\ \cite{Divotgey:2016pst,Eser:2018jqo,Divotgey:2019xea,Cichutek:2020bli}). Explicitly, the matrix $P$ reads:
		\begin{align}
			P = \tfrac{1}{\sqrt{2}}
			\begin{pmatrix}
				\frac{\eta_{N} + \pi^{0}}{\sqrt{2}}	&	\pi^{+}								&	K^{+}		\\
				\pi^{-}								&	\frac{\eta_{N} - \pi^{0}}{\sqrt{2}}	&	K^{0}		\\
				K^{-}								&	\bar{K}^{0}							&	\eta_{S}
			\end{pmatrix} \, ,	\label{eq:pseudoscalar_nonet}
		\end{align}
	where $\eta_{N} \equiv ( \bar{u} \, \mathrm{i} \gamma^5 \, u + \bar{d} \, \mathrm{i} \gamma^5 \, d ) / \sqrt{2}$ stands for the the purely nonstrange state and $\eta_{S} \equiv \bar{s} \, \mathrm{i} \gamma^5 \, s$ stands for the pure strange state. For the pions and kaons, the physical states are directly assigned to the fields in the model. In the isoscalar sector, physical and model fields are related by mixing\footnote{We refer to App.~\ref{app:masses_and_mixing} for the description of the mass terms of the corresponding Lagrangian(s) and for the derivation of the PDG-mixing formula, which was applied in the PDG \cite{Zyla:2020zbs} in the case of ground-state vector and pseudoscalar mesons as well as for the ground-state tensor mesons with $J^{P C}=2^{+ +}$ and $J^{P C}=3^{- -}$. We shall also present the link between the singlet-octet basis used in the PDG and the strange-nonstrange one employed in this work.},
		\begin{align}
			\begin{pmatrix}
				\eta									\\
				\eta^\prime (958)
			\end{pmatrix}
			=
			\begin{pmatrix}
				\cos \beta_{p}		&	\sin \beta_{p}	\\
				- \sin \beta_{p}	&	\cos \beta_{p}
			\end{pmatrix}
			\begin{pmatrix}
				\eta_{N}	\\
				\eta_{S}
			\end{pmatrix}	\, .
		\end{align}
	Here we shall use $\beta_{p} = - 43.4^{\circ}$ obtained in Ref.\ \cite{AmelinoCamelia:2010me}. Note, the rather large mixing angle and the unexpectedly high mass of the $\eta^{\prime} (958)$ result from the chiral (or axial) anomaly $U_\mathrm{A}(1)$ \cite{Feldmann:1998vh,tHooft:1986ooh}. According to a classifications introduced by two of the authors of the present work in Ref.\ \cite{Giacosa:2017pos}, pseudoscalar mesons together with their chiral partners (the scalars), belong to what we call a ``heterochiral multiplet'', which allows for the constructions of chirally anomalous mass and interaction terms.
	
\paragraph*{Vector mesons}

	Next, the second entry in Table \ref{tab:transformations} refers to the $J^{PC} = 1^{--}$ $(n^{2 S + 1} L_J = 1^{3} S_1)$ nonet with $L = 0$ and $S = 1$. These are the very well-known vector states \{$\rho (770)$, $K^{\ast} (892)$, $\omega (782)$, $\phi (1020)$\}, which are for example also included in certain extensions of ChPT \cite{Schindler:2005ke,Jenkins:1995vb,Booth:1996hk,Terschlusen:2016kje} as well as in enlarged (and realistic) versions of hadronic models \cite{GellMann:1960np,Parganlija:2012fy,Ko:1994en,Carter:1995zi,Janowski:2014ppa}. The matrix $V_1^{\mu}$ with elements $V_{1,ij}^{\mu} = ( \bar{q}_j \, \gamma^{\mu} \, q_i ) / \sqrt{2}$ has the form
		\begin{align}
			V_1^{\mu} = \tfrac{1}{\sqrt{2}}
			\begin{pmatrix}
				\frac{\omega_{1 , N}^{\mu} + \rho_1^{0 \mu}}{\sqrt{2}}	&	\rho_1^{+ \mu}										&	K_1^{\ast + \mu}	\\
				\rho_1^{- \mu}										&	\frac{\omega_{1 , N}^{\mu} - \rho_1^{0 \mu}}{\sqrt{2}}	&	K_1^{\ast 0 \mu}	\\
				K_1^{\ast - \mu}										&	\bar{K}_1^{\ast 0 \mu}								&	\omega_{1 , S}^{\mu}
			\end{pmatrix} \, ,	\label{eq:vector_nonet}
		\end{align}
	where $\omega_{N}$ and $\omega_{S}$ are purely nonstrange and strange states, respectively. Similarly as before, the physical fields arise upon mixing
		\begin{align}
			\begin{pmatrix}
				\omega (782)	\\
				\phi (1020)
			\end{pmatrix}
			=
			\begin{pmatrix}
				\cos \beta_{v_1}		&	\sin \beta_{v_1}	\\
				- \sin \beta_{v_1}	&	\cos \beta_{v_1}
			\end{pmatrix}
			\begin{pmatrix}
				\omega_{1 , N}	\\
				\omega_{1 , S}
			\end{pmatrix} \, ,
		\end{align}
	where the very small isoscalar-vector mixing angle $\beta_{v_1} = - 3.9^{\circ}$ is taken from the PDG \cite{Zyla:2020zbs}. Hence, the physical states $\omega (782)$ and $\phi (1020)$ are dominated by nonstrange and strange components, respectively. This is in agreement with the ``homochiral'' nature of these states \cite{Giacosa:2017pos}. In fact vector and axial-vector mesons form a ``homochiral multiplet'', for which the effect of the chiral anomaly is of subleading order and can safely be ignored.
	
\paragraph*{Pseudovector mesons}

	Next, we consider $L = 1$. The choice $L = 1$ and $S = 0$, leading to $J^{PC} = 1^{+-}$ $(n^{2 S + 1} L_J = 1^{1} P_1)$, contains the established pseudovector mesons \{$b_1 (1235)$, $K_{1 , B}$, $h_1 (1170)$, $h_1 (1415)$\}. The corresponding nonet with elements $B_{1 , ij}^{\mu} = ( \bar{q}_j \, \gamma^{5} \, \partial^{\mu} \, q_i ) / \sqrt{2}$ is
		\begin{align}
			B_1^{\mu} = \tfrac{1}{\sqrt{2}}
			\begin{pmatrix}
				\frac{h_{1 , N}^{\mu} + b_1^{0 \mu}}{\sqrt{2}}	&	b_1^{+ \mu}									&	K_{1 , B}^{+ \mu}	\\
				b_1^{- \mu}									&	\frac{h_{1 , N}^{\mu} - b_1^{0 \mu}}{\sqrt{2}}	&	K_{1 , B}^{0 \mu}	\\
				K_{1 , B}^{- \mu}									&	\bar{K}_{1 , B}^{0 \mu}							&	h_{1 , S}^{\mu}
			\end{pmatrix} \, .	\label{eq:pvector_nonet}
		\end{align}
	The mixing angle
		\begin{align}
			\begin{pmatrix}
				h_1 (1170)	\\
				h_1 (1415)
			\end{pmatrix}
			=
			\begin{pmatrix}
				\cos \beta_{b_1}		&	\sin \beta_{b_1}	\\
				- \sin \beta_{b_1}	&	\cos \beta_{b_1}
			\end{pmatrix}
			\begin{pmatrix}
				h_{1 , N}	\\
				h_{1 , S}
			\end{pmatrix}
		\end{align}
	is not known. Quite interestingly, since this nonet belongs to a heterochiral multiplet (the chiral partners are the orbitally excited vector mesons) the mixing angle could be nonnegligible, just as for pseudoscalar mesons. Here, we shall consider two scenarios for our model calculations: in the first, the mixing angle $\beta_{b_1}$ is set to zero, in the second we consider a large mixing similar to the pseudoscalar sector, $\beta_{b_1} \approx - 40^{\circ}$ and similar to what two of the authors found for the pseudotensor meson nonet in Ref.\ \cite{Koenigstein:2016tjw} and elaborated on in Ref.\ \cite{Giacosa:2017pos}.
	
	\paragraph*{Tensor mesons}
	
	For $L = S = 1$ three nonets are possible. The well-known $J^{PC} = 2^{++}$ $(n^{2 S + 1} L_J = 1^{3} P_{2})$ tensor states \{$a_{2} (1320)$, $K_{2}^{\ast} (1430)$, $f_{2} (1270)$, $f_{2}^{\prime} (1525)$\}, with elements $A_{2 , ij}^{\mu} = [ \bar{q}_j \, ( \mathrm{i} \gamma^{\mu} \, \partial^\nu + \ldots ) \, q_i ] / \sqrt{2}$, represent an almost ideal nonet of quark-antiquark states
		\begin{align}
			A_2^{\mu\nu} = \tfrac{1}{\sqrt{2}}
			\begin{pmatrix}
				\frac{f_{2 , N}^{\mu\nu} + a_{2}^{0 \mu\nu}}{\sqrt{2}}	&	a_{2}^{+ \mu\nu}										&	K_{2}^{\ast + \mu\nu}	\\
				a_{2}^{- \mu\nu}										&	\frac{f_{2 , N}^{\mu\nu} - a_{2}^{0 \mu\nu}}{\sqrt{2}}	&	K_{2}^{\ast 0 \mu\nu}	\\
				K_{2}^{\ast - \mu\nu}									&	\bar{K}_{2}^{\ast 0 \mu\nu}								&	f_{2 , S}^{\mu\nu}
			\end{pmatrix} \, .	\label{eq:tensor_nonet}
		\end{align}
	The physical isoscalar-tensor states are
		\begin{align}
			\begin{pmatrix}
				f_{2} (1270)			\\
				f_{2}^{\prime} (1525)
			\end{pmatrix}
			=
			\begin{pmatrix}
				\cos \beta_{a_2}		&	\sin \beta_{a_2}	\\
				- \sin \beta_{a_2}	&	\cos \beta_{a_2}
			\end{pmatrix}
			\begin{pmatrix}
				f_{2 , N}	\\
				f_{2 , S}
			\end{pmatrix} \, ,
		\end{align}
	where $\beta_{a_2} = 5.7^{\circ}$ is the small mixing angle reported in the PDG \cite{Zyla:2020zbs}, in agreement with the fact that tensor mesons belong to a homochiral multiplet. The decays of tensor mesons were studied in great detail in Refs.\ \cite{Giacosa:2005bw,Burakovsky:1997ci} and fit very well into the $\bar{q} q$ picture \cite{Godfrey:1985xj}.

\paragraph*{Axial-vector mesons}

	The choice $L = S = 1$ allows also for the $J^{PC} = 1^{++}$ $(n^{2 S + 1} L_J = 1^{3} P_1)$ axial-vector nonet $A_1$, which contains the resonances \{$a_1 (1260)$, $K_{1 , A}$, $f_1 (1285)$, $f_1 (1420)$\} which are linked to the vector mesons mentioned above by chiral transformations, see \textit{e.g.}\ Refs.\ \cite{Hatanaka:2008gu,Cheng:2011pb} (building a homochiral multiplet). The nonet matrix, whose microscopic currents are $A_{1 , ij}^{\mu} \equiv \bar{q}_j \, \gamma^{5} \gamma^{\mu} \, q_i / \sqrt{2}$, reads:
		\begin{align}
			A_1^{\mu} = \tfrac{1}{\sqrt{2}}
			\begin{pmatrix}
				\frac{f_{1 , N}^{\mu} + a_1^{0 \mu}}{\sqrt{2}}	& a_1^{+ \mu}										&	K_{1 , A}^{+ \mu}	\\
				a_1^{- \mu}									& \frac{f_{1 , N}^{\mu} - a_1^{0 \mu}}{\sqrt{2}}	&	K_{1 , A}^{0 \mu}	\\
				K_{1 , A}^{-\mu}									&	\bar{K}_{1 , A}^{0 \mu}							&	f_{1 , S}^{\mu}
			\end{pmatrix} \, .	\label{eq:avector_nonet}
		\end{align}
	For the isoscalar sector we find,
		\begin{align}
			\begin{pmatrix}
				f_1 (1285)	\\
				f_1 (1420)
			\end{pmatrix}
			=
			\begin{pmatrix}
				\cos \beta_{a_1}		&	\sin \beta_{a_1}	\\
				- \sin \beta_{a_1}	&	\cos \beta_{a_1}
			\end{pmatrix}
			\begin{pmatrix}
				f_{1 , N}	\\
				f_{1 , S}
			\end{pmatrix} \, .
		\end{align}
	The mixing angle $\beta_{a_1}$ is expected to be small, as the homochiral nature of the multiplet \cite{Giacosa:2017pos} and the decay properties \cite{Divotgey:2013jba} suggest. We set $\beta_{a_1} = 0$ for simplicity (an anyhow small mixing angle would not affect our results very much).
	
	It is important to note at this point, that the kaonic states $K_{1 , A}$ and $K_{1 , B}$ mix. For a study of the axial-vector and pseudovector nonets, with focus on the $K_1 (1270) / K_1 (1400)$ system, by using an approach similar to the one used in this work, see Ref.\ \cite{Divotgey:2013jba}. The main result concerning the kaonic mixing reads:
		\begin{align}
			\begin{pmatrix}
				K_1 (1270)	\\
				K_1 (1400)
			\end{pmatrix}
			=
			\begin{pmatrix}
				\cos \varphi_{K}				&	- \mathrm{i} \sin \varphi_{K}	\\
				- \mathrm{i} \sin \varphi_{K}	&	\cos \varphi_{K}
			\end{pmatrix}
			\begin{pmatrix}
				K_{1 , A}	\\
				K_{1 , B}
			\end{pmatrix} \, .	\label{mixk1abfields}
		\end{align}
	Equivalently, in terms of kets and the angle $\theta_{K} = \varphi_{K} + 90^{\circ}$ typically used in the literature \cite{Yang:2007zt,Hatanaka:2008gu,Cheng:2011pb}:
		\begin{align}
			\begin{pmatrix}
				| K_1^{+} (1270) \rangle	\\
				| K_1^{+} (1400) \rangle
			\end{pmatrix}
			=
			\begin{pmatrix}
				\sin \theta_{K}	&	\cos \theta_{K}		\\
				\cos \theta_{K} &	- \sin \theta_{K}
			\end{pmatrix}
			\begin{pmatrix}
				| K_{1 , A}^{+} \rangle	\\
				| K_{1 , B}^{+} \rangle
			\end{pmatrix} \, .
		\end{align}
	The final result reads $\varphi_{K} = ( 56.4 \pm 4.3 )^{\circ}$, or equivalently $\theta_{K} = ( - 33.6 \pm 4.3 )^{\circ}$, hence $K_1 (1270)$ is predominantly $K_{1 , B}$ and $K_1 (1400)$ predominantly $K_{1 , A}$, yet the mixing is still large.
	
\paragraph*{Scalar mesons}

	For completeness, we recall that $L = S = 1$ yields also the scalar states $J^{PC} = 0^{++}$ $(n^{2 S + 1} L_J = 1^{3} P_{0})$, corresponding to the chiral partners of the pseudoscalar mesons. The scalars are subject to an ongoing controversial debate on the correct assignment of measured states. The identification is still uncertain, but the set
		\begin{align}
			\{ a_{0} (1450), K_{0}^{\ast} (1430), f_{0} (1710), f_{0} (1370) \} \, ,	\label{eq:scalar_nonet}
		\end{align}
	seems to be favored (for the description of mixing with \textit{e.g.}\ the long-searched scalar glueball, see \textit{e.g.}\ Refs. \cite{Giacosa:2005zt,Giacosa:2005qr,Amsler:1995tu,Amsler:1995td,StrohmeierPresicek:1999yv,Lee:1999kv,Cheng:2006hu,Gutsche:2016wix} and Refs.\ therein). The reason why this nonet is so peculiar is due to the fact that it has the quantum numbers of the QCD vacuum, hence condensation of the scalar-isoscalar fields should take place. The corresponding nonet $S$, whose microscopic currents are $S_{ij} \equiv ( \bar{q}_j \, q_i ) / \sqrt{2}$, is not included in this work, since no decays of spin-$3$ states into scalars can be realized in our model approach, see Sub.Sec.\ \ref{subsec:remarks_on_exclusion_of_scalars}.
	
	Moreover, light scalar four-quark states (in the form and combination of tetraquark/molecular/companion poles) below $1 \, \,$GeV also exist \cite{Zyla:2020zbs}. Nevertheless, they are also not included our work.

\paragraph*{Higher-spin mesons}

	Besides the peculiar case of the scalar nonet, all other nonets mentioned above are well-established. Yet, what about nonets with $L = 2$? For $L = 2$ four nonets can be constructed, one of which is the main subject of this work. Thus, even if, besides spin-$3$ fields, they do not enter into the Lagrangian of this work, since they are too heavy, they allow us to put the nonet of mesons with $J = 3$ into the correct physical framework. Hence, we briefly discuss them.
	
	The easiest choice is $L=2$ and $S=0,$ which implies $J^{PC} = 2^{-+}$ $(n^{2 S + 1} L_J = 1^{1} D_{2})$: these so-called pseudotensor states with currents $[ \bar{q}_j \, ( \mathrm{i} \gamma^{5} \, \partial^{\mu} \partial^{\nu} + \ldots ) \, q_i ] / \sqrt{2}$ are assigned to \{$\pi_{2} (1670)$, $K_{2} (1770)$, $\eta_{2}^{\prime} (1870)$, $\eta_{2} (1645)$\} \cite{Wang:2014sea,Bing:2013fva,Kuhn:2004en,Koenigstein:2016tjw}. Moreover, as recently described in Ref.\ \cite{Giacosa:2017pos}, the chiral anomaly may also be important for this nonet, explaining a potential large mixing between strange and nonstrange components, leading to the isoscalar states $\eta_{2} (1650)$ and $\eta_{2}^{\prime} (1870)$. The details of the need for a large mixing angle can be found in Ref.\ \cite{Koenigstein:2016tjw}, in which strong decays of these mesons are studied in a similar relativistic effective model.
	
	For $L = 2$ and $S = 1$ one can construct the orbitally excited vector $J^{PC} = 1^{--}$ ($n^{2 S + 1} L_J = 1^{3} D_{1}$) states \{$\rho (1700)$, $K^{\ast} (1680)$, $\phi(????)$, $\omega(1650)$\}, which are the (hetero) chiral partners of the pseudovector mesons \cite{Giacosa:2016hrm,Giacosa:2017pos}. Microscopically, the elements are $( \bar{q}_j \mathrm{i} \partial^{\mu} q_i ) / \sqrt{2}$. The predominately strange member of the nonet $\phi(????)$ could not be identified yet. For predictions of a possible $\phi(1930)$ meson see Ref.\ \cite{Piotrowska:2017rgt} and Refs.\ therein, in which a flavor-symmetry-based QFT model is employed for two nonets of excited vector mesons.
	
	The quantum numbers $L = 2$ and $S = 1$ may also lead to a axial tensor nonet with $J^{PC} = 2^{--}$ $(n^{2 S + 1} L_J = 1^{3} D_{2})$ with currents $[ \bar{q}_j \, ( \gamma^{5} \, \gamma^{\mu} \, \partial^{\nu} + \ldots ) \, q_i ] / \sqrt{2}$, which is however poorly understood. In fact, only the state $K_{2}(1820)$ is identified and listed by the PDG \cite{Zyla:2020zbs}. This nonet builds together with the $2^{++}$ tensor mesons a homochiral multiplet.
	
	The latter two nonets are omitted in this work, because their members are too heavy.
	
\paragraph*{Spin-3 mesons}
	
	Finally and most importantly, the choice $L = 2$ and $S = 1$ can also lead the nonet of $J^{PC} = 3^{--}$ $(n^{2 S + 1} L_J = 1^{3} D_{3})$ states \{$\rho_{3} (1690)$, $K_{3}^{\ast} (1780)$, $\phi_{3} (1850)$, $\omega_{3} (1670)$\}. The nonet with elements $W_{ij}^{\mu\nu\rho} = [ \bar{q}_j \, ( \gamma^{\mu} \, \partial^{\nu} \partial^{\rho} + \ldots ) \, q_i ]/\sqrt{2}$ reads,
		\begin{align}
			W_3^{\mu\nu\rho} = \tfrac{1}{\sqrt{2}}
			\begin{pmatrix}
				\frac{\omega_{3 , N}^{\mu\nu\rho} + \rho_{3}^{0 \mu\nu\rho}}{\sqrt{2}}	&	\rho_{3}^{+ \mu\nu\rho}													&	K_{3}^{+ \mu\nu\rho}		\\
				\rho_{3}^{- \mu\nu\rho}													&	\frac{\omega_{3 , N}^{\mu\nu\rho} - \rho_{3}^{0 \mu\nu\rho}}{\sqrt{2}}	&	K_{3}^{0 \mu\nu\rho}		\\
				K_{3}^{- \mu\nu\rho}													&	\bar{K}_{3}^{0 \mu\nu\rho}												&	\omega_{3 , S}^{\mu\nu\rho}
			\end{pmatrix} \, .	\label{eq:tensor3_nonet}
		\end{align}
	In this case the mixing angle
		\begin{align}
			\begin{pmatrix}
				\omega_{3} (1670)	\\
				\phi_{3} (1850)
			\end{pmatrix}
			=
			\begin{pmatrix}
				\cos \beta_{w_3}		&	\sin \beta_{w_3}	\\
				- \sin \beta_{w_3}	&	\cos \beta_{w_3}
			\end{pmatrix}
			\begin{pmatrix}
				\omega_{3 , N}	\\
				\omega_{3 , S}
			\end{pmatrix}
		\end{align}
	is $\beta_{w_3} = 3.5^{\circ}$ \cite{Zyla:2020zbs}. The small mixing is also in agreement with the homochiral nature of the corresponding chiral multiplet in which this nonet is embedded \cite{Giacosa:2017pos}. In the next section we couple this nonet to the other nonets listed in Table \ref{tab:transformations}.
	
	\begin{table}
		\centering
		\renewcommand{\arraystretch}{1.5}
		\caption{\label{tab:resonances_states} %
			Assignment of physical resonances to $\bar{q} q$ states in the model. $^{\dagger}$The isospin $1/2$ states in the pseudovector and axial-vector sectors are strong mixtures of $K_1(1270)$ and $K_1(1400)$, see Ref.\ \cite{Zyla:2020zbs} and Refs.\ therein.
		}
		\begin{ruledtabular}
			\begin{tabular}[c]{ c | l | l | c }
				\multirow{2}{*}{$J^{PC}$}	&	\multicolumn{1}{c|}{physical}	&	\multicolumn{1}{c|}{\multirow{2}{*}{nonet $\bar{q}q$ states}}						&	mixing\\
											&	\multicolumn{1}{c|}{resonances}	&																						&	angles
				\\	
				\toprule
				\multirow{4}{*}{$0^{-+}$}	&	$\pi$							&	$\hphantom{-} \pi$																	&	\\
											&	$K$								&	$\hphantom{-} K$																	&	$\beta_{p} = - 43.4^{\circ}$	\\
											&	$\eta$					&	$\hphantom{-} \eta_{N} \cos\beta_{p} + \eta_{S} \sin \beta_{p}$								&	Ref.\ \cite{AmelinoCamelia:2010me}	\\
											&	$\eta^{\prime} (958)$			&	$- \eta_{N} \sin \beta_{p} + \eta_{S} \cos \beta_{p}$								&
				\\
				\colrule
				\multirow{4}{*}{$1^{--}$}	&	$\rho (770)$					&	$\hphantom{-} \rho_1$																&	\\
											&	$K^{\ast} (892)$				&	$\hphantom{-} K_1^{\ast}$															&	$\beta_{v_1} = - 3.9^{\circ}$	\\
											&	$\omega (782)$					&	$\hphantom{-} \omega_{1 , N} \cos \beta_{v_1} + \omega_{1 , S} \sin \beta_{v_1}$	&	Ref.\ \cite{Zyla:2020zbs}	\\
											&	$\phi (1020)$					&	$- \omega_{1 , N} \sin \beta_{v_1} + \omega_{1 , S} \cos \beta_{v_1}$				&
				\\
				\colrule
				\multirow{4}{*}{$1^{+-}$}	&	$b_1 (1235)$					&	$\hphantom{-} b_1$																	&	\\
											&	$K_{1 , B}^{\dagger}$				&	$\hphantom{-} K_{1 , B}$																&	$\beta_{b_1}$ unknown	\\
											&	$h_1 (1170)$					&	$\hphantom{-} h_{1 , N} \cos \beta_{b_1} + h_{1 , S} \sin \beta_{b_1}$				&	$0^{\circ}$ and $- 40^{\circ}$	\\
											&	$h_1 (1415)$					&	$ - h_{1 , N} \sin \beta_{b_1} + h_{1 , S} \cos \beta_{b_1}$						&	are tested
				\\
				\colrule
				\multirow{4}{*}{$1^{++}$}	&	$a_1 (1260)$					&	$\hphantom{-} a_1$																	&	\\
											&	$K_{1 , A}^{\dagger}$				&	$\hphantom{-} K_{1 , A}$																&	$\beta_{a_1} \simeq 0$	\\
											&	$f_1 (1285)$					&	$\hphantom{-} f_{1 , N} \cos \beta_{a_1} + f_{1 , S} \sin \beta_{a_1}$				&	Refs.\ \cite{Divotgey:2013jba,Giacosa:2017pos}	\\
											&	$f_1 (1420)$					&	$- f_{1 , N} \sin \beta_{a_1} + f_{1 , S} \cos \beta_{a_1}$							&
				\\
				\colrule
				\multirow{4}{*}{$2^{++}$}	&	$a_{2} (1320)$					&	$\hphantom{-} a_{2}$																&	\\
											&	$K_{2}^{\ast} (1430)$			&	$\hphantom{-} K_{2}^{\ast}$															&	$\beta_{a_2}=5.7^{\circ}$	\\
											&	$f_{2} (1270)$					&	$\hphantom{-} f_{2 , N} \cos \beta_{a_2} + f_{2 , S} \sin \beta_{a_2}$				&	Ref.\ \cite{Zyla:2020zbs}	\\
											&	$f_{2}^{\prime} (1525)$			&	$- f_{2 , N} \sin \beta_{a_2} + f_{2 , S} \cos \beta_{a_2}$							&
				\\
				\colrule
				\multirow{4}{*}{$3^{--}$}	&	$\rho_{3} (1690)$				&	$\hphantom{-} \rho_{3}$																&	\\
											&	$K_{3}^{\ast} (1780)$			&	$\hphantom{-}K_{3}^{\ast}$															&	$\beta_{w_3}=3.5^{\circ}$	\\
											&	$\omega_{3} (1670)$				&	$\hphantom{-} \omega_{3 , N} \cos \beta_{w_3} + \omega_{3 , S} \sin \beta_{w_3}$	&	Ref.\ \cite{Zyla:2020zbs}	\\
											&	$\phi_{3} (1850)$				&	$-\omega_{2 , N}\sin\beta_{w_3}+\omega_{2 , S}\cos\beta_{w_3}$						&
			\end{tabular}
		\end{ruledtabular}
	\end{table}

\subsection{Effective mesonic interactions}

	In this section, we present the effective mesonic interactions and discuss the derivation of the tree level decay widths from our model.

\subsubsection{The effective action}

	Using the nonets introduced in the previous subsection, we construct the effective Lagrangian describing the strong decays of spin-$3$ tensor mesons as follows,
		\begin{align}
			\mathcal{L}_{W,\text{total}} = \, & \mathcal{L}_{\text{mass}} + \mathcal{L}_{\text{kin}} + \mathcal{L}_{w_3 p p} + \mathcal{L}_{w_3 v_1 p} +	\vphantom{\bigg(\bigg)}	\label{lagtot}
			\\
			& + \mathcal{L}_{w_3 a_2 p} + \mathcal{L}_{w_3 a_1 p} + \mathcal{L}_{w_3 b_1 p} + \mathcal{L}_{w_3 v_1 v_1} \, ,	\vphantom{\bigg(\bigg)}	\nonumber
		\end{align}
	where $\mathcal{L}_{\text{kin}} = \frac{1}{2} \, \mathrm{tr} \, ( \partial_{\mu} P )^2 + \ldots$ contains the usual kinetic terms and $\mathcal{L}_{\text{mass}}$ contains all the quadratic mass terms describing the masses of all relevant nonets. In this work, all masses are taken from the PDG \cite{Zyla:2020zbs} and assumed to be exact, while some mixing angles are derived from the masses following the \textit{quark model review} of the PDG \cite{Zyla:2020zbs}, see also App.~\ref{app:masses_and_mixing}. The other terms describing the decays are listed explicitly in Table \ref{tab:interlag}. The quantities $g_{w_3 \circ \circ}$ are the coupling constants, which are fitted to experimental results, $\varepsilon^{\mu\nu\rho\sigma}$ is the antisymmetric Levi-Civita pseudotensor, $[ \circ , \, \circ ]_{-}$ stands for the commutator, and $\{ \circ , \, \circ \}_{+}$ for the anticommutator.

	\begin{table*}
		\centering
		\renewcommand{\arraystretch}{1.5}
		\caption{\label{tab:interlag} %
			Effective relativistic interaction terms describing the strong decays of mesons with $J^{PC} = 3^{--}$.
		}
			\begin{tabular}[c]{c @{\qquad} c }
				\toprule
				decay mode 								& interaction Lagrangians
				\\
				\colrule
				$3^{--} \rightarrow 0^{-+} + 0^{-+}$	&	$\mathcal{L}_{w_3 p p} = g_{w_3 p p} \, \mathrm{tr} \big[ W_3^{\mu\nu\rho} \, \big[ P \, , ( \partial_{\mu} \partial_{\nu} \partial_{\rho} P ) \big]_{-} \big]$
				\\
				$3^{--} \rightarrow 0^{-+} + 1^{--}$	&	$\mathcal{L}_{w_3 v_1 p} = g_{w_3 v_1 p} \, \varepsilon^{\mu\nu\rho\sigma} \, \mathrm{tr} \big[ W_{3 , \mu\alpha\beta} \, \big\{ ( \partial_\nu V_{1 , \rho} ) \, , ( \partial^{\alpha} \partial^{\beta} \partial_{\sigma} P ) \big\}_{+} \big]$
				\\
				$3^{--} \rightarrow 0^{-+} + 2^{++}$	&	$\mathcal{L}_{w_3 a_2 p} = g_{w_3 a_2 p} \, \varepsilon_{\mu\nu\rho\sigma} \, \mathrm{tr} \big[ \tensor{W}{_3^\mu_\alpha_\beta} \big[ ( \partial^{\nu} A_2^{\rho\alpha} ) \, , ( \partial^{\sigma} \partial^{\beta} P ) \big]_{-} \big]$
				\\
				$3^{--} \rightarrow 0^{-+} + 1^{+-}$	&	$\mathcal{L}_{w_3 b_1 p} = g_{w_3 b_1 p} \, \mathrm{tr} \big[ W_3^{\mu\nu\rho} \big\{ B_{1 , \mu} \, , ( \partial_{\nu} \partial_{\rho} P ) \big\}_{+} \big]$
				\\
				$3^{--} \rightarrow 0^{-+} + 1^{++}$	&	$\mathcal{L}_{w_3 a_1 p} = g_{w_3 a_1 p} \, \mathrm{tr} \big[ W_3^{\mu\nu\rho} \big[ A_{1 , \mu} \, , ( \partial_{\nu} \partial_{\rho} P ) \big]_{-} \big]$
				\\
				$3^{--} \rightarrow 1^{--} + 1^{--}$	&	$\mathcal{L}_{w_3 v_1 v_1} = g_{w_3 v_1 v_1} \, \mathrm{tr} \big[ W_3^{\mu\nu\rho} \big[ ( \partial_{\mu} V_{1 , \nu} ) \, , V_{1 , \rho} \big]_{-} \big]$
				\\
				\botrule
			\end{tabular}
	\end{table*}

	All the interaction Lagrangians in Table \ref{tab:interlag} are invariant under $CPT$- , Poincar\'{e}- and flavor transformations, listed in Table \ref{tab:transformations}. We considered only couplings involving a minimal number of derivatives within each single interaction term. This strategy turned to be successful in previous works, \textit{e.g.}\ Refs.\ \cite{Giacosa:2005bw,Divotgey:2013jba,Giacosa:2005zt,Giacosa:2005qr,Koenigstein:2016tjw,Piotrowska:2017rgt}\footnote{Note, Ref.\ \cite{Wang:2016enc} included higher derivative couplings and form factors in their approach to strong decays of spin-$3$ mesons involving charm quarks.}. Later on, in Chap.\ \ref{chap:discussion}, we shall also argue how to justify this strategy using insights from the extended linear sigma model (eLSM) and the Functional Renormalization Group (FRG) approaches to low energy QCD. The explicit form of the Lagrangians (after having carried out the traces) in Table \ref{tab:interlag} are reported in App.\ \ref{app:lagrangians}. Each Lagrangian represents the dominant contribution in the large-$N_\mathrm{c}$ expansion in the given channel, hence each coupling constant scales as $1/\sqrt{N_\mathrm{c}}$.

\subsubsection{Decay width}
	
	The tree level decay widths have the following general form \cite{Zyla:2020zbs,Peskin:1995ev},
		\begin{align}
			& \Gamma_{W_3 \rightarrow A + B} ( m_{w_3} , m_{a} , m_{b} ) =	\vphantom{\frac{\big| \vec{k}_{a , b} \big|}{8 \uppi \, m_{w_3}^2}}		\label{eq:decay}
			\\
			= \, & \frac{\big| \vec{k}_{a , b} \big|}{8 \uppi \, m_{w_3}^2} \, | \mathcal{M} |^2 \, \kappa_i \, \Theta ( m_{w_3} - m_{a} - m_{b} ) \, ,	\nonumber
		\end{align}
	where $m_{w_3}$ is the mass of a (decaying) spin-$3$ particle, while $m_{a}$ and $m_{b}$ are the masses of the decay products ``$A$'' and ``$B$'', $\Theta(x)$ denotes the Heaviside step function, and the modulus of the outgoing particles momentum has the following analytic expression,
		\begin{align}
			| \vec{k}_{a , b} | \equiv \frac{1}{2 \, m_{w_3}} \sqrt{ ( m_{w_3}^2 - m_{a}^2 - m_{b}^2 )^2 - 4 \, m_{a}^2 \, m_{b}^2 } \, .	\label{eq:kf}
		\end{align}
	We obtain the factors $\kappa_i$ in Eq.\ \eqref{eq:decay} for the $i$-th decay channel from the explicit forms of the Lagrangians in the tables in App.\ \ref{app:lagrangians} by considering the square of the coefficients for a given decay channel as well as eventual sum over members of the same isospin multiplet. The $\kappa_i$ for all interaction Lagrangians and all relevant channels are presented in App.\ \ref{app:tables}. The decay amplitudes $|\mathcal{M}|^2$ are derived via Feynman rules under the use of the polarization vectors and tensors as well as their corresponding completeness relations in App.\ \ref{app:deacyamplitudes}. The results for $\frac{1}{7} |\mathcal{M}|^2$ are listed in\footnote{Since the decaying particles are spin-$3$ mesons we multiply by the factor $\frac{1}{7}$, averaging over the different spin states.} Table \ref{tab:amplitudes}.
	
	\begin{table*}
		\centering
		\renewcommand{\arraystretch}{1.5}
		\caption{\label{tab:amplitudes} %
			Decay amplitudes for different decay modes.
		}
		\begin{tabular}[c]{ c @{\qquad} c }
			\toprule
			decay mode								&	$\frac{1}{7} \, | \mathcal{M} |^2$
			\\
			\colrule
			$3^{--} \rightarrow 0^{-+} + 0^{-+}$	&	$g_{w_3 p p}^2 \, \frac{2}{35} \, \big| \vec{k}_{p^{(1)},p^{(2)}} \big|^{6}$
			\\
			$3^{--} \rightarrow 0^{-+} + 1^{--}$	&	$g_{w_3 v_1 p}^2 \, \frac{8}{105} \, \big| \vec{k}_{v_1 , p} \big|^{6} \, m_{w_3}^2$
			\\
			$3^{--} \rightarrow 0^{-+} + 2^{++}$	&	$g_{w_3 a_2 p}^2 \, \frac{2}{105} \, \big| \vec{k}_{a_2 , p} \big|^4 \, \frac{m_{w_3}^2}{m_{a_2}^2} \, \big( 2 \, \big| \vec{k}_{a_2 , p}	\big|^2 + 7 \, m_{a_2}^2 \big)$
			\\
			$3^{--} \rightarrow 0^{-+} + 1^{+-}$	&	$g_{w_3 b_1 p}^2 \, \frac{2}{105} \, \big| \vec{k}_{b_1 , p} \big|^4 \, \Big( 7 + 3 \, \frac{| \vec{k}_{b_1 , p} |^2}{m_{b_1}^2} \Big)$
			\\
			$3^{--} \rightarrow 0^{-+} + 1^{++}$	&	$g_{w_3 a_1 p}^2 \, \frac{2}{105} \, \big| \vec{k}_{a_1 , p} \big|^4 \, \Big( 7 + 3 \, \frac{| \vec{k}_{a_1 , p} |^2}{m_{a_1}^2} \Big)$
			\\
			$3^{--} \rightarrow 1^{--} + 1^{--}$	&	$g_{w_3 v_1 v_1}^{2} \, \frac{1}{105} \, \big( m_{v_1^{(1)}}^{2} \, m_{v_1^{(2)}}^{2} \big)^{-1} \, \big| \vec{k}_{v_1^{(1)} , v_2^{(2)}} \big|^{2} \, \big[ 6 \, | \vec{k}_{v_1^{(1)} , v_1^{(2)}} \big|^{4} + 35 \, m_{v_1^{(1)}}^{2} \,  m_{v_1^{(2)}}^{2} + 14 \, \big| \vec{k}_{v_1^{(1)} , v_1^{(2)}} \big|^{2} \, \big( m_{v_1^{(1)}}^{2} + m_{v_1^{(2)}}^{2} \big)  \big]$
			\\
			\botrule
		\end{tabular}
	\end{table*}

\subsubsection{Remarks on the exclusion of scalar mesons}
\label{subsec:remarks_on_exclusion_of_scalars}

	As a final side remark, we note that we cannot couple the spin-$3$ nonet \eqref{eq:tensor3_nonet} to scalar mesons \eqref{eq:scalar_nonet}. In fact, if we try to couple them to scalars and pseudoscalars, the only possible CPT- and flavor-invariant interaction term,
		\begin{align}
			\varepsilon_{\mu\nu\rho\sigma} \, \mathrm{tr} \big( \partial^{\mu} \tensor{W}{_3^\nu_\alpha_\beta} \, \big[ ( \partial^{\rho} S ) , \, ( \partial^{\sigma} \partial^{\alpha} \partial^{\beta} P ) \big]_{-} \big) = 0 \, ,	\vphantom{\bigg(\bigg)}
		\end{align}
	vanishes identically since it involves the contraction of antisymmetric and symmetric tensors. Correspondingly, scalar mesons do not enter our study, which agrees with the experimental results reported in the PDG \cite{Zyla:2020zbs}. This is also of direct advantage for our work, since we do not have to deal with the identification of the scalar states, which, as already mentioned, is a long-standing and yet unsolved issue of low-energy QCD \cite{Amsler:1995tu,Amsler:1995td,StrohmeierPresicek:1999yv,Lee:1999kv,Cheng:2006hu,Gutsche:2016wix}.
	
\subsubsection{Suppression of next to leading order terms}

	The Lagrangian terms considered in Table~\ref{tab:interlag} are large-$N_{\mathrm{c}}$ (and thus Okubo-Zweig-Iizuka) dominant and symmetric under $U_\mathrm{V} ( 3 )$. As we shall discuss later on, this approximation is expected to be acceptable in view of the precision of the experimental data presently available. Yet, by considering the $W_3 B_1 P$ interaction as an example, the Lagrangian $\mathcal{L}_{w_3 b_1 p}$ can be regarded as the first term of an expansion in large-$N_{\mathrm{c}}$ and/or symmetry breaking terms that takes the form:
		\begin{align}
			& \mathcal{L}_{w_3 b_1 p}^{\text{full}} =	\vphantom{\bigg(\bigg)}
			\\
			= \, & g_{w_3 b_1 p} \, \mathrm{tr} \big[ W^{\mu \nu \rho} \big\{ B_{1 \mu} \, , ( \partial _\mu \partial_\rho P ) \big\}_+ \big] +	\vphantom{\bigg(\bigg)}	\nonumber
			\\
			& + g_{w_3 b_1 p}^{(2)} \, \mathrm{tr} \big[ W^{\mu \nu \rho} \big] \, \mathrm{tr} \big[ B_{1 \mu } \, ( \partial _\mu \partial_\rho P ) \big] +	\vphantom{\bigg(\bigg)}	\nonumber
			\\
			& + g_{w_3 b_1 p}^{(3)} \, \mathrm{tr} \big[ W^{\mu \nu \rho} \big] \, \mathrm{tr} \big[ B_{1 \mu } \big] \, \mathrm{tr} \big[ \partial _\mu \partial_\rho P \big] +	\vphantom{\bigg(\bigg)}	\nonumber
			\\
			& + g_{w_3 b_1 p}^{(4)} \, \mathrm{tr} \big[ \hat{\delta} \, W^{\mu \nu \rho} \, \big\{ B_{1 \mu } \, , ( \partial_\mu \partial_\rho P ) \big\}_+ \big] +	\vphantom{\bigg(\bigg)}	\nonumber
			\\
			& + g_{w_3 b_1 p}^{(5)} \, \mathrm{tr} \big[ \hat{\delta} \, W^{\mu \nu \rho} \big] \, \mathrm{tr} \big[ B_{1 \mu } \, ( \partial_\mu \partial_\rho P ) \big] +	\vphantom{\bigg(\bigg)}	\nonumber
			\\
			& + g_{w_3 b_1 p}^{(6)} \, \mathrm{tr} \big[ W^{\mu \nu \rho} \big] \, \mathrm{tr} \big[ \hat{\delta} \, B_{1 \mu } \, ( \partial_\mu \partial_\rho P ) \big] +	\vphantom{\bigg(\bigg)}	\nonumber
			\\
			& + \ldots	\vphantom{\bigg(\bigg)}	\nonumber
		\end{align}
	The first term, reported in Table~\ref{tab:interlag}, is a flavor symmetric term and scales as $g_{w_3 b_1 p} \propto N_{\mathrm{c}}^{-1/2},$ hence it is the dominant term in a large-$N_{\mathrm{c}}$ expansion. The second and the third term are also flavor symmetric, but scale as $g_{w_3 b_1 p}^{(2)}\propto N_{\mathrm{\mathrm{c}}}^{- \frac{3}{2}}$ and $g_{w_3 b_1 p}^{(3)} \propto N_{\mathrm{c}}^{- \frac{5}{2}}$, respectively. They involve gluon exchanges and are therefore suppressed. The fourth term is not large-$N_{\mathrm{c}}$ suppressed, but it breaks flavor symmetry via the matrix
		\begin{align}
			\hat{\delta} = \mathrm{diag} \{ 0, \delta_{d}, \delta_s \} \, .
		\end{align}
	Isospin violation is proportional to $\delta_{d} \propto m_{d} - m_{u}$ and is expected to be very small, yet the breaking due to the $s$ quark $\delta_s \propto m_s - m_{u}$ can be nonnegligible following Ref.~\cite{Amsler:1995td,Amsler:1995tu}, it could be along the order of $g_{w_3 b_1 p}^{(4)} \simeq g_{w_3 b_1 p} \, \mathrm{diag} \{ 0, 0, 0.1 \, - \, 0.2 \}$, but the actual value for $J = 3$ mesons should be determined by an independent fit to data. It is however expected to be sufficiently small to be neglected in this work. At this stage, the first large-$N_{\mathrm{c}}$ correction $g_{w_3 b_1 p}^{(2)}$ and the first flavor symmetric correction $g_{w_3 b_1 p}^{(4)}$ are expected to be of the same intensity and should be the first to be included. Further terms $g_{w_3 b_1 p}^{( n \geq 5 )}$ are both large-$N_{\mathrm{c}}$ subdominant and flavor suppressed, thus are regarded to be very small.
		
	The same analysis can be carried out for all the interaction terms, yet it is interesting to observe that the second and the third term would vanish whenever the commutator is present. (Flavor-symmetry violation is expected to be the main next-to-leading-order contribution for those interaction terms).

\section{Phenomenology of the \texorpdfstring{$J^{PC} = 3^{--}$}{JPC 3--} nonet}
\label{sec:results}

	In this section we present our results for decay rates and branching ratios of the $J^{PC} = 3^{--}$ mesons. In each subsection and for each interaction Lagrangian term, we compare the experimental data to our theoretical results.
	
	We recall that the total decay widths of the $J^{PC} = 3^{--}$ mesons under consideration are \cite{Zyla:2020zbs},
		\begin{align}
			\Gamma_{\rho_{3} (1690)}^{\text{tot}} = \, & ( 161 \pm 10 ) \, \, \text{MeV} \, ,	\vphantom{\bigg(\bigg)}
			\\
			\Gamma_{K_{3}^{\ast} (1780) }^{\text{tot}} = \, & ( 159 \pm 21 ) \, \, \text{MeV} \, ,	\vphantom{\bigg(\bigg)}
			\\
			\Gamma_{\omega_{3} (1670)}^{\text{tot}} = \, & ( 168 \pm 10 ) \, \, \text{MeV} \, ,	\vphantom{\bigg(\bigg)}
			\\
			\Gamma_{\phi_{3} (1850)}^{\text{tot}} = \, & ( 87_{-23}^{+28} ) \, \, \text{MeV} \, .	\vphantom{\bigg(\bigg)}
		\end{align}
	We shall verify that the sum of all the single decay channels from our theoretical calculations never overshoots these values.
	
	For what concerns the accuracy of our results, whenever possible we determine the coupling and its error via a simple fit of experimental decay width and ratios. We ignore the experimental uncertainties for all masses of the particles of the model and assume the masses to be exact, because their errors are small and they are of minor importance for the overall errors. The experimental errors for the decay widths and branching ratios as well as the systematic errors in the model are much larger, which justifies this approximation. Yet, the quoted errors represents a lower(!) limit of the actual error of this work, since other indeterminacy features are inherent in the approximations, that lead us to the effective action, where only those terms are included, which are expected to be dominant in terms of large-$N_\mathrm{c}$, flavor- and momentum-space expansions. Moreover, the decay widths are calculated at tree level. Since the width/mass ratio for the decaying resonances is rather small, contributions due to loops are expected to be negligible \cite{Giacosa:2007bn,Schneitzer:2014rsa}, if seen from a perturbative perspective. Otherwise, if the interaction terms are interpreted as effective couplings in a full quantum effective IR action, calculations have to be performed at tree level anyhow.
	
\subsection{Decay process \texorpdfstring{$W_3 \rightarrow P + P$}{W3 -> P + P}}

	The effective interaction term describing the decay of spin-$3$ tensor mesons into two pseudoscalar mesons has the following form
		\begin{align}
			\mathcal{L}_{w_3 p p} = g_{w_3 p p} \, \mathrm{tr} \big[ W_3^{\mu\nu\rho} \, \big[ P , \, ( \partial_{\mu} \partial_{\nu} \partial_{\rho} P ) \big]_{-} \big] \, .
		\end{align}
	Its extended version is listed in App.\ \ref{app:lagrangians} in Eq.\ \eqref{eq:extended_lagrangian_w3_p_p}.	Correspondingly, the tree level decay rate is
		\begin{align}
			& \Gamma_{W_3 \rightarrow P^{(1)} + P^{(2)}} ( m_{w_3} , m_{p^{(1)}} , m_{p^{(2)}} ) =	\vphantom{\Bigg(\Bigg)}	\label{eq:dec-wpp}
			\\
			= \, & g_{w_3 p p}^2 \, \frac{\big| \vec{k}_{p^{(1)} , p^{(2)}} \big|^{7}}{140 \uppi \,m_{w_3}^2} \, \kappa_i \, \Theta( m_{w_3} - m_{p^{(1)}} - m_{p^{(2)}} ) \, ,	\vphantom{\Bigg(\Bigg)}	\nonumber
		\end{align}
	compare Eq.\ \eqref{eq:amplitude_square_w3_p_p}. The factors $\kappa_i$ are reported in Table \ref{tab:kappa_w3_p_p}. In order to determine the coupling constant $g_{w_3 p p}$, we use the above formula and the following experimental data:
	\begin{enumerate}
		\item For $\rho_{3} (1690) \rightarrow \pi \, \pi$ one finds $\kappa_1 = 1$. Using the experimental result
			\begin{align}
				\Gamma^\mathrm{exp}_{\rho_{3} (1690) \rightarrow \pi \, \pi} = ( 38.0 \pm 3.2 ) \, \, \text{MeV} \, ,
			\end{align}
		one obtains a first determination of the coupling constant squared and its error. We shall denote them as $\tilde{g}_1^2$ and $\Delta \tilde{g}_1^2$.
		
		\item The experimental value
			\begin{align}
				\Gamma^\text{exp}_{\rho_{3} (1690) \rightarrow \bar{K} \, K} = ( 2.54 \pm 0.45 ) \, \, \text{MeV} \, ,
			\end{align}
		together with $\kappa_{2} = 2 \, \big( \frac{1}{2} \big)^{2}$ yields an independent determinations of $\tilde{g}_{2}^{2}$ and $\Delta \tilde{g}_{2}^{2}$.
		
		\item From
			\begin{align}
				\Gamma^\text{exp}_{K_{3}^{\ast} (1780) \rightarrow \pi \, \bar{K}} = ( 29.9 \pm 4.3 ) \, \, \text{MeV}
			\end{align}
		and $\kappa_{3} = \big( \frac{1}{2} \big)^{2} + \big(\frac{\sqrt{2}}{2} \big)^{2}$ we get $\tilde{g}_{3}^{2}$ and $\Delta \tilde{g}_{3}^{2}$.
		
		\item Finally, from
			\begin{align}
				\Gamma^\text{exp}_{K_{3}^{\ast} (1780) \rightarrow \bar{K} \, \eta} = ( 48 \pm 22 )\, \, \text{MeV}
			\end{align}
		and $\kappa_{4} = \big[ \frac{1}{2} \, \big( - \cos \beta_{p} + \sqrt{2} \sin \beta_{p} \big) \big]^{2}$ we obtain $\tilde{g}_{4}$ and $\Delta \, \tilde{g}_{4}$.
	\end{enumerate}
	The coupling constant and its indeterminacy are evaluated using a simple $\chi^{2}$-approach,\footnote{We are aware of the fact that some of the data might be correlated, because it is taken from similar measurements at the same experiments. Within the level of accuracy of our approach, there is no need to take care of these effects.}
		\begin{align}
			&	g_{w_3 p p}^{2} = \frac{\sum_{i = 1}^{4} \frac{\tilde{g}_i}{\Delta \tilde{g}_i^{2}}}{\sum_{j = 1}^{4} \frac{1}{\Delta \tilde{g}_j^{2}}} \, ,	&&	\Delta g_{w_3 p p}^{2} = \sqrt{\frac{1}{\sum_{j = 1}^{4} \frac{1}{\Delta g_j^{2}}}} \, ,
		\end{align}
	which results in
		\begin{equation}
			g_{w_3 p p}^{2} = ( 1.5 \pm 0.1 ) \cdot 10^{-10} \, \, \text{MeV}^{-4} \, .
		\end{equation}
	The comparison of theoretical and experimental results, which is obtained by using this value for the coupling constant, is reported in Table \ref{tab:w3pp}. A good overall agreement is obtained, but there is also a sizable mismatch: the experimental value for $K_{3}^{\ast} (1780) \rightarrow \bar{K} \, \eta$ is much larger than our theoretical prediction. Still, the experimental error is large and a better experimental determination would be interesting. Moreover, a noteworthy prediction concerning $\phi_{3} (1850) \rightarrow \bar{K} \, K$ is obtained. From the theoretical large prediction, we conclude that an experimental determination should be feasible.
	
	\begin{table}
		\centering
		\renewcommand{\arraystretch}{1.5}
		\caption{\label{tab:w3pp} %
			Decays of $J^{PC} = 3^{--}$ mesons into two pseudoscalars. Experimental data is taken from Ref.\ \cite{Zyla:2020zbs}.
		}
		\begin{ruledtabular}
			\begin{tabular}[c]{ l c c }
				\multicolumn{1}{c}{ \multirow{2}{*}{ decay process } }				&	theory				&	experiment
				\\
																					&	$\Gamma/$MeV		&	$\Gamma/$MeV
				\\
				\toprule
				$\rho_{3} (1690) \rightarrow \pi \, \pi$							&	$32.7 \pm 2.3$		&	$38.0 \pm 3.2$
				\\
				$\rho_{3} (1690) \rightarrow \bar{K} \, K$							&	$4.0 \pm 0.3$		&	$2.54 \pm 0.45$
				\\
				\colrule
				$K_{3}^{\ast} (1780) \rightarrow \pi \, \bar{K}$					&	$18.5 \pm 1.3$		&	$29.9 \pm 4.3$
				\\
				$K_{3}^{\ast} (1780) \rightarrow \bar{K} \, \eta$					&	$7.4 \pm 0.5$		&	$48 \pm 22$
				\\
				$K_{3}^{\ast} (1780) \rightarrow \bar{K} \, \eta^{\prime} (958)$	&	$0.021 \pm 0.001$	&
				\\
				\colrule
				$\omega_{3} (1670) \rightarrow\bar{K} \, K$							&	$3.0 \pm 0.2$		&
				\\
				\colrule
				$\phi_{3} (1850) \rightarrow \bar{K} \, K$							&	$18.8 \pm 1.3$		&	seen
			\end{tabular}
		\end{ruledtabular}
	\end{table}

\subsection{Decay process \texorpdfstring{$W_3 \rightarrow V_1 + P$}{W -> V + P}}

	The interaction Lagrangian for the vector and pseudoscalar decay mode reads
		\begin{align}
			& \mathcal{L}_{w_3 v_1 p} =	\vphantom{\bigg(\bigg)}	\label{eq:lag-wvp}
			\\
			= \, & g_{w_3 v_1 p} \, \varepsilon^{\mu\nu\rho\sigma} \, \mathrm{tr} \big[ W_{3 , \mu\alpha\beta} \, \big\{ ( \partial_\nu V_{1 , \rho} ) , \, ( \partial^{\alpha} \partial^{\beta} \partial_{\sigma} P ) \big\}_{+} \big] \, .	\vphantom{\bigg(\bigg)}	\nonumber
		\end{align}
	An extended version is given by Eq.\ \eqref{eq:extended_lagrangian_w3_v1_p} in App.\ \ref{app:lagrangians}.	In this case, the tree level decay rate formula has the form
		\begin{align}
			& \Gamma_{W_3 \rightarrow V_1 + P} ( m_{w_3} , m_{v_1} , m_{p} ) =	\vphantom{\Bigg(\Bigg)}	\label{eq:decay-gvp}
			\\
			= \, & g_{w_3 v_1 p}^{2} \, \frac{\big| \vec{k}_{v_1 , p} \big|^{7}}{105 \uppi} \, \kappa_i \, \Theta( m_{w_3} - m_{v_1} - m_{p} ) \, ,	\vphantom{\Bigg(\Bigg)}	\nonumber
		\end{align}
	see also Eq.\ \eqref{eq:amplitude_square_w3_v1_p}, where the factors $\kappa_i$ are reported in Table \ref{tab:kappa_w3_v1_p}. In order to define the coupling constant we proceed as in the pseudoscalar-pseudoscalar case. We use
		\begin{align*}
			\Gamma^\text{exp}_{\rho_{3} (1690) \rightarrow \omega (782) \, \pi} = \, & ( 25.8 \pm 9.8 ) \, \, \text{MeV} \, ,	\vphantom{\bigg(\bigg)}
			\\
			\Gamma^\text{exp}_{K_{3}^{\ast} (1780) \rightarrow \rho (770) \, K} = \, & ( 49.3 \pm 15.7 ) \, \, \text{MeV} \, ,	\vphantom{\bigg(\bigg)}
			\\
			\Gamma^\text{exp}_{K_{3}^{\ast} (1780) \rightarrow \bar{K}^{\ast} (892) \, \pi} = \, & ( 31.8 \pm 9.0 ) \, \, \text{MeV} \, .	\vphantom{\bigg(\bigg)}
		\end{align*}
	The coupling constant and its error are
		\begin{align}
			&	g_{w_3 v_1 p}^{2} = \frac{\sum_{i = 1}^{3} \frac{\tilde{g}_i}{\Delta \tilde{g}_i^{2}}}{\sum_{j = 1}^{3} \frac{1}{\Delta \tilde{g}_j^{2}}} \, ,	&&	\Delta g_{w_3 v_1 p}^{2} = \sqrt{\frac{1}{\sum_{j = 1}^{3} \frac{1}{\Delta g_j^{2}}}} \, ,
		\end{align}
	hence
		\begin{align}
			g_{w_3 v_1 p}^{2} = ( 9.2 \pm 1.9 ) \cdot 10^{-16} \, \, \text{MeV}^{-6}	\label{eq:coupling_w3_v1_p}
		\end{align}
	This value leads to the results listed in Table \ref{tab:w3v1p}. We observe that an acceptable agreement is reached, although the $\Gamma_{K_{3}^{\ast} (1780) \rightarrow \rho (770) \, K}$ mode is theoretically underestimated (the experimental error is nevertheless large). Quite remarkably, the two theoretically sizable and dominant decays $\omega_{3} (1670) \rightarrow \rho (770) \, K$ and $\phi_{3} (1850) \rightarrow K^{\ast} (892) \, K$ have been indeed seen in experiments, though they could not be quantified. Their future determination would represent a test of our approach -- at least on a qualitative level.
	
	\begin{table}
		\centering
		\renewcommand{\arraystretch}{1.5}
		\caption{\label{tab:w3v1p} %
			Decays of $J^{PC} = 3^{--}$ mesons into a pseudoscalar-vector pair. Experimental data taken from Ref.\ \cite{Zyla:2020zbs}.
		}
		\begin{ruledtabular}
			\begin{tabular}[c]{ l c c }
				\multicolumn{1}{c}{ \multirow{2}{*}{ decay process } }							&	theory							&	experiment
				\\
																									&	$\Gamma/$MeV					&	$\Gamma/$MeV
				\\
				\colrule
				$\rho_{3} (1690) \rightarrow \rho (770) \, \eta$									&	$3.8 \pm 0.8$					&	seen
				\\
				$\rho_{3} (1690) \rightarrow \bar{K}^{\ast} (892) \, K$								&	$3.4 \pm 0.7$					&
				\\
				$\rho_{3} (1690) \rightarrow \omega (782) \, \pi$									&	$35.8 \pm 7.4$					&	$25.8 \pm 9.8$
				\\
				$\rho_{3} (1690) \rightarrow \phi (1020) \, \pi$									&	$0.036 \pm 0.007$					&
				\\
				\colrule
				$K_{3}^{\ast} (1780) \rightarrow \rho(770) \, K$									&	$16.8 \pm 3.5$					&	$49.3 \pm 15.7$
				\\
				$K_{3}^{\ast} (1780) \rightarrow \bar{K}^{\ast} (892) \, \pi$						&	$27.2 \pm 5.6$					&	$31.8 \pm 9.0$
				\\
				$K_{3}^{\ast} (1780) \rightarrow \bar{K}^{\ast} (892) \, \eta$						&	$0.09 \pm 0.02$ 				&
				\\
				$K_{3}^{\ast} (1780) \rightarrow \omega (782) \, \bar{K}$							&	$4.3 \pm 0.9$					&
				\\
				$K_{3}^{\ast} (1780) \rightarrow \phi (1020) \, \bar{K}$							&	$1.2 \pm 0.3$					&
				\\
				\colrule
				$\omega_{3} (1670) \rightarrow \rho (770) \, \pi$									&	$97 \pm 20$						&	seen
				\\
				$\omega_{3} (1670) \rightarrow \bar{K}^{\ast} (892) \, K$							&	$2.9 \pm 0.6$					&
				\\
				$\omega_{3} (1670) \rightarrow \omega (782) \, \eta$								&	$2.8 \pm 0.6$					&
				\\
				$\omega_{3} (1670) \rightarrow \phi (1020) \, \eta$									&	$( 7.6 \pm 1.6 ) \cdot 10^{-6}$	&
				\\
				\colrule
				$\phi_{3} (1850) \rightarrow \rho (770) \, \pi$										&	$1.1 \pm 0.2$					&
				\\
				$\phi_{3} (1850) \rightarrow \bar{K}^{\ast} (892) \, K$								&	$35.5 \pm 7.3$					&	seen
				\\
				$\phi_{3} (1850) \rightarrow \omega (782) \, \eta$									&	$0.015 \pm 0.003$				&
				\\
				$\phi_{3} (1850) \rightarrow \omega (782) \, \eta^{\prime} (958)$					&	$0.003 \pm 0.001$				&
				\\
				$\phi_{3} (1850) \rightarrow \phi(1020) \, \eta$									&	$3.8 \pm 0.8$					&
			\end{tabular}
		\end{ruledtabular}
	\end{table}

	In addition, we present the following ratio
		\begin{align}
			\frac{ \Gamma_{\phi_{3} (1850) \rightarrow \bar{K}^{\ast} (892) \, K}}{\Gamma_{\phi_{3} (1850) \rightarrow \bar{K} \, K}} = 1.9 \pm 0.4 \, ,
		\end{align}
	which does not totally contradict the PDG \cite{Zyla:2020zbs} average taken from Ref.\ \cite{Aston:1988rf},
		\begin{align}
			\frac{ \Gamma^\text{exp}_{\phi_{3} (1850) \rightarrow \bar{K}^{\ast} (892) \, K}}{\Gamma^\text{exp}_{\phi_{3} (1850) \rightarrow \bar{K} \, K}} = 0.55_{-0.45}^{+0.85} \, .
		\end{align}
	Interestingly Ref.\ \cite{AlHarran:1981tk} even reports a slightly larger ratio of
		\begin{align}
			\frac{ \Gamma^\text{exp}_{\phi_{3} (1850) \rightarrow \bar{K}^{\ast} (892) \, K}}{\Gamma^\text{exp}_{\phi_{3} (1850) \rightarrow \bar{K} \, K}} = 0.8 \pm 0.4 \, .
		\end{align}
	Furthermore, we would like to mention a recent lattice QCD study \cite{Johnson:2020ilc}, which also confirms our overall predictions on dominant and less dominant vector-pseudoscalar decay channels. Reference \cite{Johnson:2020ilc} predicts (without providing explicit errors due to large uncertainties),
		\begin{align}
			\Gamma^\text{J,D}_{\rho_3 (1690) \rightarrow \omega (782) \, \pi} = \, & 22 \, \, \text{MeV} \, ,	\vphantom{\bigg(\bigg)}
			\\
			\Gamma^\text{J,D}_{\rho_3 (1690) \rightarrow \bar{K}^\ast (892) \, K} = \, & 2 \, \, \text{MeV} \, ,	\vphantom{\bigg(\bigg)}
			\\
			\Gamma^\text{J,D}_{\omega_3 (1670) \rightarrow \rho (770) \, \pi} = \, & 62 \, \, \text{MeV} \, ,	\vphantom{\bigg(\bigg)}
			\\
			\Gamma^\text{J,D}_{\omega_3 (1670) \rightarrow \bar{K}^\ast (892) \, K} = \, & 2 \, \, \text{MeV} \, ,	\vphantom{\bigg(\bigg)}
			\\
			\Gamma^\text{J,D}_{\omega_3 (1670) \rightarrow \omega (782) \, \eta} = \, & 1 \, \, \text{MeV} \, ,	\vphantom{\bigg(\bigg)}
			\\
			\Gamma^\text{J,D}_{\phi_3 (1850) \rightarrow \bar{K}^\ast (892) \, K} = \, & 20 \, \, \text{MeV} \, ,	\vphantom{\bigg(\bigg)}
			\\
			\Gamma^\text{J,D}_{\phi_3 (1850) \rightarrow \phi (1020) \, \eta} = \, & 3 \, \, \text{MeV} \, .	\vphantom{\bigg(\bigg)}
		\end{align}
	It is quite remarkable that our (quite simple) model is able to predict at least qualitatively rather similar results to such an advanced and comprehensive lattice QCD study. This leads us to the conclusion, that even for high-spin mesons with masses above $1\, \,$GeV like the conventional $3^{--}$-nonet chiral symmetry (breaking) is the decisive guiding principle for their phenomenology.
	
\subsection{Decay process \texorpdfstring{$W_3 \rightarrow \gamma + P$}{W -> gamma + P}}
	
	As a next step, we also present the results for the radioactive decays $W_3 \rightarrow \gamma \, P$, where $\gamma$ represents the photon. These can be obtained by using vector meson dominance \cite{Ebert:1982pk,OConnell:1995nse,Kim:1996nq}, which takes into account of the photon-vector-meson mixing through the shift
		\begin{align}
			V_{\mu\nu} \mapsto V_{\mu\nu} + \tfrac{e Q}{g_{\rho}} \, F_{\mu\nu} \, ,	\label{shiftvmd}
		\end{align}
	where $V_{\mu\nu} \equiv \partial_{\mu} V_{\nu} - \partial_{\nu} V_{\mu}$ and $Q = \mathrm{diag} \big( \frac{2}{3} , - \frac{1}{3} , \, - \frac{1}{3} \big)$ is the charge quark matrix, which includes the charges of the up, down, and strange quark. The electromagnetic field tensor is denoted as $F_{\mu\nu} = \partial_\mu a_\nu - \partial_\nu a_\mu$, with $a_\mu$ being the photon field, while $e = \sqrt{4 \uppi \, \alpha}$ is the electric coupling constant, and $g_{\rho} \simeq 5.5$ parametrizes the photon-vector-meson transition. By applying the shift of Eq.\ \eqref{shiftvmd} into Eq.\ \eqref{eq:lag-wvp} we obtain the Lagrangian for radioactive decays
		\begin{align}
			& \mathcal{L}_{w_3 \gamma p} =	\vphantom{\bigg(\bigg)}
			\\
			= \, & g_{w_3 v_1 p} \, \tfrac{e}{g_\rho} \, \varepsilon^{\mu\nu\rho\sigma} \, ( \partial_\nu a_\rho ) \, \mathrm{tr} \big[ W_{3 , \mu\alpha\beta} \, \big\{ Q , \, ( \partial^{\alpha} \partial^{\beta} \partial_{\sigma} P ) \big\}_{+} \big] \, .	\vphantom{\bigg(\bigg)}	\nonumber
		\end{align}
	The extended version of this Lagrangian is provided by Eq.\ \eqref{eq:extended_lagrangian_w3_gamma_p} in App.\ \ref{app:lagrangians}. For the tree level decay rate formula, we obtain via Eq.\ \eqref{eq:amplitude_square_w3_gamma_p},
		\begin{align}
				\Gamma_{W_3 \rightarrow \gamma + P} ( m_{w_3} , m_{p} ) = \, & g_{w_3 v_1 p}^{2} \, \big( \tfrac{e}{g_\rho} \big)^2 \, \frac{\big| \vec{k}_{\gamma , p} \big|^{7}}{105 \pi} \, \kappa_i^\gamma \, .	\vphantom{\Bigg(\Bigg)}	\label{eq:decay_w3_gamma_p}
		\end{align}
	where one of the masses $m_a$ and $m_b$ in Eq.\ \eqref{eq:kf} is set to zero, because the photon is massless. The factors $\kappa_i^\gamma$ are listed in Table \ref{tab:kappa_w3_gamma_p}. (The Heaviside function is not needed, because all pseudoscalars are lighter than the conventional spin-$3$ mesons and the photons can be arbitrarily soft.)
	
	Various predictions for the radiative decays $W_3 \rightarrow \gamma \, P$ are calculated and presented in Table \ref{tab:w3gammap}. Because of the rather large errors in the coupling constant \eqref{eq:coupling_w3_v1_p}, we round all results to integers in units of keV. Quite large radiative decay channels are $\omega_{3} (1670) \rightarrow \gamma \, \pi^0$ as well as $K_{3}^{0} (1780) \rightarrow \gamma \, K^{0}$. 
	
	In general, these processes imply that a photoproduction of mesons with spin $J^{PC}=3^{--}$ can take place at the ongoing \textit{GlueX} \cite{Ghoul:2015ifw,Zihlmann:2010zz,Proceedings:2014joa} and \textit{CLAS12} \cite{Rizzo:2016idq} experiments at \textit{Jefferson Lab}.

	\begin{table}
		\centering
		\renewcommand{\arraystretch}{1.5}
		\caption{\label{tab:w3gammap} %
			Theoretical predictions for the radiative decays $W_3 \rightarrow \gamma \, P$.
		}
		\begin{ruledtabular}
			\begin{tabular}[c]{ l c }
				\multicolumn{1}{c}{ \multirow{2}{*}{ decay process } }			&	theory
				\\
																				&	$\Gamma/$keV
				\\
				\colrule
				$\rho_{3}^{\pm/0} (1690) \rightarrow \gamma \, \pi^{\pm/0}$		&	$69 \pm 14$
				\\
				$\rho_{3}^{0} (1690) \rightarrow \gamma \,\eta$					&	$157 \pm 32$
				\\
				$\rho_{3}^{0} (1690) \rightarrow \gamma \, \eta^{\prime} (958)$	&	$20 \pm 4$
				\\
				\colrule
				$K_{3}^{\pm} (1780) \rightarrow \gamma \, K^{\pm}$				&	$58 \pm 12$
				\\
				$K_{3}^{0} (1780) \rightarrow \gamma \, K^{0}$					&	$231 \pm 48$
				\\
				\colrule
				$\omega_{3} (1670) \rightarrow \gamma \, \pi^0$					&	$560 \pm 120$
				\\
				$\omega_{3} (1670) \rightarrow \gamma \, \eta$					&	$19 \pm 4$
				\\
				$\omega_{3} (1670) \rightarrow \gamma \, \eta^{\prime} (958)$	&	$1.4 \pm 0.3$
				\\
				\colrule
				$\phi_{3} (1850) \rightarrow \gamma \, \pi^0$					&	$4 \pm 1$
				\\
				$\phi_{3} (1850) \rightarrow \gamma \, \eta$					&	$129 \pm 26$
				\\
				$\phi_{3} (1850) \rightarrow \gamma \, \eta^{\prime} (958)$		&	$35 \pm 7$
			\end{tabular}
		\end{ruledtabular}
	\end{table}

\subsection{Decay process \texorpdfstring{$W_3 \rightarrow A_2 + P$}{W3 -> A2 + P}}

	The interaction Lagrangian describing the decay of the spin-$3$ tensor mesons into a spin-$2$ tensor and a pseudoscalar mesons has the following form
		\begin{align}
			& \mathcal{L}_{w_3 a_2 p} =	\vphantom{\bigg(\bigg)}
			\\
			= \, & g_{w_3 a_2 p} \, \varepsilon_{\mu\nu\rho\sigma} \, \mathrm{tr} \big[ \tensor{W}{_3^\mu_\alpha_\beta} \, \big[ ( \partial^{\nu} A_2^{\rho\alpha} ) , \, ( \partial^{\sigma} \partial^{\beta} P ) \big]_{-} \big] \, .	\vphantom{\bigg(\bigg)}	\nonumber
		\end{align}
	The extended version is provided in App. \ref{app:lagrangians} in Eq.\ \eqref{eq:extended_lagrangian_w3_a2_p}. Using Eq.\ \eqref{eq:amplitude_square_w3_a2_p} the corresponding decay formula is
		\begin{align}
			& \Gamma_{W_3 \rightarrow A_2 + P} ( m_{w_3} , m_{a_2} , m_{p} ) =	\vphantom{\Bigg(\Bigg)}
			\\
			= \, & g_{w_3 a_2 p}^{2} \, \frac{\big| \vec{k}_{a_2 , p} \big|^{5}}{420 \uppi \, m_{a_2}^{2}} \, \big( 2 \, \big| \vec{k}_{a_2 , p} \big|^{2} + 7 \, m_{a_2}^{2} \big) \, \kappa_i \times	\vphantom{\Bigg(\Bigg)}	\nonumber
			\\
			& \times \Theta( m_{w_3} - m_{a_2} - m_{p} ) \, ,	\vphantom{\Bigg(\Bigg)}	\nonumber
		\end{align}
	where the $\kappa_i$ are listed in Table \ref{tab:kappa_w3_a2_p}.	For this channel, there are no experimental reported branching ratios that allow for a direct determination of the coupling constant. Yet, we can use the experimental ratio listed by the PDG \cite{Zyla:2020zbs} and taken from Ref.\ \cite{Amelin:2000nm},
		\begin{align}
			\frac{\Gamma^\text{exp}_{\rho_{3} (1690) \rightarrow a_{2} (1320) \, \pi}}{\Gamma^\text{exp}_{\rho_{3} (1690) \rightarrow \rho (770) \, \eta}} = 5.5 \pm 2.0 \, . \label{eq:exp-wxp}
		\end{align}
	together with the previously determined theoretical value $\Gamma_{\rho_{3} (1690) \rightarrow \rho (770) \, \eta} = ( 3.8 \pm 0.8 ) \, \,$MeV reported in Table \ref{tab:w3v1p}. The corresponding value for the coupling constant is
		\begin{align}
			g_{w_3 a_2 p}^{2} = ( 2.8 \pm 1.2 ) \cdot 10^{-9} \, \, \text{MeV}^{-4} \, .
		\end{align}
	Once the coupling constant is fixed,  we get the results for the decay rates reported in Table \ref{tab:w3a2p}. The decay width $\Gamma_{K_{3}^{\ast} (1780) \rightarrow \bar{K}^{\ast}_{2} (1430) \, \pi}$ is safely smaller the experimental upper limit \cite{Zyla:2020zbs,Aston:1986jb}. Moreover, the quite large mode $\rho_{3} (1690) \rightarrow a_{2} (1320) \, \pi$ is seen in experiments, but also in this case no branching ratio is listed in the PDG \cite{Zyla:2020zbs}.

	\begin{table}
		\centering
		\renewcommand{\arraystretch}{1.5}
		\caption{\label{tab:w3a2p} %
			Decays of $J^{PC} = 3^{--}$ mesons into a pseudoscalar-tensor pair. Experimental data taken from Ref.\ \cite{Zyla:2020zbs}.
		}
		\begin{ruledtabular}
			\begin{tabular}[c]{ l c c }
				\multicolumn{1}{c}{ \multirow{2}{*}{ decay process } }				&	theory							&	experiment
				\\
																					&	$\Gamma/$MeV					&	$\Gamma/$MeV
				\\
				\colrule
				$\rho_{3} (1690) \rightarrow a_{2} (1320) \, \pi$					&	$20.9 \pm 8.7$					&	seen
				\\
				\colrule
				$K_{3}^{\ast} (1780) \rightarrow \bar{K}^{\ast}_{2} (1430) \, \pi$	&	$5.8 \pm 2.4$					&	$< 25.4 \pm 3.4$
				\\
				$K_{3}^{\ast} (1780) \rightarrow f_{2} (1270) \, \bar{K}$			&	$( 5.4 \pm 2.2 ) \cdot 10^{-5}$	&
			\end{tabular}
		\end{ruledtabular}
	\end{table}

\subsection{Decay process \texorpdfstring{$W_3 \rightarrow B_1 + P$}{W3 -> B1 + P}}

	The interaction Lagrangian describing the decay into a pseudovector and pseudoscalar meson is given by
		\begin{align}
			\mathcal{L}_{w_3 b_1 p} = g_{w_3 b_1 p} \, \mathrm{tr} \big[ W_3^{\mu\nu\rho} \, \big\{ B_{1 , \mu} , \, ( \partial_{\nu} \partial_{\rho} P ) \big\}_{+} \big] \, ,	\label{eq:lag-wsp}
		\end{align}
	where the extended form is given in Eq.\ \eqref{eq:extended_lagrangian_w3_b1_p} in App.\ \ref{app:lagrangians}. The tree level decay rate takes the form,
		\begin{align}
			& \Gamma_{W_3 \rightarrow B_1 + P} ( m_{w_3} , m_{b_1} , m_{p} ) =	\vphantom{\Bigg(\Bigg)}	\label{eq:decay_formula_w3_b1_p}
			\\
			= \, & g_{w_3 b_1 p}^{2} \, \frac{| \vec{k}_{b_1 , p} |^{5}}{420 \uppi \, m_{w_3}^{2}} \, \Bigg( 7 + 3 \, \frac{\big| \vec{k}_{b_1 , p} \big|^{2}}{m_{b_1}^{2}} \Bigg) \, \kappa_i \times	\vphantom{\Bigg(\Bigg)}	\nonumber
			\\
			& \times \Theta ( m_{w_3} - m_{b_1} - m_{p} ) \, .	\vphantom{\Bigg(\Bigg)}	\nonumber
		\end{align}
	Here, we use Eq.\ \eqref{eq:amplitude_square_w3_b1_p} and the $\kappa_i$ can be found in Table \ref{tab:kappa_w3_b1_p}. Also in this case, a direct determination of the coupling constant $g_{w_3 b_1 p}$ is not possible due to the lack of experimental information. In order to estimate it, we proceed as it follows.
	\begin{enumerate}
		\item	According to the PDG \cite{Zyla:2020zbs}, there exists a lower limit for the ratio
			\begin{align}
				\frac{\Gamma_{\omega_{3} (1670) \rightarrow b_1 (1235) \, \pi}}{\Gamma_{\omega_{3} (1670) \rightarrow \omega (782) \, \pi \, \pi}} > 0.75 \, .
			\end{align}
		The PDG \cite{Zyla:2020zbs} extracts this lower bound for the ratio from Ref.\ \cite{Baubillier:1979ve} and does not list this bound as confirmed data. In Ref.\ \cite{Baubillier:1979ve} the authors in fact report
			\begin{align}
				\frac{\Gamma_{\omega_{3} (1670) \rightarrow b_1 (1235) \, \pi}}{\Gamma_{\omega_{3} (1670) \rightarrow \omega (782) \, \pi \, \pi}} = 1.0_{-0.25}^{+0.0} \, .
			\end{align}
		They further argue that the decay mode $\omega_{3} (1670) \rightarrow b_1 (1235) \, \pi$ is the dominant contribution to the $\omega (782) \, \pi \, \pi$ final state via the subsequent decay $b_1 (1235) \, \rightarrow \omega (782) \, \pi$. They claim that one can assume that
			\begin{align}
				\Gamma_{\omega_{3} (1670) \rightarrow b_1 (1235) \, \pi} \approx \Gamma_{\omega_{3} (1670) \rightarrow \omega (782) \, \pi \, \pi}	\label{saturation}
			\end{align}
		and deviations are expected to be smaller than $10\%$.

		\item Based on this assumption we use the following approximation
			\begin{align}
				& \frac{\Gamma_{\omega_{3} (1670) \rightarrow b_1 (1235) \, \pi}}{\Gamma_{\omega_{3} (1670) \rightarrow \rho (770) \, \pi}} \overset{\text{Eq.\ \eqref{saturation}}}{\approx}	\label{eq:branching_b1_pi_rho1_pi}
				\\
				\overset{\text{Eq.\ \eqref{saturation}}}{\approx} \, & \frac{\Gamma_{\omega_{3} (1670) \rightarrow \omega \, \pi \, \pi}}{\Gamma_{\omega_{3} (1670) \rightarrow \rho (770) \, \pi}} \overset{\text{PDG}}{=} 0.71 \pm 0.27 \, .	\nonumber
			\end{align}
		Here, one should mention that the PDG \cite{Zyla:2020zbs} cites Ref.\ \cite{Diaz:1974bq} for the second ratio, but still excludes it from their confirmed data. Additionally, Ref.\ \cite{Diaz:1974bq} even lists a different branching ratio,
			\begin{align}
				\frac{\Gamma_{\omega_{3} (1670) \rightarrow \omega \, \pi \, \pi}}{\Gamma_{\omega_{3} (1670) \rightarrow \rho (770) \, \pi}} = 0.47 \pm 0.18 \, ,
			\end{align}
		as well as
			\begin{align}
				\frac{\Gamma_{\omega_{3} (1670) \rightarrow b_1 (1235) \, \pi}}{\Gamma_{\omega_{3} (1670) \rightarrow \rho (770) \, \pi}} = 0.32 \pm 0.16 \, ,
			\end{align}
		It is not clear to the authors of this work, how the PDG \cite{Zyla:2020zbs} extracted their ratio from Ref.\ \cite{Diaz:1974bq}. Nevertheless, we will stick to the data reported in the PDG \cite{Zyla:2020zbs} and assume that there was some reasonable reanalysis of the data of Ref.\ \cite{Diaz:1974bq}. In this context, it might be worth mentioning, that all experimental data concerning the $\omega_3 (1670)$ is rather old and its mass and total width could not be determined up to high accuracy at the time when Refs.\ \cite{Baubillier:1979ve,Diaz:1974bq} were published. Definitely, there is a need for future experimental investigations of this state.
	\end{enumerate}
	For what concerns our work, we use the result $\Gamma_{\omega_{3} (1670) \rightarrow \rho (770) \, \pi} = ( 97 \pm 20 ) \, \,$MeV presented in the previous section as well as the ratio \eqref{eq:branching_b1_pi_rho1_pi}. We obtain the following estimate for the coupling constant
		\begin{align}
			g_{w_3 b_1 p}^{2} \approx ( 0.008 \pm 0.003 ) \, \, \text{MeV}^{-2} \, .
		\end{align}
	We are aware that this value and the corresponding results, which are listed in Table \ref{tab:w3b1p}, are only first rough estimates. Still, they might again help identifying dominant decay channels. Additionally, we consider the following aspects for the determination and interpretation of our results.

	\begin{table}
		\centering
		\renewcommand{\arraystretch}{1.5}
		\caption{\label{tab:w3b1p} %
			Decays of $J^{PC} = 3^{--}$ mesons into a pseudoscalar-pseudovector pair obtained by using the mixing angle $\beta_{b_1} = 0$ and $\beta_{b_1} = - 40^{\circ}$, respectively. No experimental value is listed in the PDG \cite{Zyla:2020zbs}. The decay channel $\omega_{3} (1670) \rightarrow b_1 (1235) \, \pi$ is possibly seen.
		}
		\begin{ruledtabular}
			\begin{tabular}[c]{ l c c }
				\multicolumn{1}{c}{ \multirow{3}{*}{ decay process } }							&	theory for						&	theory for
				\\
																								&	$\beta_{b_1} = 0$				&	$\beta_{b_1} = -40^{\circ}$
				\\
																								&	$\Gamma/$MeV					&	$\Gamma/$MeV
				\\
				\hline\hline
				$\rho_{3} (1690) \rightarrow h_1 (1170) \, \pi$									&	$53 \pm 23$						&	$31 \pm 13$
				\\
				$\rho_{3} (1690) \rightarrow h_1 (1415) \, \pi$									&	$0$								&	$0.73 \pm 0.31$
				\\
				\colrule
				$K_{3}^{\ast} (1780) \rightarrow b_1 (1235) \, K$								&	$0.55 \pm 0.24$					&	$0.55 \pm 0.24$
				\\
				$K_{3}^{\ast} (1780) \rightarrow \bar{K}_{1 , B} \, \pi$						&	$37 \pm 16$						&	$37 \pm 16$
				\\
				$K_{3}^{\ast} (1780) \rightarrow h_1 (1170) \, \bar{K}$							&	$1.61 \pm 0.70$					&	$0.03 \pm 0.01$
				\\
				\colrule
				$\omega_{3} (1670) \rightarrow b_1 (1235) \, \pi$								&	$69 \pm 30$						&	$69 \pm 30$
				\\
				\colrule
				$\phi_{3} (1850) \rightarrow b_1 (1235) \, \pi$									&	$1.17 \pm 0.51$					&	$1.17 \pm 0.51$
				\\
				$\phi_{3} (1850) \rightarrow \bar{K}_{1 , B} \, K + \mathrm{c}. \mathrm{c}.$	&	$9.1 \pm 3.9$					&	$9.1 \pm 3.9$
				\\
				$\phi_{3} (1850) \rightarrow h_1 (1170) \, \eta$								&	$0.006 \pm 0.003$				&	$1.1 \pm 0.5$
			\end{tabular}
		\end{ruledtabular}
	\end{table}

	\begin{enumerate}
		\item	We consider $K_{1 , B} \approx K_1 (1270)$, since this is the dominant contribution; then, the decay rate $\Gamma_{\phi_{3} (1850) \rightarrow \bar{K}_{1 , B} \, K}$ is rather small. A disclaimer is in order: due to the large mixing between $K_{1 , A}$ and $K_{1 , B}$ the results involving the identification $K_{1 , B} \approx K_1 (1270)$ can be only considered as a first approximation. The more correct procedure should be to consider the full mixing in Eq.\ \eqref{mixk1abfields}, hence $K_{1 , B}$ should be expressed as a superposition of $K_1 (1270)$ and $K_1 (1400)$. Yet, this is not an easy task, since the interaction in Eq.\ \eqref{eq:lag-wsp} alone is not enough. One should also take into account the interaction terms for the decays $W_3 \rightarrow A_1 \, P$ in Eq.\ \eqref{lagwa1p}, which	includes $K_{1 , A}$, see next section. Then, the decays into $K_1 (1270)$ and $K_1 (1400)$ should be calculated by the joint Lagrangians \eqref{eq:lag-wsp} and \eqref{lagwa1p} together with the mixing \eqref{mixk1abfields}. Furthermore, interference between the two Lagrangians is expected. This calculation is however not possible, since the coupling constant $g_{w_3 a_1 p}$ cannot be determined by the present experimental data (see next subsection). Hence, we must limit our study to the dominant assignment $K_{1 , B} \approx K_1 (1270)$ (in this subsection) and $K_{1 , A} \approx	K_1 (1400)$ (in the next subsection).
		
		\item	We present the results in Table \ref{tab:w3b1p} for two values of the unknown mixing angle in the isoscalar sector. In one case, we assume that the mixing angle vanishes, $\beta_{b_1} \approx 0$, in the second we study $\beta_{b_1} \approx - 40^{\circ}$, which is a quite large and negative value similar to the mixing angle for the pseudoscalar-isoscalars. Note, the mixing angle $\beta_{b_1}$ appears in the decay channels $\rho_{3} (1690) \rightarrow h_1 (1170) \, \pi$, $\rho_{3} (1690) \rightarrow h_1 (1415) \, \pi$, $K_{3}^{\ast} (1780) \rightarrow h_1 (1170) \, \bar{K}$, and $\phi_{3} (1850) \rightarrow h_1 (1170) \, \eta$, see Table \ref{tab:kappa_w3_b1_p}. The only experimentally possibly seen decay $\omega_{3} (1670) \rightarrow b_1 (1235) \, \pi$, see Ref.\ \cite{Diaz:1974bq}, corresponds to a quite large theoretical partial decay widths and is independent of any assumption on $\beta_{b_1}$.
		
		\item The decay $\rho_{3} (1690) \rightarrow h_1 (1170) \,\pi$ is also quite large for both choices of the mixing angle, hence it is a potentially interesting channel for future search. Moreover, it is also an interesting decay channel in order to determine the value of the mixing angle $\beta_{b_1}$. The remaining decay channels are pretty small, which might be an explanation why they could not be observed in experiment.
	\end{enumerate}

\subsection{Decay process \texorpdfstring{$W_3 \rightarrow A_1 + P$}{W3 -> A1 + P}}

	The interaction Lagrangian and the decay formula for the decay into an axial-vector and a pseudoscalar meson are given by
		\begin{align}
			\mathcal{L}_{w_3 a_1 p} = g_{w_3 a_1 p} \, \mathrm{tr} \big[ W_3^{\mu\nu\rho} \, \big[ A_{1 , \mu} , \, ( \partial_{\nu} \partial_{\rho} P ) \big]_{-} \big] \, ,	\label{lagwa1p}
		\end{align}
	and
		\begin{align}
			& \Gamma_{W_3 \rightarrow A_1 + P} ( m_{w_3} , m_{a_1} , m_{p} ) =	\vphantom{\Bigg(\Bigg)}
			\\
			= \, & g_{w_3 a_1 p}^{2} \, \frac{\big| \vec{k}_{a_1 , p} \big|^{5}}{420 \uppi \, m_{w_3}^{2}} \, \Bigg( 7 + 3 \, \frac{\big| \vec{k}_{a_1 , p} \big|^{2}}{m_{a_1}^{2}} \Bigg) \, \kappa_i \times	\vphantom{\Bigg(\Bigg)}	\nonumber
			\\
			& \times \Theta ( m_{w_3} - m_{a_1} - m_{p} ) \, .	\vphantom{\Bigg(\Bigg)}	\nonumber
		\end{align}
	The extended version of the Lagrangian is given by Eq.\ \eqref{eq:extended_lagrangian_w3_a1_p} in App.\ \ref{app:lagrangians}. The $\kappa_i$-values for each channel are taken from Table \ref{tab:kappa_w3_a1_p} and the decay formula was derived by using Eq.\ \eqref{eq:amplitude_square_w3_a1_p}. As explained previously, for a first rough estimate, we assume here that $K_{1 , A} \approx K_1 (1400)$.
	
	Since we do not have enough experimental information for obtaining the coupling constant $g_{w_3 a_1 p}$, we can only get some theoretical predictions for ratios among different decay channels, that are reported in Table \ref{tab:w3a1p}. Note, the mixing angle $\beta_{a_1}$ does not appear in any of the nonzero reported decay channel, as it can be read from Table \ref{tab:kappa_w3_a1_p}. Anyway, the mixing is expected to be small \cite{Giacosa:2017pos}.
	\begin{table}
		\centering
		\renewcommand{\arraystretch}{1.8}
		\caption{\label{tab:w3a1p} %
			Predictions for the nonzero branching ratios of $W_3 \rightarrow A_1 \, P$. We choose as a reference the $a_1(1260) \, \pi$ decay mode.
		}
		\begin{ruledtabular}
			\begin{tabular}[c]{ c c }
				branching ratio																												&	theory	\\
				\colrule
				$\frac{\Gamma_{K_{3}^{\ast} (1780) \rightarrow K_{1 , A} \, \pi}}{\Gamma_{\rho_{3} (1690) \rightarrow a_1 (1260) \, \pi}}$		&	$0.12$
				\\
				$\frac{\Gamma_{K_{3}^{\ast} (1780) \rightarrow a_1 (1260)\, K}}{\Gamma_{\rho_{3} (1690) \rightarrow a_1 (1260) \, \pi}}$	&	$0.01$
			\end{tabular}
		\end{ruledtabular}
	\end{table}

	In the end, it is interesting to mention that there might be an option to size the coupling constant by linking the present model to an underlying chiral model, \textit{e.g.}\ the eLSM. This approximately predicts $2\, \,$MeV for $\rho_{3} (1690) \rightarrow a_1 (1260) \, \pi$, see Chap.\ \ref{chap:discussion} for an estimate of $g_{w_3 a_1 p}$. It seems therefore that the decays of the type $W_3 \rightarrow A_1 + P$ are suppressed.
	
\subsection{Decay process \texorpdfstring{$W_3 \rightarrow V_1 + V_1$}{W3 -> V1 + V1}}

	The interaction Lagrangian describing the decay into two vector mesons is given by
		\begin{align}
			\mathcal{L}_{w_3 v_1 v_1} = g_{w_3 v_1 v_1} \, \mathrm{tr} \big[ W_3^{\mu\nu\rho} \, \big[ ( \partial_{\mu} V_{1 , \nu}) , \, V_{1 , \rho} \big]_{-} \big] \, ,
		\end{align}
	see Eq.\ \eqref{eq:extended_lagrangian_w3_v1_v1} in App.\ \ref{app:lagrangians} for the extended version . The corresponding decay rate reads
		\begin{align}
			& \Gamma_{W_3 \rightarrow V_1^{(1)} + V_1^{(2)}} ( m_{w_3} , m_{v_1^{(1)}} , m_{v_1^{(2)}} ) =	\vphantom{\Bigg(\Bigg)}
			\\
			= \, & g_{w_3 v_1 v_1}^{2} \, \frac{\big| \vec{k}_{v_1^{(1)} , v_1^{(2)}} \big|^{3}}{840 \uppi \, m_{v_1^{(1)}}^{2} \, m_{v_1^{(2)}}^{2} \, m_{w_3}^{2}} \, \Big[ 6 \, | \vec{k}_{v_1^{(1)} , v_1^{(2)}} \big|^{4} +	\vphantom{\Bigg(\Bigg)}	\nonumber
			\\
			& + 35 \, m_{v_1^{(1)}}^{2} \,  m_{v_1^{(2)}}^{2} + 14 \, \big| \vec{k}_{v_1^{(1)} , v_1^{(2)}} \big|^{2} \, \big( m_{v_1^{(1)}}^{2} + m_{v_1^{(2)}}^{2} \big)  \Big] \times	\vphantom{\Bigg(\Bigg)}	\nonumber
			\\
			& \times \kappa_i \, \Theta ( m_{w_3} - m_{v_1^{(1)}} - m_{v_1^{(2)}} ) \, ,	\vphantom{\Bigg(\Bigg)}	\nonumber
		\end{align}
	where the $\kappa_i$'s are listed in Table \ref{tab:kappa_w3_v1_v1}. For the derivation of the decay formula, we use Eq.\ \eqref{eq:amplitude_square_w3_v1_v1}. Also in this case, we cannot determine the coupling constant, but we can present the branching ratios reported in Table \ref{tab:w3v1v1_ratios}.

	\begin{table}
		\centering
		\renewcommand{\arraystretch}{1.8}
		\caption{\label{tab:w3v1v1_ratios} %
			Predictions for the branching ratios of $W_3 \rightarrow V_1 + V_1$. We choose as a reference the $\rho \, \rho$ decay mode.
		}
		\begin{ruledtabular}
			\begin{tabular}[c]{ c c }
				branching ratio																																	&	theory
				\\
				\colrule
				$\frac{\Gamma_{K_{3}^{\ast} (1780) \rightarrow \rho (770) \, K^{\ast} (892)}}{\Gamma_{\rho_{3} (1690) \rightarrow \rho (770) \, \rho (770)}}$	&	$0.49$
				\\
				$\frac{\Gamma_{K_{3}^{\ast} (1780) \rightarrow K^{\ast} (892) \, \omega (782)}}{\Gamma_{\rho_{3} (1690) \rightarrow \rho (770) \, \rho (770)}}$	&	$0.17$
				\\
				$\frac{\Gamma_{\phi_{3} (1850) \rightarrow K^{\ast} (892) \, K^{\ast} (892)}}{\Gamma_{\rho_{3} (1690) \rightarrow \rho (770) \, \rho (770)}}$	&	$0.35$
			\end{tabular}
		\end{ruledtabular}
	\end{table}
	
	Moreover, we can find an upper limit for the coupling constant by considering
		\begin{align}
			& \frac{\Gamma_{\rho_{3} (1690) \rightarrow \pi \, \pi}}{\Gamma_{\rho_{3} (1690) \rightarrow \rho (770) \, \rho(770)}} \gtrapprox	\label{ratvv1}
			\\
			\gtrapprox \, & \frac{\Gamma_{\rho_{3} (1690) \rightarrow \pi \, \pi}}{\Gamma_{\rho_{3} (1690) \rightarrow \pi^{\pm} \, \pi^{+} \, \pi^{-} \, \pi^{0}}} \overset{\text{PDG}}{=} 0.35 \pm 0.11 \, ,	\nonumber
		\end{align}
	which follows from the fact that the $\rho_{3} (1690) \rightarrow \rho (770) \, \rho (770)$ mode is part of the $\pi^{\pm} \, \pi^{+} \, \pi^{-} \, \pi^{0}$ decay mode (and eventually it is one of its dominating contributors). Since the experimental value for $\Gamma_{\rho_{3} (1690) \rightarrow \pi \, \pi} = ( 38.0 \pm 3.2 ) \, \,$MeV is known, we find
		\begin{align}
			g_{w_3 v_1 v_1}^{2} \lesssim 535 \pm 174 \, .	\label{ratvv2}
		\end{align}
	Based on this information we can predict upper limits for the decay rates in Table \ref{tab:w3v1v1}.
	\begin{table}
		\centering
		\renewcommand{\arraystretch}{1.5}
		\caption{\label{tab:w3v1v1} %
			Upper limits for the decays of $J^{PC} = 3^{--}$ mesons into a vector-vector pair. Two different estimates are used, see text.
		}
		\begin{ruledtabular}
			\begin{tabular}[c]{ l c c }
				\multicolumn{1}{c}{ \multirow{3}{*}{ decay process } }					&	theory using						&	theory using
				\\
																						&	\eqref{ratvv1} \& \eqref{ratvv2}	&	\eqref{totvv1} \& \eqref{totvv2}
				\\
																						&	$\Gamma/$MeV						&	$\Gamma/$MeV
				\\
				\colrule
				$\rho_{3} (1690) \rightarrow \rho (770) \, \rho (770)$					&	$\lesssim 108.6 \pm 35.3$			&	$\lesssim 30$
				\\
				\colrule
				$K_{3}^{\ast} (1780) \rightarrow \rho (770) \, \bar{K}^{\ast} (892)$	&	$\lesssim 53.6 \pm 17.4$				&	$\lesssim 15$
				\\
				$K_{3}^{\ast} (1780) \rightarrow \bar{K}^{\ast} (892) \, \omega (782)$	&	$\lesssim 19.0 \pm 6.2$				&	$\lesssim 5$
				\\
				\colrule
				$\phi_{3} (1850) \rightarrow \bar{K}^{\ast} (892) \, K^{\ast} (892)$	&	$\lesssim 38.3 \pm 12.5$				&	$\lesssim 11$
			\end{tabular}
		\end{ruledtabular}
	\end{table}	
	The theoretically largest decay is $\rho_{3} (1690) \rightarrow \rho (770) \, \rho (770)$, which has been experimentally seen. Yet, the decay width of about $108\, \,$ MeV is surely too large, since using a  value close to this upper limit would imply that the the sum of all the decays of the state $\rho_{3} (1690)$ would overshoot the experimental total width of $( 161 \pm 10 ) \, \,$MeV. This is in agreement with the ratio
		\begin{align}
			\frac{\Gamma_{\rho_3 (1690) \rightarrow \rho(770) \, \rho(770)}}{\Gamma_{\rho_3 (1690) \rightarrow \pi^\pm \, \pi^+ \, \pi^- \, \pi^0}} \, ,
		\end{align}
	which is smaller than $1$ for all experimental measurements \cite{Baltay:1977bq,Bartsch:1970fk,Kliger:1974nea,Thompson:1974zm}, which are listed by the PDG \cite{Zyla:2020zbs}.
	
	However, we can then obtain a second, more realistic estimate of these decay channels by summing up the largest decay channels:
		\begin{align}
			& \Gamma_{\rho_{3}}^\text{tot} = 161 \, \, \text{MeV} \approx	\vphantom{\bigg(\bigg)}	\label{totvv1}
			\\
			= \, & \Gamma_{\rho_{3} (1690) \rightarrow \pi \, \pi} + \Gamma_{\rho_{3} (1690) \rightarrow \bar{K} \, K} +	\vphantom{\bigg(\bigg)}	\nonumber
			\\
			& + \Gamma_{\rho_{3} (1690) \rightarrow \rho (770) \, \eta} + \Gamma_{\rho_{3} (1690) \rightarrow \bar{K}^{\ast} (892) \, K} +	\vphantom{\bigg(\bigg)}	\nonumber
			\\
			& + \Gamma_{\rho_{3} (1690) \rightarrow \omega (782) \, \pi} + \Gamma_{\rho_{3} (1690) \rightarrow \phi (1020) \, \pi} +	\vphantom{\bigg(\bigg)}	\nonumber
			\\
			& + \Gamma_{\rho_{3} (1690) \rightarrow a_2 (1320) \, \pi} + \Gamma_{\rho_{3} (1690) \rightarrow a_1 (1260) \, \pi} +	\vphantom{\bigg(\bigg)}	\nonumber
			\\
			& + \Gamma_{\rho_{3} (1690) \rightarrow h_1 (1170) \, \pi} + \Gamma_{\rho_{3} (1690) \rightarrow h_1 (1415) \, \pi} +	\vphantom{\bigg(\bigg)}	\nonumber
			\\
			& + \Gamma_{\rho_{3} (1690) \rightarrow \rho (770) \, \rho (770)} + \Gamma_{\rho_{3} (1690) \rightarrow \text{other}} \, ,	\vphantom{\bigg(\bigg)}	\nonumber
		\end{align}
	where $\Gamma_{\rho_{3} \rightarrow \text{other}}$ refers to all other decay channels not listed above. Anyhow, these should be small. For $\Gamma_{\rho_{3} (1690) \rightarrow h_1 (1170) \, \pi}$ we use the smaller value presented in Table \ref{tab:w3b1p}, assuming a large mixing $\beta_{b_1} = - 40^\circ$. In this way we estimate the more reasonable upper limit
		\begin{align}
			&	\Gamma_{\rho_{3} (1690) \rightarrow \rho (770) \, \rho (770)} \lesssim 30 \, \, \text{MeV} \, ,
		\end{align}
	from which follows,
		\begin{align}
			g_{w_3 v_1 v_1}^{2} \lesssim 148 \, .	\label{totvv2}
		\end{align}
	The corresponding estimates for the other channels can be found in Table \ref{tab:w3v1v1}.
	
	Following this estimate, we perform the sum of the predicted and sizable decay channels for the other spin-$3$ mesons, obtaining:
		\begin{align}
			\Gamma^\mathrm{sum}_{K_{3}^{\ast} (1780)} \approx \, & 146 \, \, \text{MeV}	\, ,	\vphantom{\bigg(\bigg)}
			\\
			\Gamma^\mathrm{sum}_{\omega_{3} (1670)} \approx \, & 175 \, \, \text{MeV} \, ,	\vphantom{\bigg(\bigg)}
			\\
			\Gamma^\mathrm{sum}_{\phi_{3} (1850)} \approx \, & 80 \, \, \text{MeV} \, ,	\vphantom{\bigg(\bigg)}
		\end{align}
	which are in agreement with  the experimental values from the PDG \cite{Zyla:2020zbs},
		\begin{align}
			\Gamma_{K_{3}^{\ast} (1780) }^{\text{tot}} = \, & ( 159 \pm 21 ) \, \, \text{MeV} \, ,	\vphantom{\bigg(\bigg)}
			\\
			\Gamma_{\omega_{3} (1670)}^{\text{tot}} = \, & ( 168 \pm 10 ) \, \, \text{MeV} \, ,	\vphantom{\bigg(\bigg)}
			\\
			\Gamma_{\phi_{3} (1850)}^{\text{tot}} = \, & ( 87_{-23}^{+28} ) \, \, \text{MeV} \, .	\vphantom{\bigg(\bigg)}
		\end{align}
	These results demonstrate that the model is internally consistent.

\section{Phenomenology of the \texorpdfstring{$J^{PC} = 3^{--}$}{JPC = 3--} glueball}
\label{sec:glueball}

	In this chapter, we study the branching ratios of the decays of a hypothetical glueball with $J^{PC} = 3^{--}$. Similar to all other bound states of gluons, this glueball is not yet experimentally detected. However, lattice QCD calculations in the quenched approximation predict its mass of approximately $4.13\, \,$GeV \cite{Morningstar:1999rf}, $4.33\, \,$GeV \cite{Meyer:2004gx}, or $4.20\, \,$GeV \cite{Chen:2005mg}. For general works on glueballs, see \textit{e.g.}\ Refs.\ \cite{Gregory:2012hu,Ochs:2013gi,Mathieu:2008me,Crede:2008vw}. For model approaches to the phenomenology of glueballs, which are similar to this one, see for example previous publications by some of the authors \cite{Giacosa:2005bw,Koenigstein:2016tjw,Janowski:2014ppa,Giacosa:2016hrm,Eshraim:2012jv}.
	
	Before we start our discussion, we explicitly state that all branching ratios should be considered as first indicative results.

\subsection{The effective model for glueball decays}

	 The Lagrangian describing the decays of this glueball can be obtained by making use of the fact that each glueball is a flavor singlet -- an object, which is invariant under $SU_\mathrm{V}(3)$ transformations. Hence, we can perform the simple replacement
		\begin{align}
			W^{\mu\nu\rho} \rightarrow G_3^{\mu\nu\rho} \cdot \openone_{3 \times 3}
		\end{align}
	in the effective interaction terms,	where $\openone_{3 \times 3}$ is the flavor-space-identity matrix and $G_3^{\mu\nu\rho}$ is the glueball field \cite{Giacosa:2005bw,Eshraim:2012jv}. As a consequence, the interaction terms that come with a commutator vanish and the only nonzero contributions to a residual IR-effective action for the $3^{--}$ glueball can be derived from the former interaction terms involving the anticommutator. Thus,
		\begin{align}
			&	\Gamma_{G_3 (4200) \rightarrow V_1 + P} \neq 0 \, ,	&&	\Gamma_{G_3 (4200) \rightarrow B_1 + P} \neq 0 \, .
		\end{align}
	The corresponding Lagrangian terms are reported in Table \ref{tab:glueball_decay_lagrangians}.
	\begin{table*}
		\centering
		\renewcommand{\arraystretch}{1.5}
		\caption{\label{tab:glueball_decay_lagrangians} %
			Strong interaction Lagrangian terms for the hypothetical glueball state $G_3 (4200)$ [in comparison with Eq.\ \eqref{lagtot}].
		}
		\begin{tabular}[c]{ l @{\qquad} c }
			\toprule
			\multicolumn{1}{c}{decay mode}	&	interaction Lagrangians
			\\
			\colrule
			$G_3 \rightarrow V_1 + P$		&	$\mathcal{L}_{g_3 v_1 p} = c_{g_3 v_1 p} \, G_{3 , \mu\alpha\beta} \, \varepsilon^{\mu\nu\rho\sigma} \, \mathrm{tr} \big[ \big\{ ( \partial_{\nu} V_{1 , \rho} ) , \, ( \partial^{\alpha} \partial^{\beta} \partial_{\sigma} P ) \big\}_{+} \big]$
			\\
			$G_3 \rightarrow V_1 + A_1$		&	$\mathcal{L}_{g_3 v_1 a_1} = c_{g_3 v_1 a_1} \, G_{3 , \mu\alpha\beta} \, \varepsilon^{\mu\nu\rho\sigma} \, \mathrm{tr} \big[ \big\{ ( \partial_{\nu} V_{1 , \rho} ) , \, ( \partial^{\alpha} \partial^{\beta} A_{1 , \sigma} ) \big\}_{+} \big]$
			\\
			$G_3 \rightarrow B_1 + P$		&	$\mathcal{L}_{g_3 b_1 p} = c_{g_3 b_1 p} \, G_3^{\mu\nu\rho} \, \mathrm{tr} \big[ \big\{ B_{1 , \mu} , \, ( \partial_{\nu} \partial_{\rho} P ) \big\}_{+} \big]$
			\\
			$G_3 \rightarrow B_1 + A_1$		&	$\mathcal{L}_{g_3 b_1 a_1} = c_{g_3 b_1 a_1} \, G_3^{\mu\nu\rho} \, \mathrm{tr} \big[ \big\{ B_{1 , \mu} , \, ( \partial_{\nu} A_{1 , \rho} ) \big\}_{+} \big]$
			\\
			\botrule
		\end{tabular}
	\end{table*}
	In addition, there are also decays of the type $G_3 (4200) \rightarrow V_1 + A_1$ and $G_3 (4200) \rightarrow B_1 + A_1$ that were not included in for the lightest conventional mesons with $J^{PC} = 3^{--}$ because they were kinematically forbidden. They are however possible for the glueball $G_3 (4200)$ since it is expected to be much heavier than the $3^{--}$ nonet \eqref{eq:spin_3--_nonet}. Below we present the theoretical predictions for ratios of decays, which can be easily calculated following the same steps of the previous sections. The only ingredients that are needed for a determination of the branching ratios are the masses of the particles, which are for the sake of simplicity again assumed to be exact. For the $3^ {--}$ glueball mass, we use $m_{g_3} = 4.2\, \,$GeV \cite{Chen:2005mg}.

\subsection{Decay process \texorpdfstring{$G_3 \rightarrow V_1 + P$}{G3 -> V1 + P}}

	The interaction Lagrangian for decaying spin-$3$ tensor glueball into a vector and a pseudoscalar meson reads
		\begin{align}
			& \mathcal{L}_{g_3 v_1 p} =	\vphantom{\bigg(\bigg)}
			\\
			= \, & c_{g_3 v_1 p} \, G_{3 , \mu\alpha\beta} \, \varepsilon^{\mu\nu\rho\sigma} \, \mathrm{tr} \big[ \big\{ ( \partial_\nu V_{1 , \rho} ) , \, ( \partial^{\alpha} \partial^{\beta} \partial_{\sigma} P ) \big\}_{+} \big] \, .	\vphantom{\bigg(\bigg)}	\nonumber
		\end{align}
	An extended version is given by Eq.\ \eqref{eq:extended_lagrangian_g3_v1_p} in App.\ \ref{app:lagrangians}. The corresponding decay formula is analogous to Eq.\ \eqref{eq:decay-gvp},
		\begin{align}
			& \Gamma_{G_3 \rightarrow V_1 + P} ( m_{g_3} , m_{v_1} , m_{p} ) =	\vphantom{\Bigg(\Bigg)}	\label{eq:decay-gvp_2}
			\\
			= \, & c_{g_3 v_1 p}^{2} \, \frac{\big| \vec{k}_{v_1 , p} \big|^{7}}{105 \uppi} \, \kappa_i \, \Theta( m_{g_3} - m_{v_1} - m_{p} ) \, .	\vphantom{\Bigg(\Bigg)}	\nonumber
		\end{align}
	Here, the $\kappa_i$'s can be taken from Table \ref{tab:kappa_g3_v1_p}. We use again Eq.\ \eqref{eq:kf} for the momenta of the outgoing particles throughout this chapter, while replacing the mass of the conventional spin-$3$ mesons $m_{w_3}$ with the glueball mass $m_{g_3} = 4200 \, \,$MeV. Since the coupling constant $c_{g_3 v_1 p}$ is unknown, we calculate the nonzero ratios in Table \ref{tabgvp}. This is an interesting decay channel to search for this glueball. Since the vector mesons further decay in two (or three) pseudoscalar mesons, one should search for final channels with three (or four) pseudoscalar mesons.

		\begin{table}
			\centering
			\renewcommand{\arraystretch}{1.8}
			\caption{\label{tabgvp} %
				Predictions for the branching ratios of glueball $G_3 (4200) \rightarrow V_1 + P$.
			}
			\begin{ruledtabular}
				\begin{tabular}[c]{ l c }
					\multicolumn{1}{c}{branching ratio}																								&	theory
					\\
					\colrule
					$\frac{\Gamma_{G_3 (4200)	\rightarrow	\bar{K}^{\ast} (892) \, K}}{\Gamma_{G_3 (4200) \rightarrow \rho (770) \, \pi}}$			&	$1.11$
					\\
					$\frac{\Gamma_{G_3 (4200) \rightarrow \omega (782) \, \eta}}{\Gamma_{G_3 (4200) \rightarrow \rho (770) \, \pi}}$				&	$0.17$
					\\
					$\frac{\Gamma_{G_3 (4200) \rightarrow \omega (782) \, \eta^{\prime} (958)}}{\Gamma_{G_3 (4200) \rightarrow \rho (770) \, \pi}}$	&	$0.089$
					\\
					$\frac{\Gamma_{G_3 (4200) \rightarrow \phi (1020) \, \eta}}{\Gamma_{G_3 (4200) \rightarrow \rho (770) \, \pi}}$					&	$0.098$
					\\
					$\frac{\Gamma_{G_3 (4200) \rightarrow \phi (1020) \, \eta^{\prime} (958)}}{\Gamma_{G_3 (4200) \rightarrow \rho (770) \, \pi}}$	&	$0.11$
				\end{tabular}
			\end{ruledtabular}
		\end{table}

\subsection{Decay process \texorpdfstring{$G_3 \rightarrow \gamma + P$}{G3 -> gamma + P}}

	Using vector meson dominance via the shift of Eq.\ \eqref{shiftvmd} and analogous replacements, we can also construct a radioactive decay term for the spin-$3$ glueball,
		\begin{align}
			& \mathcal{L}_{g_3 \gamma p} =	\vphantom{\bigg(\bigg)}
			\\
			= \, & c_{g_3 \gamma p} \, \tfrac{e}{g_\rho} \, G_{3 , \mu \alpha \beta} \, ( \partial_\nu a_\rho ) \, \varepsilon^{\mu\nu\rho\sigma} \, \mathrm{tr} \big[ \big\{ Q , \, ( \partial^{\alpha} \partial^{\beta} \partial_{\sigma} P ) \big\}_{+} \big] \, .	\vphantom{\bigg(\bigg)}	\nonumber
		\end{align}
	An extended version of this Lagrangian can be found in Eq.\ \eqref{eq:extended_lagrangian_g3_gamma_p} in App.\ \ref{app:lagrangians}. For the decay formula one derives
		\begin{align}
			\Gamma_{G_3 \rightarrow \gamma + P} ( m_{g_3} , m_{p} ) = \, & c_{g_3 v_1 p}^{2} \, \big( \tfrac{e}{g_\rho} \big)^2 \, \frac{\big| \vec{k}_{\gamma , p} \big|^{7}}{105 \pi} \, \kappa_i^\gamma \, ,	\vphantom{\Bigg(\Bigg)}	\nonumber
		\end{align}
	where the corresponding factors $\kappa_i^\gamma$ are listed in Table \ref{tab:kappa_g3_gamma_p}. Similar to Eq.\ \eqref{eq:decay_w3_gamma_p} one of the masses $m_a$ and $m_b$ in Eq.\ \eqref{eq:kf} has to be set to zero, due to the masslessness of the photon.

	From these formula, we calculate ratios for the radioactive decays $\Gamma_{G_3 \rightarrow \gamma \, P}$ w.r.t.\ the strong decay channel $\Gamma_{G_3 (4200) \rightarrow \rho (770) \, \pi}$. The results are listed in Table \ref{tabggammap}.
		\begin{table}
			\centering
			\renewcommand{\arraystretch}{1.8}
			\caption{\label{tabggammap} %
				Predictions for the branching ratios of glueball $G_3 (4200) \rightarrow \gamma + P$ with the $G_3 (4200) \rightarrow \rho \, \pi$ reference decay mode.
			}
			\begin{ruledtabular}
				\begin{tabular}[c]{ l c}
					\multicolumn{1}{c}{branching ratio}																							&	theory
					\\
					\colrule
					$\frac{\Gamma_{G_3 (4200) \rightarrow \gamma \, \pi^0}}{\Gamma_{G_3 (4200) \rightarrow \rho (770) \, \pi}}$					&	$129 \cdot 10^{-5}$
					\\
					$\frac{\Gamma_{G_3 (4200) \rightarrow \gamma \, \eta}}{\Gamma_{G_3 (4200) \rightarrow \rho (770) \, \pi}}$					&	$37 \cdot 10^{-5}$
					\\
					$\frac{\Gamma_{G_3 (4200) \rightarrow \gamma \,\eta^{\prime} (958)}}{\Gamma_{G_3 (4200) \rightarrow \rho (770) \, \pi}}$	&	$1 \cdot 10^{-5}$
				\end{tabular}
			\end{ruledtabular}
		\end{table}

\subsection{Decay process \texorpdfstring{$G_3 \rightarrow V_1 + A_1$}{G3 -> V1 + A1}}
	
	We also present the glueball decays into vectors and axial-vectors by considering the Lagrangian in which $G_3$ couples to $V_1 \, A_1$
		\begin{align}
				& \mathcal{L}_{g_3 v_1 a_1} =	\vphantom{\bigg(\bigg)}
				\\
				= \, & c_{g_3 v_1 a_1} \, G_{3 , \mu\alpha\beta} \, \varepsilon^{\mu\nu\rho\sigma} \, \mathrm{tr} \big[ \big\{ ( \partial_\nu V_{1 , \rho} ) , \, ( \partial^{\alpha} \partial^{\beta} A_{1 , \sigma} ) \big\}_{+} \big] \, .	\vphantom{\bigg(\bigg)}	\nonumber
		\end{align}
	The extended version is provided in App.\ \ref{app:lagrangians} in Eq.\ \eqref{eq:gva1lagexp}. The tree level decay formula is given by
		\begin{align}
			& \Gamma_{G_3 \rightarrow V_1 + A_1} ( m_{g_3} , m_{v_1} , m_{a_1} ) =	\vphantom{\Bigg(\Bigg)}	\label{eq:decay_g3_v1_a_1}
			\\
			= \, & c_{g_3 v_1 a_1}^{2} \, \frac{\big| \vec{k}_{v_1 , a_1} \big|^5}{210 \uppi \, m_{g_3}^2} \, \bigg[ \big| \vec{k}_{v_1 , a_1} \big|^2 \, \bigg( 3 + 2 \, \frac{m_{g_3}^2}{m_{a_1}^2} \bigg) + 7 \, m_{v_1}^2 \bigg] \times	\vphantom{\Bigg(\Bigg)}	\nonumber
			\\
			& \times \kappa_i \, \Theta( m_{g_3} - m_{v_1} - m_{a_1} ) \, .	\vphantom{\Bigg(\Bigg)}	\nonumber
		\end{align}
	where we used Eq.\ \eqref{eq:amplitude_square_w3_v1_a1}. The coefficients $\kappa_i$ are listed in Table \ref{tab:kappa_g3_v1_a1}. Results are listed in Table \ref{tabgva}.
		\begin{table}
			\centering
			\renewcommand{\arraystretch}{1.8}
			\caption{\label{tabgva} %
				Predictions for the branching ratios of glueball $G_3 (4200) \rightarrow V_1 + A_1$.
			}
			\begin{ruledtabular}
				\begin{tabular}[c]{ l c }
					\multicolumn{1}{c}{branching ratio}																										&	theory
					\\
					\colrule
					$\frac{\Gamma_{G_3 (4200) \rightarrow \bar{K}^{\ast} (892) \, K_{1 , A}}}{\Gamma_{G_3 (4200) \rightarrow \rho (770) \, a_1 (1260)}}$	&	$0.78$
					\\
					$\frac{\Gamma_{G_3 (4200) \rightarrow \omega (782) \, f_1 (1285)}}{\Gamma_{G_3 (4200) \rightarrow \rho (770) \, a_1 (1260)}}$			&	$0.29$
					\\
					$\frac{\Gamma_{G_3 (4200) \rightarrow \omega (782) \, f_1 (1420)}}{\Gamma_{G_3 (4200) \rightarrow \rho (770) \, a_1 (1260)}}$ 			&	$0.0009$
					\\
					$\frac{\Gamma_{G_3 (4200) \rightarrow \phi (1020) \, f_1 (1285)}}{\Gamma_{G_3 (4200) \rightarrow \rho (770) \, a_1 (1260)}}$			&	$0.0011$
					\\
					$\frac{\Gamma_{G_3 (4200) \rightarrow \phi (1020) \, f_1 (1420)}}{\Gamma_{G_3 (4200) \rightarrow \rho (770) \, a_1 (1260)}}$			&	$0.16$
				\end{tabular}
			\end{ruledtabular}
		\end{table}

\subsection{Decay process \texorpdfstring{$G_3 \rightarrow B_1 + P$}{G3 -> B1 + P}}

	The second dominant decay of the $J^{PC}=3^{--}$ glueball is the one into a pseudovector and a pseudoscalar meson. The interaction Lagrangian is given by
		\begin{align}
			\mathcal{L}_{g_3 b_1 p} = c_{g_3 b_1 p} \, G_3^{\mu\nu\rho} \, \mathrm{tr} \big[ \big\{ B_{1 , \mu} , \, ( \partial_{\nu} \partial_{\rho} P ) \big\}_{+} \big] \, ,	\vphantom{\bigg(\bigg)}	\label{eq:glueball_lagrangian}
		\end{align}
	and its extended version can be found in Eq.\ \eqref{eq:extended_lagrangian_g3_b1_p} in App.\ \ref{app:lagrangians}. This implies that the formula for the decay width is identical to Eq.\ \eqref{eq:decay_formula_w3_b1_p},
		\begin{align}
			& \Gamma_{G_3 \rightarrow B_1 + P} ( m_{g_3} , m_{b_1} , m_{p} ) =	\vphantom{\Bigg(\Bigg)}	\label{eq:decay-gsp_2}
			\\
			= \, & c_{g_3 b_1 p}^{2} \, \frac{\big| \vec{k}_{b_1 , p} \big|^{5}}{420 \uppi \, m_{g_3}^{2}} \, \Bigg( 7 + 3 \, \frac{\big| \vec{k}_{b_1 , p} \big|^{2}}{m_{b_1}^{2}} \Bigg) \times	\vphantom{\Bigg(\Bigg)}	\nonumber
			\\
			& \times \kappa_i \, \Theta( m_{g_3} - m_{b_1} - m_{p} ) \, .	\vphantom{\Bigg(\Bigg)}	\nonumber
		\end{align}
	The factors $\kappa_i$ are listed in Table \ref{tab:kappa_g3_b1_p}. Again, we study two possible choices for the mixing angle ($\beta_{b_1} \approx 0^{\circ}$ and $\beta_{b_1} \approx - 40^{\circ})$ and upon using $m_{K_{1 , B}} \approx m_{K_1 (1270)}$, we present the results in Table \ref{tabgbp}.
		\begin{table}
			\centering
			\renewcommand{\arraystretch}{1.8}
			\caption{\label{tabgbp} %
				Predictions for the branching ratios of glueball $G_3 (4200) \rightarrow B_1 + P$.
			}
			\begin{ruledtabular}
				\begin{tabular}[c]{ l c c }
					\multicolumn{1}{c}{ \multirow{2}{*}{ branching ratio } }																		&	theory for					&	theory for
					\\
					&	$\beta_{b_1} = 0^{\circ}$	&	$\beta_{b_1}=-40^{\circ}$
					\\
					\colrule
					$\frac{\Gamma_{G_3 (4200) \rightarrow K_{1 , B} \, K}}{\Gamma_{G_3 (4200) \rightarrow b_1 (1235) \, \pi}}$						&	$1.15$						&	$1.15$
					\\
					$\frac{\Gamma_{G_3 (4200) \rightarrow h_1 (1170) \, \eta}}{\Gamma_{G_3 (4200) \rightarrow b_1 (1235) \, \pi}}$					&	$0.17$						&	$0.33$
					\\
					$\frac{\Gamma_{G_3 (4200) \rightarrow h_1 (1170) \, \eta^{\prime} (958)}}{\Gamma_{G_3 (4200) \rightarrow b_1 (1235) \, \pi}}$	&	$0.12$						&	$0.001$
					\\
					$\frac{\Gamma_{G_3 (4200) \rightarrow h_1 (1415) \, \eta}}{\Gamma_{G_3 (4200) \rightarrow b_1 (1235) \, \pi}}$					&	$0.10$						&	$0.001$
					\\
					$\frac{\Gamma_{G_3 (4200) \rightarrow h_1 (1415) \, \eta^{\prime} (958)}}{\Gamma_{G_3 (4200) \rightarrow b_1 (1235) \, \pi}}$	&	$0.08$						&	$0.16$
				\end{tabular}
			\end{ruledtabular}
		\end{table}

\subsection{Decay process \texorpdfstring{$G_3 \rightarrow B_1 + A_1$}{G3 -> B1 + A1}}
	
	Moreover, we present the results for the glueball into a pseudovector and an axial-vector, $G_3 \rightarrow B_1 + A_1$ in Table \ref{tabgba} for $\beta_{a_1} \approx 0^{\circ}$. Namely, the Lagrangian for this decay can be obtained from Eq.\ \eqref{eq:glueball_lagrangian} as
		\begin{align}
			\mathcal{L}_{g_3 b_1 a_1} = c_{g_3 b_1 a_1} \, G_3^{\mu\nu\rho} \, \mathrm{tr} \big[ \big\{ B_{1 , \mu} , \, ( \partial_{\nu} A_{1 , \rho} ) \big\}_{+} \big] \, .	\vphantom{\bigg(\bigg)}
		\end{align}
	The extended form is presented in Eq.\ \eqref{eq:extended_lagrangian_g3_b1_a1} in App.\ \ref{app:lagrangians}. The corresponding decay formula is given by
		\begin{align}
			& \Gamma_{G_3 \rightarrow B_1 + A_1} ( m_{g_3} , m_{b_1} , m_{a_1} ) =	\vphantom{\Bigg(\Bigg)}
			\\
			= \, & c_{g_3 b_1 a_1}^{2} \, \frac{\big| \vec{k}_{b_1 , a_1} \big|^{3}}{840 \uppi \, m_{b_1}^{2} \, m_{a_1}^{2} \, m_{g_3}^{2}} \, \Big[ 6 \, | \vec{k}_{b_1 , a_1} \big|^{4} +	\vphantom{\Bigg(\Bigg)}	\nonumber
			\\
			& + 35 \, m_{b_1}^{2} \,  m_{a_1}^{2} + 14 \, \big| \vec{k}_{b_1 , a_1} \big|^{2} \, \big( m_{b_1}^{2} + m_{a_1}^{2} \big)  \Big] \times	\vphantom{\Bigg(\Bigg)}	\nonumber
			\\
			& \times \kappa_i \, \Theta ( m_{g_3} - m_{b_1} - m_{a_1} ) \, ,	\vphantom{\Bigg(\Bigg)}	\nonumber
		\end{align}
	where the $\kappa_i$'s are listed in Table \ref{tab:kappa_g3_b1_a1}.
	\begin{table}
		\centering
		\renewcommand{\arraystretch}{1.8}
		\caption{\label{tabgba} %
			Predictions for the branching ratios of glueball $G_3 (4200) \rightarrow B_1 + A_1$.
		}
		\begin{ruledtabular}
			\begin{tabular}[c]{ l c c }
				\multicolumn{1}{c}{ \multirow{2}{*}{ branching ratio } }																	&	theory for					&	theory for
				\\
																																			&	$\beta_{b_1} = 0^{\circ}$	&	$\beta_{b_1}=-40^{\circ}$
				\\
				\colrule
				$\frac{\Gamma_{G_3 (4200) \rightarrow K_{1 , B} \, K_{1 , A}}}{\Gamma_{G_3 (4200) \rightarrow b_1 (1235) \, a_1 (1260)}}$	&	$0.96$						&	$0.96$
				\\
				$\frac{\Gamma_{G_3 (4200) \rightarrow h_1 (1170) \, f_1 (1285)}}{\Gamma_{G_3 (4200) \rightarrow b_1 (1235) \, a_1 (1260)}}$	&	$0.34$						&	$0.20$
				\\
				$\frac{\Gamma_{G_3 (4200) \rightarrow h_1 (1170) \, f_1 (1420)}}{\Gamma_{G_3 (4200) \rightarrow b_1 (1235) \, a_1 (1260)}}$	&	$0$							&	$0.11$
				\\
				$\frac{\Gamma_{G_3 (4200) \rightarrow h_1 (1415) \, f_1 (1285)}}{\Gamma_{G_3 (4200) \rightarrow b_1 (1235) \, a_1 (1260)}}$	&	$0$							&	$0.092$
				\\
				$\frac{\Gamma_{G_3 (4200) \rightarrow h_1 (1415) \, f_1 (1420)}}{\Gamma_{G_3 (4200) \rightarrow b_1 (1235) \, a_1 (1260)}}$	&	$0.17$						&	$0.10$
			\end{tabular}
		\end{ruledtabular}
	\end{table}

\section{Discussion of the construction of the model}
\label{chap:discussion}

	In this section we discuss the interpretation of Eq.\ \eqref{lagtot}, the interaction Lagrangian $\mathcal{L}_{W,\text{total}}$, which was constructed by keeping only the lowest possible number of derivatives for a given interaction term. We justify this choice by following two different lines of argumentation. First, our effective model naturally emerges, if the Lagrangian $\mathcal{L}_{W,\text{total}}$ is interpreted as being part of a more general and complete chiral hadronic model in the vacuum, such as the (extended) linear sigma model \cite{Koch:1997ei}, which is aforementioned in the introduction. Second, a (Functional) Renormalization Group perspective also indicates that terms with the lowest derivatives might be retained as the dominant contributions within our effective hadronic model.\\


	The eLSM is built under the assumption of chiral symmetry, thus
		\begin{align*}
			U_\mathrm{L} (3) \times U_\mathrm{R} (3) = SU_\mathrm{L} (3) \times SU_\mathrm{R} (3) \times U_\mathrm{L} (1) \times U_\mathrm{R} (1) \, ,
		\end{align*}
	and the assumption of scale invariance. Besides the small explicit breaking of both of these symmetries, because of the nonzero bare quark masses, which derive from the electro-weak sector, the symmetries are also broken dynamically. The $SU_\mathrm{L} (3) \times SU_\mathrm{R} (3)$ breaks down to a $SU_\mathrm{V} (3)$ symmetry spontaneously, which was already discussed before. Also the scale invariance breaks down via gluonic quantum fluctuations and the emergence of a gluon condensate, which is also called the \textit{conformal/trace anomaly} \cite{Collins:1976yq,Peskin:1995ev,Greiner:2007zz}.
	Additionally, a second anomaly -- the chiral anomaly -- breaks $U_\mathrm{A}(1)$ and explains the heavy mass of $\eta^{\prime} (958)$ as well as the large pseudoscalar mixing angle \cite{tHooft:1986ooh}. As discussed in Ref.\ \cite{Giacosa:2017pos} and commented in various part of this work, this anomaly may affect also the masses and the decays of other nonets (those being part of homochiral multiplets). In the chiral limit and ignoring the chiral anomaly all the interaction terms of the eLSM are invariant under scale/dilatation transformations. Consequently, they are parametrized by dimensionless coupling constants.
	
	The Lagrangian $\mathcal{L}_{w_3 , \text{total}}$ \eqref{lagtot}, that we use in this work, contains various decay constants, see Table \ref{tab:interlag}. With the exception of the last entry, all of them are dimensionful. At a first sight, this feature seems in disagreement with the requirement that should descend from a chiral model such as the eLSM \cite{Parganlija:2012fy,Janowski:2014ppa}. Yet, this is not the case and a closer inspection also shows why the lowest number of derivatives should be kept. First, let us recall that the spontaneous symmetry breaking implies a peculiar scalar-axial-vector mixing, that is removed by the a shift of the axial-vector nonet $A_1$. In the simplified case with vanishing bare quark masses, where $U_\mathrm{V} (3)$ is exact (this is enough for our illustrative purposes), it takes the form \cite{Parganlija:2012fy,Eshraim:2020ucw}:
		\begin{align}
			A_1^{\mu} \mapsto A_1^{\mu} + Z \, w \, \partial^{\mu} P \, ,	\label{eq:shiftelsm}
		\end{align}
	where $Z \approx 1.6$ and $w \approx \frac{g_1 \phi_{N}}{m_{a_1}^{2}}$ where $g_1 \approx g_{\rho} \approx 5.5$, $m_{a_1} \simeq 1.4 \, \,$GeV, and $\phi_{N} \approx Z f_{\pi}$ ($f_{\pi} = 92.4 \, \,$MeV is the pion decay constant). 
	
	Now, the eLSM was studied in a variety of frameworks: Besides (pseudo)scalar and (axial-)vector mesons \cite{Parganlija:2012fy} and the aforementioned calculation of glueballs \cite{Janowski:2014ppa,Giacosa:2016hrm,Eshraim:2012jv}, it was also applied to study hybrid mesons \cite{Eshraim:2020ucw} and excited scalar mesons \cite{Parganlija:2016yxq}. Even if various mesonic nonets were not included in the eLSM yet, it is clear, which features such terms should have.
	
	Within this framework, the $W_3 P P$-Lagrangian,
		\begin{align}
			\mathcal{L}_{w_3 p p} = g_{w_3 p p} \, \mathrm{tr} \big[ W^{\mu\nu\rho} \, \big[ P , \, ( \partial_{\mu} \partial_{\nu} \partial_{\rho} P ) \big]_{-} \big] \, ,
		\end{align}
	where $g_{w_3 p p}$ has dimension $E^{-2}$ and the $W_3 A_1 P$-Lagrangian,
		\begin{align}
			\mathcal{L}_{w_3 a_1 p} = g_{w_3 a_1 p} \, \mathrm{tr} \big[ W^{\mu\nu\rho} \, \big[ A_{1 , \mu} , \, ( \partial_{\nu} \partial_{\rho} P ) \big]_{-} \big] \, ,
		\end{align}
	where $g_{w_3 a_1 p}$ has dimension $E^{-1},$ can be seen as the result of the interaction Lagrangian
		\begin{align}
			\mathcal{L}_{w_3 a_1 a_1} = g_{w_3 a_1 a_1} \, \mathrm{tr} \big( W^{\mu\nu\rho} \, \big[ A_{1 , \mu} , \, \partial_{\nu} A_{1 , \rho} \big]_{-} \big) \, ,
		\end{align}
	involving solely the dimensionless coupling $g_{w_3 a_1 a_1}$ via the shift of Eq.\ \eqref{eq:shiftelsm} applied once/twice. In other words, only the Lagrangian $\mathcal{L}_{w_3 a_1 a_1}$ is part of a generalized eLSM Lagrangian that includes fields with $J = 3$, in agreement with dilatation invariance. Then, upon spontaneous symmetry breaking and mixing/shifts, the Lagrangians $\mathcal{L}_{w_3 a_1 p}$ and $\mathcal{L}_{w_3 p p}$ are a consequence of $\mathcal{L}_{w_3 a_1 a_1}$, thus explaining how dimensional couplings appear in the model, even if one starts solely with dimensionless ones.
	
	While the decay $W_3 \rightarrow A_1 \, A_1$ is not kinematically allowed, the corresponding term explains how the first and the fifth entries of Table \ref{tab:interlag} emerge. The simple relations
		\begin{align}
			g_{w_3 p p} \approx Z^{2} \, w^{2} \, g_{w_3 a_1 a_1 }
		\end{align}
	and
		\begin{align}
			g_{w_3 a_1 p} \approx Z \, w \, g_{w_3 a_1 a_1}
		\end{align}
	follow. Moreover, the ratio $g_{w_3 a_1 p} / g_{w_3 p p} \approx (  Z \, w )^{-1}$, out of which we may estimate that
		\begin{align}
			g_{w_3 a_1 p}^{2} \approx 3 \cdot 10^{-4} \, \, \text{MeV}^{-2}
		\end{align}
	is obtained (it should be however stressed that the present heuristic discussion cannot provide a precise determination of ratios of couplings, but it solely offers a guide to understand the origin of the model). This value leads to $\Gamma_{\rho_{3} (1690) \rightarrow a_1 (1260) \, \pi} \approx 2 \, \,$MeV, as previously mentioned.
	
	Next, let us consider the $W_3 B_1 P$-Lagrangian,
	\begin{align}
			\mathcal{L}_{w_3 b_1 p} = g_{w_3 b_1 p} \, \mathrm{tr} \big( W^{\mu\nu\rho} \, \big\{ B_{1 , \mu} , \, ( \partial_{\nu} \partial_{\rho} P ) \big\}_{+} \big) \, ,
		\end{align}
	with $g_{w_3 b_1 p}$, which can be seen as emerging from
		\begin{align}
			\mathcal{L}_{w_3 b_1 a_1} = g_{w_3 b_1 a_1} \, \mathrm{tr} \big[ W^{\mu\nu\rho} \, \big\{ B_{1 , \mu} , \, ( \partial_{\nu} A_{1 , \rho} ) \big\}_{+} \big] \, ,
		\end{align}
	with a dimensionless coupling $g_{w_3 b_1 a_1}$. The $A_1$-shift introduced above implies that
		\begin{align}
			g_{w_3 b_1 p} \approx Z \, w \, g_{w_3 b_1 a_1} \, .
		\end{align}
	Next, let us study the remaining terms. The $W_3 V_1 V_1$-Lagrangian is already dilatation invariant, while the two other Lagrangians $\mathcal{L}_{w_3 v_1 p}$ and $\mathcal{L}_{w_3 a_2 p}$ are not, since they involve the coupling $g_{w_3 v_1 p}$ and $g_{w_3 a_2 p}$, having dimensions $E^{-3}$ and $E^{-2}$. In this case, one could build analogous Lagrangians $\mathcal{L}_{w_3 v_1 a_1}$ and $\mathcal{L}_{w_3 a_2 a_1}$, but their coupling coupling constants $g_{w_3 v_1 a_1} $ and $g_{w_3 a_2 a_1}$ still have dimension $E^{-2}$ and $E^{-1}$. This is however understandable, since these terms involve the Levi-Civita pseudotensor $\varepsilon_{\mu\nu\rho\sigma}$ and hence are a manifestation of the the chiral anomaly.
	
	In conclusion, terms being part of the model are (i) dilatation invariant or (ii) can be obtained from dilatation invariant terms of a more general underlying chiral model via the shift of the axial-vector nonet or (iii) they are linked to the chiral anomaly. Now, one also understand why only terms with the lowest number of derivatives appear.
	
	For the effective interaction terms of the $3^{--}$ glueball, the argument is completely analogous.\\
	
	Still, we also provide a brief alternative motivation for the construction of the model from a renormalization group perspective. In this framework, we know that the UV limit of all hadronic physics is QCD, where the degrees of freedom are quarks and gluons. However, we want to describe low-energy QCD and the effective hadronic degrees of freedom. By integrating out quantum fluctuations of gluons and quarks via renormalization group transformations from the UV scales to the IR scales including dynamical hadronization, one finally ends up with an effective hadronic theory, were hadrons are the effective degrees of freedom. While integrating out quantum fluctuations momentum shell after momentum shell, all kinds of effective couplings/vertices, which are in accordance with the UV symmetries of the theory -- the symmetries of QCD -- will be generated. This  process is effectively described as the RG flow of the coupling constants of the theory, where effective couplings of hadrons are initialy zero and only generated dynamically during the RG flow. The same applies to the effective fermionic and hadronic quantum fields.
	
	An efficient and modern framework to describe this process, which effectively is an implementation of Wilsons idea of the RG \cite{Wilson:1971bg,Wilson:1971dh,Polchinski:1983gv}, is the \textit{Functional Renormalization Group} (FRG) in its formulation via the \textit{Exact Renormalization Group Equation} \cite{Wetterich:1992yh,Morris:1993qb}, see for example Refs.\ \cite{Gies:2006wv,Pawlowski:2005xe,Polonyi:2001se,Gies:2001nw,Braun:2007bx,Braun:2014ata} and Ref.\ \cite{Dupuis:2020fhh} for an comprehensive up-to-date review. For our purpose, it is solely important to note that the FRG is most efficiently formulated on the level of the quantum effective IR action $\Gamma [ \Phi ]$, which is the generating functional of all $1$PI-$n$-point-correlation functions. The FRG allows to calculate $\Gamma [ \Phi ]$ -- at least within certain truncations -- from the UV theory, which here is QCD. The interesting part is that the quantum effective IR action contains all kinds of effective vertices, that were generated during the RG flow and contain all information about the higher energy scales. For the calculation of decays \textit{etc.}, however, only these effective couplings are relevant and calculations are performed at tree level, because all loop contributions are already contained in the couplings and because we are working with the $1$PI-$n$-point-correlation functions. Dynamical symmetry breaking is realized in this framework as a nontrivial vacuum/ground-state, which realizes as a minimum of the effective action $\Gamma [\Phi]$. All $1$PI-$n$-point functions must be extracted at this physical point, which yields shifts like in Eq.\ \eqref{eq:shiftelsm}.
	
	In practice, it is almost impossible to really do this calculation from first principles, especially concerning couplings for mesons with higher spin and large masses. Nevertheless, we know from the previous argumentation that a low IR-effective action $\Gamma [ \Phi ]$ must contain all possible interaction terms, which are in accordance with the symmetry of the theory and if shifted to the physical point, then it must contain all symmetries of the vacuum of the theory.	This also applies to the effective couplings of our spin-$3$ mesons. Thus, we can use the following argument: instead of calculating all interaction terms and couplings, which are in accordance with the symmetries of QCD, via an RG flow from first principles (which is anyhow impossible), we just write down all possible effective interaction terms, that are invariant under the residual IR vacuum-symmetries of QCD and simply fit the effective coupling constants to experimental data. This approach additionally justifies the use of tree level-calculations, because experimental measurements for decay widths \textit{etc.}\ already contain all information about the high energy scales of QCD and so does an effective IR action $\Gamma [ \Phi ]$, from which information is extracted at tree level. We therefore interpret all effective Lagrangians $\mathcal{L}_{w_3 \circ \circ}$ as being part of an effective IR action $\Gamma [\Phi]$ for QCD.
	
	Last, we argue, that as long as the momentum exchange in processes like three-point-interactions (decays) is small, we can expand an effective IR action in powers of momenta of the effective hadronic fields. In the context of the FRG, this approach is denoted as a \textit{derivative expansion} and turned out as a decent approximation in a lot of low-energy models of QCD \cite{Grahl:2013pba,Eser:2015pka,Rennecke:2016tkm,Pawlowski:2014zaa,Eser:2018jqo,Divotgey:2019xea,Eser:2019pvd,Cichutek:2020bli,Otto:2019zjy,Otto:2020hoz,Jung:2019nnr,Heller:2015box}.\\
	
	All in all, both approaches complement each other  and explain that we only retain terms that are of the lowest orders in the derivatives of the fields, that are of leading order in a large-$N_\mathrm{c}$ expansion and that are flavor invariant.

\section{Conclusions}
\label{sec:conclusions}

	In this paper, we have studied the decays of the lightest mesonic $\bar{q} q$ nonet with quantum numbers $J^{PC}=3^{--}$ using an effective QFT model based on flavor symmetry. Our model retained only the dominant terms in an large-$N_\mathrm{c}$ expansion and the lowest possible number of derivatives. By comparing the theoretical results with the current status of experimental data for decay width and some known branching ratios, which are all reported by PDG \cite{Zyla:2020zbs}, we conclude that the $\bar{q} q$ assignment works quite well. Still, we remark that there are also other rather successful approaches different from the $\bar{q} q$ picture towards a coherent description of the nature of the  $J^{PC}=3^{--}$ states, see \textit{e.g.}\ via multi-$\rho (770)$ resonances \cite{Roca:2010tf,YamagataSekihara:2010qk}. However, also in our work some of the decay channels deserve deeper theoretical and experimental investigation. Additionally, we presented decay ratios of a putative and not yet detected $J^{PC} = 3^{--}$ glueball by considering decays into the vector-pseudoscalar and pseudovector-pseudoscalar mesonic pairs. In summary, we provide many qualitative predictions of strong and radiative decays of conventional $J^{PC} = 3^{--}$ mesons and the $J^{PC} = 3^{--}$ glueball for future experimental tests.

\begin{acknowledgments}
	S.J.\ acknowledges financial support through the project \textit{Development Accelerator of the Jan Kochanowski University of Kielce}, co-financed by the \textit{European Union} under the \textit{European Social Fund}, with no. POWR.03.05. 00-00-Z212 / 18.
	
	A.K.\ acknowledges the support of the \textit{Deutsche Forschungsgemeinschaft} (DFG, German Research Foundation) through the collaborative research center trans-regio  CRC-TR 211 \textit{Strong-interaction matter under extreme conditions} -- project number 315477589 -- TRR 211. A.K.\ also acknowledges the support of the \textit{Friedrich-Naumann-Foundation for Freedom}, the \textit{Giersch Foundation} and the \textit{Helmholtz Graduate School for Hadron and Ion Research}.
	
	F.G.\ acknowledges support from the \textit{Polish National Science Centre} (NCN) through the \textit{OPUS} projects 2019/33/B/ST2/00613 and no. 2018/29/B/ST2/02576.

	The authors thank J.~Braun, F.~Divotgey, J.~Eser, M.~Piotrowska, D.~H.~Rischke, and M.~J.~Steil for useful discussions. 
\end{acknowledgments}

\appendix

\section{Masses and strange-nonstrange mixing angles}
\label{app:masses_and_mixing}

	For a certain nonet, we denote as $\vec{a}$ the triplet of fields describing the isospin $I=1$ elements $\bar{d} u$, $\bar{u} d$, ${\frac{1}{\sqrt{2}} \, (\bar{u} u - \bar{d} d)}$, with $K^{\pm}$ the complex kaonic fields representing the doublet $\bar{s} u$ and $\bar{u} s$, and $\bar{K}^0$ the analogous fields $\bar{d} s$ and $\bar{s} d$ for the neutral kaonic elements, and as $f_8$ and $f_0$ the octet and the singlet elements, whose quark-antiquark configurations are $\frac{1}{\sqrt{6}} \, ( \bar{u} u + \bar{d} d - 2 \bar{s} s )$ and $\frac{1}{\sqrt{3}} \, ( \bar{u} u + \bar{d} d + \bar{s} s )$, respectively. The corresponding $SU_\mathrm{V}(3)$-invariant mass term reads (omitting possible Lorentz indices for simplicity),	
		\begin{align}
			& \tfrac{1}{2} \, m_{a}^2 \, \vec{a}^{\, 2} + m_{K}^2 \, ( K^+ \, K^- + \bar{K}^0 \, K^0 ) +	\vphantom{\bigg(\bigg)}
			\\
			& + \tfrac{1}{2} \, m_8^2 \, f_8^2 + \tfrac{1}{2} \, m_0^2 \, f_0^2 + m_{08} \, f_0 \, f_8 \, ,	\vphantom{\bigg(\bigg)}	\nonumber
		\end{align}
	where also a singlet-octet mixing parameter $m_{08}$ has been introduced. The masses of the fields $\vec{a}$, $\bar{K}^{0/\pm}$, and $f_8$ are expressed as
		\begin{align}
			m_a^2 = \, & m_{\mathrm{nonet}}^2 + 2 \, \delta_n^2 \, ,	\vphantom{\bigg(\bigg)}
			\\
			m_K^2 = \, & m_{\mathrm{nonet}}^2 + \delta_n^2 + \delta_s^2 \, ,	\vphantom{\bigg(\bigg)}
			\\
			m_8^2 = \, & m_{\mathrm{nonet}}^2 + \tfrac{2}{3} \, \delta_n^2 + \tfrac{4}{3} \, \delta_s^2 \, ,	\vphantom{\bigg(\bigg)}
		\end{align}
	where $m_{\mathrm{nonet}}$ is the mass of the nonet in the chiral limit as a consequence of spontaneous symmetry breaking (see \textit{e.g.}\ Ref.~\cite{Parganlija:2012fy}), and $\delta_n^2$ represents the (almost negligible) contribution of the bare nonstrange quarks $u$ and $d$ and $\delta_s^2$ the (on hadronic scales) small but nonnegligible contribution of the bare quark $s$. (Besides the pseudoscalar nonet, one simply approximates $\delta_n \propto m_u \simeq m_d$ and $\delta_s \propto m_s$.) From the expressions above it follows that
		\begin{align}
			m_8^2 = \, & \tfrac{4}{3} \, m_{K}^2 - \tfrac{1}{3} \, m_{a}^2  =	\vphantom{\bigg(\bigg)}
			\\
			= \, & m_{\mathrm{nonet}}^2 + \tfrac{2}{3} \, \delta_n^2 + \tfrac{4}{3} \, \delta_s^2 \, .	\vphantom{\bigg(\bigg)}	\nonumber
		\end{align}
	The mass of the singlet member of the nonet may contain an additional unknown contribution (due to the anomaly and/or conversion to gluons), parametrized by:
		\begin{align}
			m_0^2 = \, & m_{\mathrm{nonet}}^2 + \tfrac{4}{3} \, \delta_n^2 + \tfrac{2}{3} \, \delta_s^2 + \alpha =	\vphantom{\bigg(\bigg)}	\label{singletmass}
			\\
			= \, & \tfrac{2}{3} \, m_{K}^2 + \tfrac{1}{3} \, m_{a}^2 + \alpha \, .	\vphantom{\bigg(\bigg)}	\nonumber
		\end{align}
	Next, we diagonalize the system in the isoscalar sector. We then introduce the physical fields $f$ and $f^{\prime }$ as
		\begin{align}
			\begin{pmatrix}
				- f^{\prime }
				\\
				f
			\end{pmatrix}
			=
			\begin{pmatrix}
				\cos \beta_{\text{\tiny PDG}}	&	- \sin \beta_{\text{\tiny PDG}}
				\\
				\sin \beta_{\text{\tiny PDG}}	&	\cos \beta_{\text{\tiny PDG}}
			\end{pmatrix}
			\begin{pmatrix}
				f_8
				\\
				f_0
			\end{pmatrix}	\label{mixffp}
		\end{align}
	where $\beta_{\text{\tiny PDG}}$ is the corresponding mixing angle,
		\begin{align}
			\tan \big( 2 \beta_{\text{\tiny PDG}} \big) = \, & \frac{2 m_{08}}{m_0^2 - m_8^2} \, ,	\label{betapdg1}
		\end{align}
	which is obtained via diagonalization of the mass matrix together with the relation,
		\begin{align}
			m_8^2 = \, & m_{f^{\prime }}^2 \cos^2 \beta_{\text{\tiny PDG}} + m_f^2 \sin^2 \beta_{\text{\tiny PDG}} \, ,	\vphantom{\bigg(\bigg)}	\label{m8}
			\\
			m_0^2 = \, & m_f^2 \cos^2 \beta_{\text{\tiny PDG}} + m_{f^{\prime }}^2 \sin^2 \beta_{\text{\tiny PDG}} \, .	\vphantom{\bigg(\bigg)}	\label{m0}
		\end{align}
	In Eq.~\eqref{mixffp} the extra minus sign in front of $f^{\prime }$ is introduced for later convenience. Using Eq.~\eqref{m8} one gets
		\begin{align}
			& \frac{3 m_8^2 - 3 m_{f^\prime}^2}{ - 3 m_8^2 + 3 m_f^2} =	\vphantom{\bigg(\bigg)}
			\\
			= \, & \frac{3 m_{f^\prime}^2 \cos^2 \beta_{\text{\tiny PDG}} + 3 m_f^2 \sin^2 \beta_{\text{\tiny PDG}} - 3 m_{f^\prime}^2}{- 3 m_{f^\prime}^2 \cos^2 \beta_{\text{\tiny PDG}} - 3 m_f^2 \sin^2 \beta_{\text{\tiny PDG}} + 3 m_f^2} =	\vphantom{\bigg(\bigg)}	\nonumber
			\\
			= \, & \tan^2 \beta_{\text{\tiny PDG}} \, ,	\vphantom{\bigg(\bigg)}	\nonumber
		\end{align}
	but on the other hand using the above expressions for the masses the \textit{l.h.s.}\ can be written as
		\begin{align}
			\frac{3 m_8^2 - 3 m_{f^\prime}^2}{- 3 m_8^2 + 3 m_f^2} = \frac{4 m_K^2 - m_a^2 - 3 m_{f^\prime}^2}{- 4 m_K^2 + m_a + 3 m_f^2} \, ,
		\end{align}
	leading to expressions used in the PDG \cite{Zyla:2020zbs} to calculate the mixing angle in the singlet-octet basis:
		\begin{align}
			\tan^2 \beta_{\text{\tiny PDG}} = \frac{4 m_k^2 - m_a^2 - 3 m_{f^\prime}^2}{- 4 m_K^2 + m_a + 3 m_f^2} \, .
		\end{align}
	In the last step we relate this to the (for our purposes) more convenient nonstrange-strange basis $f_N$ and $f_S$ that corresponds to the configurations $\frac{1}{\sqrt{2}} \, ( \bar{u} u + \bar{d} d )$ and $\bar{s} s$. Via
		\begin{align}
			\begin{pmatrix}
				f_8
				\\
				f_0
			\end{pmatrix}
			=
			\tfrac{1}{\sqrt{3}}
			\begin{pmatrix}
				- \sqrt{2}	&	1
				\\
				1			&	\sqrt{2}
			\end{pmatrix}
			\begin{pmatrix}
				f_S
				\\
				f_N
			\end{pmatrix} \, ,
		\end{align}
	we introduce the strange-nonstrange mixing $\beta $ as:
		\begin{align}
			& \begin{pmatrix}
				- f^\prime
				\\
				f
			\end{pmatrix} =	\label{twobetas}
			\\
			= \, &
			\tfrac{1}{\sqrt{3}}
			\begin{pmatrix}
				\cos \beta_{\text{\tiny PDG}}	&	- \sin \beta_{\text{\tiny PDG}}
				\\
				\sin \beta_{\text{\tiny PDG}}	&	\cos \beta_{\text{\tiny PDG}}
			\end{pmatrix}
			\begin{pmatrix}
				- \sqrt{2}	&	1
				\\
				1			&	\sqrt{2}
			\end{pmatrix}
			\begin{pmatrix}
				f_S
				\\
				f_N
			\end{pmatrix} =	\nonumber
			\\
			= \, &
			\begin{pmatrix}
				- \cos \beta	&	\sin \beta
				\\
				\sin \beta		&	\cos \beta
			\end{pmatrix}
			\begin{pmatrix}
				f_S
				\\
				f_N
			\end{pmatrix} \, ,	\nonumber
		\end{align}
	or, as reported in the main text, as:
		\begin{align}
			\begin{pmatrix}
				f
				\\
				f^\prime
			\end{pmatrix}
			=
			\begin{pmatrix}
				\cos \beta		&	\sin \beta
				\\
				- \sin \beta	&	\cos \beta
			\end{pmatrix}
			\begin{pmatrix}
				f_N
				\\
				f_S
			\end{pmatrix}
		\end{align}
	The strange-nonstrange mixing angle $\beta$ and the angle $\beta_{\text{\tiny PDG}}$ can be obtained by Eq.~\eqref{twobetas} as:
		\begin{align}
			& \sqrt{\tfrac{1}{3}} \cos \beta_{\text{\tiny PDG}} - \sqrt{\tfrac{2}{3}} \sin \beta_{\text{\tiny PDG}} =	\vphantom{\bigg(\bigg)}
			\\
			= \, & \sin \beta \equiv \sin ( - \beta_{\text{\tiny PDG}} + \beta_0 ) \, ,	\vphantom{\bigg(\bigg)}	\nonumber
		\end{align}
	thus
		\begin{align}
			\beta = -\beta_{\text{\tiny PDG}} + \beta_0
		\end{align}
	with
		\begin{equation}
			\beta_0 = \arccos \sqrt{\tfrac{2}{3}} \simeq 35.3^\circ \, .
		\end{equation}
	The mixing angle $\beta_{\text{\tiny PDG}}$ (and hence $\beta $) has been calculated in the PDG for certain well-known mesonic nonets (pseudoscalar, vector, as well as $J^{P C} = 2^{+ +}$ and $J^{P C} = 3^{- -}$). While for the pseudoscalar case we employ the result of Ref.~\cite{AmelinoCamelia:2010me} (where not only masses but also decays are used to get $\beta$), in the remaining three cases we used the value $\beta_{\text{\tiny PDG}}$ to determine the strange-nonstrange mixing angle. With the exception of the peculiar pseudoscalar nonet, where the axial anomaly is large, the other values of $\beta$ calculated from $\beta_{\text{\tiny PDG}}$ are quite small -- in agreement with the theoretical expectations presented in Ref.~\cite{Giacosa:2017pos}. As a consequence, the parameter $\alpha$ of Eq.~\eqref{singletmass} is negligible. In particular, for $J = 3$ the numerical value $\beta = \beta_{w_3} = 3.5^{\circ}$ has been used throughout our calculations. When future experimental data will be more accurate, one could also include $\beta_{w_3}$ as a fit parameter in an improved model. In this respect, better experimental data for radiative decays would be useful.

\section{Extended form of the Lagrangians of the model}
\label{app:lagrangians}

	In this appendix, we provide the explicit form of the interaction Lagrangians in Table \ref{tab:interlag} and in Table \ref{tab:glueball_decay_lagrangians}.
	
		\begin{align}
			& \mathcal{L}_{w_3 p p} =	\vphantom{\bigg(\bigg)}	\label{eq:extended_lagrangian_w3_p_p}
			\\
			= \, & g_{w_3 p p} \, \mathrm{tr} \big( W_3^{\mu\nu\rho} \big[ P ,\, (\partial_\mu \partial_\nu \partial_\rho P) \big]_{-} \big) =	\vphantom{\bigg(\bigg)}	\nonumber
			\\
			= \, & \frac{g_{w_3 p p}}{4} \, \Big(	\vphantom{\bigg(\bigg)}	\nonumber
			\\
			& + \rho^{0,\mu\nu\rho}_3 \Big\{ + \bar{K}^0 (\partial_\mu \partial_\nu \partial_\rho K^0) - K^0 (\partial_\mu \partial_\nu \partial_\rho \bar{K}^0) +	\vphantom{\bigg(\bigg)}	\nonumber
			\\
			& \quad + K^+ (\partial_\mu \partial_\nu \partial_\rho K^-) - K^- (\partial_\mu \partial_\nu \partial_\rho K^+) +	\vphantom{\bigg(\bigg)}	\nonumber
			\\
			& \quad + 2 \, \big[ \pi^+ (\partial_\mu \partial_\nu \partial_\rho \pi^-) - \pi^- (\partial_\mu \partial_\nu \partial_\rho \pi^+) \big] \Big\} +	\vphantom{\bigg(\bigg)}	\nonumber
			\\
			& + \rho^{+,\mu\nu\rho}_3 \Big\{ \sqrt{2} \, \big[ - K^- (\partial_\mu \partial_\nu \partial_\rho K^0) + K^0 (\partial_\mu \partial_\nu \partial_\rho K^-) \big] +	\vphantom{\bigg(\bigg)}	\nonumber
			\\
			& \quad + 2 \, \big[ \pi^- (\partial_\mu \partial_\nu \partial_\rho \pi^0) - \pi^0 (\partial_\mu \partial_\nu \partial_\rho \pi^-) \big] \Big\} +	\vphantom{\bigg(\bigg)}	\nonumber
			\\
			& + \rho^{-,\mu\nu\rho}_3 \Big\{ \sqrt{2} \, \big[ K^+ (\partial_\mu \partial_\nu \partial_\rho \bar{K}^0) - \bar{K}^0 (\partial_\mu \partial_\nu \partial_\rho K^+) \big] +	\vphantom{\bigg(\bigg)}	\nonumber
			\\
			& \quad + 2 \, \big[ - \pi^+ (\partial_\mu \partial_\nu \partial_\rho \pi^0) + \pi^0 (\partial_\mu \partial_\nu \partial_\rho \pi^+) \big] \Big\} +	\vphantom{\bigg(\bigg)}	\nonumber
			\\
			& + \bar{K}^{\ast 0, \mu\nu\rho}_3 \Big\{ + K^0 (\partial_\mu \partial_\nu \partial_\rho \pi^0) - \pi^0 (\partial_\mu \partial_\nu \partial_\rho K^0) +	\vphantom{\bigg(\bigg)}	\nonumber
			\\
			& \quad + \sqrt{2} \, \big[ - K^+ (\partial_\mu \partial_\nu \partial_\rho \pi^-) + \pi^- (\partial_\mu \partial_\nu \partial_\rho K^+) \big] +	\vphantom{\bigg(\bigg)}	\nonumber
			\\
			& \quad + K^0 \big[ (\partial_\mu \partial_\nu \partial_\rho \eta) ( - \cos \beta_p + \sqrt{2} \sin \beta_p ) +	\vphantom{\bigg(\bigg)}	\nonumber
			\\
			& \qquad + (\partial_\mu \partial_\nu \partial_\rho \eta^\prime) ( \sin \beta_p + \sqrt{2} \cos \beta_p ) \big] -	\vphantom{\bigg(\bigg)}	\nonumber
			\\
			& \quad - \big[ \eta ( - \cos \beta_p + \sqrt{2} \sin \beta_p ) +	\vphantom{\bigg(\bigg)}	\nonumber
			\\
			& \qquad + \eta^\prime ( \sin \beta_p + \sqrt{2} \cos \beta_p ) \big] (\partial_\mu \partial_\nu \partial_\rho K^0) \Big\} +	\vphantom{\bigg(\bigg)}	\nonumber
			\\
			& + K^{\ast 0, \mu\nu\rho}_3 \Big\{ - \bar{K}^0 (\partial_\mu \partial_\nu \partial_\rho \pi^0) + \pi^0 (\partial_\mu \partial_\nu \partial_\rho \bar{K}^0 ) +	\vphantom{\bigg(\bigg)}	\nonumber
			\\
			& \quad + \sqrt{2} \, \big[ K^- (\partial_\mu \partial_\nu \partial_\rho \pi^+ - \pi^+ (\partial_\mu \partial_\nu \partial_\rho K^-) \big] -	\vphantom{\bigg(\bigg)}	\nonumber
			\\
			& \quad - \bar{K}^0 \big[ (\partial_\mu \partial_\nu \partial_\rho \eta) ( - \cos \beta_p + \sqrt{2} \sin \beta_p ) +	\vphantom{\bigg(\bigg)}	\nonumber
			\\
			& \qquad + (\partial_\mu \partial_\nu \partial_\rho \eta^\prime) ( \sin \beta_p + \sqrt{2} \cos \beta_p ) \big] +	\vphantom{\bigg(\bigg)}	\nonumber
			\\
			& \quad + \big[ \eta ( - \cos \beta_p + \sqrt{2} \sin \beta_p ) +	\vphantom{\bigg(\bigg)}	\nonumber
			\\
			& \qquad + \eta^\prime ( \sin \beta_p + \sqrt{2} \cos \beta_p ) \big] (\partial_\mu \partial_\nu \partial_\rho \bar{K}^0) \Big\} +	\vphantom{\bigg(\bigg)}	\nonumber
			\\
			& + K^{\ast +, \mu\nu\rho}_3 \Big\{ + K^- (\partial_\mu \partial_\nu \partial_\rho \pi^0) - \pi^0 (\partial_\mu \partial_\nu \partial_\rho K^-) +	\vphantom{\bigg(\bigg)}	\nonumber
			\\
			& \quad + \sqrt{2} \, \big[ \bar{K}^0 (\partial_\mu \partial_\nu \partial_\rho \pi^-) - \pi^- (\partial_\mu \partial_\nu \partial_\rho \bar{K}^0) \big] -	\vphantom{\bigg(\bigg)}	\nonumber
			\\
			& \quad - K^- \big[ (\partial_\mu \partial_\nu \partial_\rho \eta) ( - \cos \beta_p + \sqrt{2} \sin \beta_p ) +	\vphantom{\bigg(\bigg)}	\nonumber
			\\
			& \qquad + (\partial_\mu \partial_\nu \partial_\rho \eta^\prime) ( \sin \beta_p + \sqrt{2} \cos \beta_p ) \big] +	\vphantom{\bigg(\bigg)}	\nonumber
			\\
			& \quad + \big[ \eta ( - \cos \beta_p + \sqrt{2} \sin \beta_p ) +	\vphantom{\bigg(\bigg)}	\nonumber
			\\
			& \qquad + \eta^\prime ( \sin \beta_p + \sqrt{2} \cos \beta_p ) \big] (\partial_\mu \partial_\nu \partial_\rho K^-) \Big\} +	\vphantom{\bigg(\bigg)}	\nonumber
			\\
			& + K^{\ast -, \mu\nu\rho}_3 \Big\{ - K^+ (\partial_\mu \partial_\nu \partial_\rho \pi^0) + \pi^0 (\partial_\mu \partial_\nu \partial_\rho K^+) +	\vphantom{\bigg(\bigg)}	\nonumber
			\\
			& \quad + \sqrt{2} \, \big[ - K^0 (\partial_\mu \partial_\nu \partial_\rho \pi^+) + \pi^+ (\partial_\mu \partial_\nu \partial_\rho K^0) \big] +	\vphantom{\bigg(\bigg)}	\nonumber
			\\
			& \quad + K^+ \big[ (\partial_\mu \partial_\nu \partial_\rho \eta) ( - \cos \beta_p + \sqrt{2} \sin \beta_p ) +	\vphantom{\bigg(\bigg)}	\nonumber
			\\
			& \qquad + (\partial_\mu \partial_\nu \partial_\rho \eta^\prime) ( \sin \beta_p + \sqrt{2} \cos \beta_p ) \big] -	\vphantom{\bigg(\bigg)}	\nonumber
			\\
			& \quad - \big[ \eta ( - \cos \beta_p + \sqrt{2} \sin \beta_p ) +	\vphantom{\bigg(\bigg)}	\nonumber
			\\
			& \qquad + \eta^\prime ( \sin \beta_p + \sqrt{2} \cos \beta_p ) \big] (\partial_\mu \partial_\nu \partial_\rho K^+) \Big\} +	\vphantom{\bigg(\bigg)}	\nonumber
			\\
			& + \omega^{\mu\nu\rho}_3 \Big\{ ( - \cos \beta_{w_3} + \sqrt{2} \sin \beta_{w_3} ) \times	\vphantom{\bigg(\bigg)}	\nonumber
			\\
			& \quad \times \big[ \bar{K}^0 (\partial_\mu \partial_\nu \partial_\rho K^0) - K^0 (\partial_\mu \partial_\nu \partial_\rho \bar{K}^0) -	\vphantom{\bigg(\bigg)}	\nonumber
			\\
			& \qquad - K^+ (\partial_\mu \partial_\nu \partial_\rho K^-) + K^- (\partial_\mu \partial_\nu \partial_\rho K^+) \big] \Big\} +	\vphantom{\bigg(\bigg)}	\nonumber
			\\
			& + \phi^{\mu\nu\rho}_3 \Big\{ ( \sin \beta_{w_3} + \sqrt{2} \cos \beta_{w_3} ) \times	\vphantom{\bigg(\bigg)}	\nonumber
			\\
			& \quad \times \big[ \bar{K}^0 (\partial_\mu \partial_\nu \partial_\rho K^0) - K^0 (\partial_\mu \partial_\nu \partial_\rho \bar{K}^0) -	\vphantom{\bigg(\bigg)}	\nonumber
			\\
			& \qquad - K^+ (\partial_\mu \partial_\nu \partial_\rho K^-) + K^- (\partial_\mu \partial_\nu \partial_\rho K^+) \big] \Big\} \Big) \, .	\vphantom{\bigg(\bigg)}	\nonumber
		\end{align}
	
		\begin{align}
			& \mathcal{L}_{w_3 v_1 p} =	\vphantom{\bigg(\bigg)}	\label{eq:extended_lagrangian_w3_v1_p}
			\\
			= \, & g_{w_3 v_1 p} \, \varepsilon^{\mu\nu\rho\sigma}\, \mathrm{tr} \big[ W_{3 , \mu\alpha\beta} \big\{ (\partial_\nu V_{1 , \rho}) , \, (\partial^\alpha \partial^\beta \partial_\sigma P) \big\}_{+} \big] =	\vphantom{\bigg(\bigg)}	\nonumber
			\\
			= \, & \frac{g_{w_3 v_1 p}}{4} \, \varepsilon^{\mu\nu\rho\sigma} \, \Big(	\vphantom{\bigg(\bigg)}	\nonumber
			\\
			& + \rho^0_{3 , \mu\alpha\beta} \Big\{ - (\partial_\nu \bar{K}^{\ast 0}_\rho) (\partial^\alpha \partial^\beta \partial_\sigma K^0) -	\vphantom{\bigg(\bigg)}	\nonumber
			\\
			& \quad - (\partial_\nu K^{\ast 0}_\rho) (\partial^\alpha \partial^\beta \partial_\sigma \bar{K}^0) + (\partial_\nu K^{\ast +}_\rho) (\partial^\alpha \partial^\beta \partial_\sigma K^-) +	\vphantom{\bigg(\bigg)}	\nonumber
			\\
			& \quad + (\partial_\nu K^{\ast -}_\rho) (\partial^\alpha \partial^\beta \partial_\sigma K^+) +	\vphantom{\bigg(\bigg)}	\nonumber
			\\
			& \quad + 2 \, \big[ ( \partial_\nu \omega_\rho ) \cos \beta_v - ( \partial_\nu \phi_\rho ) \sin \beta_v \big] ( \partial^\alpha \partial^\beta \partial_\sigma \pi^0 ) + \vphantom{\bigg(\bigg)} \nonumber
			\\
			& \quad + 2 \, ( \partial_\nu \rho_\rho^0 ) \big[ ( \partial^\alpha \partial^\beta \partial_\sigma \eta ) \cos \beta_p -	\vphantom{\bigg(\bigg)}	\nonumber
			\\
			& \qquad - ( \partial^\alpha \partial^\beta \partial_\sigma \eta^\prime ) \sin \beta_p \big] \Big\} +	\vphantom{\bigg(\bigg)}	\nonumber
			\\
			& + \rho^+_{3 , \mu\alpha\beta} \Big\{ \sqrt{2} \, \big[ ( \partial_\nu K^{\ast -}_\rho ) ( \partial^\alpha \partial^\beta \partial_\sigma K^0 ) +	\vphantom{\bigg(\bigg)}	\nonumber
			\\
			& \qquad + ( \partial_\nu K^{\ast 0}_\rho ) ( \partial^\alpha \partial^\beta \partial_\sigma K^- ) \big] +	\vphantom{\bigg(\bigg)}	\nonumber
			\\
			& \quad + 2 \, \big[ ( \partial_\nu \omega_\rho ) \cos \beta_v - ( \partial_\nu \phi_\rho ) \sin \beta_v \big] ( \partial^\alpha \partial^\beta \partial_\sigma \pi^- ) +	\vphantom{\bigg(\bigg)}	\nonumber
			\\
			& \quad + 2 \, ( \partial_\nu \rho_\rho^- ) \big[ ( \partial^\alpha \partial^\beta \partial_\sigma \eta ) \cos \beta_p -	\vphantom{\bigg(\bigg)}	\nonumber
			\\
			& \qquad - ( \partial^\alpha \partial^\beta \partial_\sigma \eta^\prime ) \sin \beta_p \big] \Big\} +	\vphantom{\bigg(\bigg)}	\nonumber
			\\
			& + \rho^-_{3 , \mu\alpha\beta} \Big\{ \sqrt{2} \, \big[ ( \partial_\nu K^{\ast +}_\rho ) ( \partial^\alpha \partial^\beta \partial_\sigma \bar{K}^0 ) +	\vphantom{\bigg(\bigg)}	\nonumber
			\\
			& \qquad + ( \partial_\nu \bar{K}^{\ast 0}_\rho ) ( \partial^\alpha \partial^\beta \partial_\sigma K^+ ) \big] +	\vphantom{\bigg(\bigg)}	\nonumber
			\\
			& \quad + 2 \, \big[ ( \partial_\nu \omega_\rho ) \cos \beta_v - ( \partial_\nu \phi_\rho ) \sin \beta_v \big] ( \partial^\alpha \partial^\beta \partial_\sigma \pi^+ ) +	\vphantom{\bigg(\bigg)}	\nonumber
			\\
			& \quad + 2 \, ( \partial_\nu \rho_\rho^+ ) \big[ ( \partial^\alpha \partial^\beta \partial_\sigma \eta ) \cos \beta_p -	\vphantom{\bigg(\bigg)}	\nonumber
			\\
			& \qquad - ( \partial^\alpha \partial^\beta \partial_\sigma \eta^\prime ) \sin \beta_p \big] \Big\} +	\vphantom{\bigg(\bigg)}	\nonumber
			\\
			& + \bar{K}^{\ast 0}_{3 , \mu\alpha\beta} \Big\{ - ( \partial_\nu K^{\ast 0}_\rho ) ( \partial^\alpha \partial^\beta \partial_\sigma \pi^0 ) +	\vphantom{\bigg(\bigg)}	\nonumber
			\\
			& \quad + \sqrt{2} \, ( \partial_\nu K^{\ast +}_\rho ) ( \partial^\alpha \partial^\beta \partial_\sigma \pi^- ) -	\vphantom{\bigg(\bigg)}	\nonumber
			\\
			& \quad - ( \partial_\nu \rho^{0}_\rho ) ( \partial^\alpha \partial^\beta \partial_\sigma K^0 ) + \sqrt{2} \, ( \partial_\nu \rho^{-}_\rho ) ( \partial^\alpha \partial^\beta \partial_\sigma K^+ ) +	\vphantom{\bigg(\bigg)}	\nonumber
			\\
			& \quad + ( \partial_\nu K^{*0}_\rho ) \big[ ( \partial^\alpha \partial^\beta \partial_\sigma \eta) ( \cos \beta_p + \sqrt{2} \sin \beta_p ) +	\vphantom{\bigg(\bigg)}	\nonumber
			\\
			& \qquad + ( \partial^\alpha \partial^\beta \partial_\sigma \eta^\prime) ( - \sin \beta_p + \sqrt{2} \cos \beta_p ) \big] +	\vphantom{\bigg(\bigg)}	\nonumber
			\\
			& \quad + \big[ ( \partial_\nu \omega_\rho ) ( \cos \beta_v + \sqrt{2} \sin \beta_v ) +	\vphantom{\bigg(\bigg)}	\nonumber
			\\
			& \qquad + ( \partial_\nu \phi_\rho ) ( - \sin \beta_v + \sqrt{2} \cos \beta_v ) \big] ( \partial^\alpha \partial^\beta \partial_\sigma K^0 ) \Big\} +	\vphantom{\bigg(\bigg)}	\nonumber
			\\
			& + K^{\ast 0}_{3 , \mu\alpha\beta} \Big\{ - ( \partial_\nu \bar{K}^{\ast 0}_\rho ) ( \partial^\alpha \partial^\beta \partial_\sigma \pi^0 ) +	\vphantom{\bigg(\bigg)}	\nonumber
			\\
			& \quad + \sqrt{2} \, ( \partial_\nu K^{\ast -}_\rho ) ( \partial^\alpha \partial^\beta \partial_\sigma \pi^+ ) -	\vphantom{\bigg(\bigg)}	\nonumber
			\\
			& \quad - ( \partial_\nu \rho^{0}_\rho ) ( \partial^\alpha \partial^\beta \partial_\sigma \bar{K}^0 ) + \sqrt{2} \, ( \partial_\nu \rho^{+}_\rho ) ( \partial^\alpha \partial^\beta \partial_\sigma K^- ) +	\vphantom{\bigg(\bigg)}	\nonumber
			\\
			& \quad + ( \partial_\nu \bar{K}^{*0}_\rho ) \big[ ( \partial^\alpha \partial^\beta \partial_\sigma \eta) ( \cos \beta_p + \sqrt{2} \sin \beta_p ) +	\vphantom{\bigg(\bigg)}	\nonumber
			\\
			& \qquad + ( \partial^\alpha \partial^\beta \partial_\sigma \eta^\prime) ( - \sin \beta_p + \sqrt{2} \cos \beta_p ) \big] +	\vphantom{\bigg(\bigg)}	\nonumber
			\\
			& \quad + \big[ ( \partial_\nu \omega_\rho ) ( \cos \beta_v + \sqrt{2} \sin \beta_v ) +	\vphantom{\bigg(\bigg)}	\nonumber
			\\
			& \qquad + ( \partial_\nu \phi_\rho ) ( - \sin \beta_v + \sqrt{2} \cos \beta_v ) \big] ( \partial^\alpha \partial^\beta \partial_\sigma \bar{K}^0 ) \Big\} +	\vphantom{\bigg(\bigg)}	\nonumber
			\\
			& + K^{\ast +}_{3 , \mu\alpha\beta} \Big\{ + ( \partial_\nu K^{\ast -}_\rho ) ( \partial^\alpha \partial^\beta \partial_\sigma \pi^0 ) +	\vphantom{\bigg(\bigg)}	\nonumber
			\\
			& \quad + \sqrt{2} \, ( \partial_\nu \bar{K}^{\ast 0}_\rho ) ( \partial^\alpha \partial^\beta \partial_\sigma \pi^- ) -	\vphantom{\bigg(\bigg)}	\nonumber
			\\
			& \quad + ( \partial_\nu \rho^{0}_\rho ) ( \partial^\alpha \partial^\beta \partial_\sigma K^- ) + \sqrt{2} \, ( \partial_\nu \rho^{-}_\rho ) ( \partial^\alpha \partial^\beta \partial_\sigma \bar{K}^0 ) +	\vphantom{\bigg(\bigg)}	\nonumber
			\\
			& \quad + ( \partial_\nu K^{*-}_\rho ) \big[ ( \partial^\alpha \partial^\beta \partial_\sigma \eta) ( \cos \beta_p + \sqrt{2} \sin \beta_p ) +	\vphantom{\bigg(\bigg)}	\nonumber
			\\
			& \qquad + ( \partial^\alpha \partial^\beta \partial_\sigma \eta^\prime) ( - \sin \beta_p + \sqrt{2} \cos \beta_p ) \big] +	\vphantom{\bigg(\bigg)}	\nonumber
			\\
			& \quad + \big[ ( \partial_\nu \omega_\rho ) ( \cos \beta_v + \sqrt{2} \sin \beta_v ) +	\vphantom{\bigg(\bigg)}	\nonumber
			\\
			& \qquad + ( \partial_\nu \phi_\rho ) ( - \sin \beta_v + \sqrt{2} \cos \beta_v ) \big] ( \partial^\alpha \partial^\beta \partial_\sigma K^- ) \Big\} +	\vphantom{\bigg(\bigg)}	\nonumber
			\\
			& + K^{\ast -}_{3 , \mu\alpha\beta} \Big\{ + ( \partial_\nu K^{\ast +}_\rho ) ( \partial^\alpha \partial^\beta \partial_\sigma \pi^0 ) +	\vphantom{\bigg(\bigg)}	\nonumber
			\\
			& \quad + \sqrt{2} \, ( \partial_\nu K^{\ast 0}_\rho ) ( \partial^\alpha \partial^\beta \partial_\sigma \pi^+ ) -	\vphantom{\bigg(\bigg)}	\nonumber
			\\
			& \quad + ( \partial_\nu \rho^{0}_\rho ) ( \partial^\alpha \partial^\beta \partial_\sigma K^+ ) + \sqrt{2} \, ( \partial_\nu \rho^{+}_\rho ) ( \partial^\alpha \partial^\beta \partial_\sigma K^0 ) +	\vphantom{\bigg(\bigg)}	\nonumber
			\\
			& \quad + ( \partial_\nu K^{*+}_\rho ) \big[ ( \partial^\alpha \partial^\beta \partial_\sigma \eta) ( \cos \beta_p + \sqrt{2} \sin \beta_p ) +	\vphantom{\bigg(\bigg)}	\nonumber
			\\
			& \qquad + ( \partial^\alpha \partial^\beta \partial_\sigma \eta^\prime) ( - \sin \beta_p + \sqrt{2} \cos \beta_p ) \big] +	\vphantom{\bigg(\bigg)}	\nonumber
			\\
			& \quad + \big[ ( \partial_\nu \omega_\rho ) ( \cos \beta_v + \sqrt{2} \sin \beta_v ) +	\vphantom{\bigg(\bigg)}	\nonumber
			\\
			& \qquad + ( \partial_\nu \phi_\rho ) ( - \sin \beta_v + \sqrt{2} \cos \beta_v ) \big] ( \partial^\alpha \partial^\beta \partial_\sigma K^+ ) \Big\} +	\vphantom{\bigg(\bigg)}	\nonumber
			\\
			& + \omega_{3 , \mu\alpha\beta} \Big\{ ( \cos \beta_{w_3} + \sqrt{2} \sin \beta_{w_3} ) \times	\vphantom{\bigg(\bigg)}	\nonumber
			\\
			& \quad \times \big[ ( \partial_\nu \bar{K}^{\ast 0}_\rho ) ( \partial^\alpha \partial^\beta \partial_\sigma K^0 ) + ( \partial_\nu K^{\ast 0}_\rho ) ( \partial^\alpha \partial^\beta \partial_\sigma \bar{K}^0 ) +	\vphantom{\bigg(\bigg)}	\nonumber
			\\
			& \qquad + ( \partial_\nu K^{\ast +}_\rho ) ( \partial^\alpha \partial^\beta \partial_\sigma K^- ) + ( \partial_\nu K^{\ast -}_\rho ) ( \partial^\alpha \partial^\beta \partial_\sigma K^+ ) \big] +	\vphantom{\bigg(\bigg)}	\nonumber
			\\
			& \quad + 2 \cos \beta_{w_3} \big[ ( \partial_\nu \rho^0_\rho ) ( \partial^\alpha \partial^\beta \partial_\sigma \pi^0 ) +	\vphantom{\bigg(\bigg)}	\nonumber
			\\
			& \qquad + ( \partial_\nu \rho^+_\rho ) ( \partial^\alpha \partial^\beta \partial_\sigma \pi^- ) + ( \partial_\nu \rho^-_\rho ) ( \partial^\alpha \partial^\beta \partial_\sigma \pi^+ ) \big] +	\vphantom{\bigg(\bigg)}	\nonumber
			\\
			& \quad + 2 \cos \beta_{w_3} \big[ ( \partial_\nu \omega_\rho ) \cos \beta_v - ( \partial_\nu \phi_\rho ) \sin \beta_v \big] \times	\vphantom{\bigg(\bigg)}	\nonumber
			\\
			& \qquad \times \big[ ( \partial^\alpha \partial^\beta \partial_\sigma \eta ) \cos \beta_p -  ( \partial^\alpha \partial^\beta \partial_\sigma \eta^\prime ) \sin \beta_p  \big] +	\vphantom{\bigg(\bigg)}	\nonumber
			\\
			& \quad + 2 \sqrt{2} \sin \beta_{w_3} \big[ ( \partial_\nu \omega_\rho ) \sin \beta_v + ( \partial_\nu \phi_\rho ) \cos \beta_v \big] \times	\vphantom{\bigg(\bigg)}	\nonumber
			\\
			& \qquad \times \big[ ( \partial^\alpha \partial^\beta \partial_\sigma \eta ) \sin \beta_p +  ( \partial^\alpha \partial^\beta \partial_\sigma \eta^\prime ) \cos \beta_p  \big] \Big\} +	\vphantom{\bigg(\bigg)}	\nonumber
			\\
			& + \phi_{3 , \mu\alpha\beta} \Big\{ ( -\sin \beta_{w_3} + \sqrt{2} \cos \beta_{w_3} ) \times	\vphantom{\bigg(\bigg)}	\nonumber
			\\
			& \quad \times \big[ ( \partial_\nu \bar{K}^{\ast 0}_\rho ) ( \partial^\alpha \partial^\beta \partial_\sigma K^0 ) + ( \partial_\nu K^{\ast 0}_\rho ) ( \partial^\alpha \partial^\beta \partial_\sigma \bar{K}^0 ) +	\vphantom{\bigg(\bigg)}	\nonumber
			\\
			& \qquad + ( \partial_\nu K^{\ast +}_\rho ) ( \partial^\alpha \partial^\beta \partial_\sigma K^- ) + ( \partial_\nu K^{\ast -}_\rho ) ( \partial^\alpha \partial^\beta \partial_\sigma K^+ ) \big] +	\vphantom{\bigg(\bigg)}	\nonumber
			\\
			& \quad - 2 \sin \beta_{w_3} \big[ ( \partial_\nu \rho^0_\rho ) ( \partial^\alpha \partial^\beta \partial_\sigma \pi^0 ) +	\vphantom{\bigg(\bigg)}	\nonumber
			\\
			& \qquad + ( \partial_\nu \rho^+_\rho ) ( \partial^\alpha \partial^\beta \partial_\sigma \pi^- ) + ( \partial_\nu \rho^-_\rho ) ( \partial^\alpha \partial^\beta \partial_\sigma \pi^+ ) \big] +	\vphantom{\bigg(\bigg)}	\nonumber
			\\
			& \quad - 2 \sin \beta_{w_3} \big[ ( \partial_\nu \omega_\rho ) \cos \beta_v - ( \partial_\nu \phi_\rho ) \sin \beta_v \big] \times	\vphantom{\bigg(\bigg)}	\nonumber
			\\
			& \qquad \times \big[ ( \partial^\alpha \partial^\beta \partial_\sigma \eta ) \cos \beta_p -  ( \partial^\alpha \partial^\beta \partial_\sigma \eta^\prime ) \sin \beta_p  \big] +	\vphantom{\bigg(\bigg)}	\nonumber
			\\
			& \quad + 2 \sqrt{2} \cos \beta_{w_3} \big[ ( \partial_\nu \omega_\rho ) \sin \beta_v + ( \partial_\nu \phi_\rho ) \cos \beta_v \big] \times	\vphantom{\bigg(\bigg)}	\nonumber
			\\
			& \qquad \times \big[ ( \partial^\alpha \partial^\beta \partial_\sigma \eta ) \sin \beta_p +  ( \partial^\alpha \partial^\beta \partial_\sigma \eta^\prime ) \cos \beta_p  \big] \Big\} \Big)  \, .	\vphantom{\bigg(\bigg)}	\nonumber
		\end{align}

		\begin{align}
			& \mathcal{L}_{w_3 \gamma p} =	\vphantom{\bigg(\bigg)}	\label{eq:extended_lagrangian_w3_gamma_p}	\vphantom{\bigg(\bigg)}
			\\
			= \, & g_{w_3 v_1 p} \, \tfrac{e}{g_\rho} \, \varepsilon^{\mu \nu \rho \sigma} \, ( \partial_\nu a_\rho ) \, \mathrm{tr} \big[ W_{3 , \mu\alpha\beta} \, \big\{ Q , \, ( \partial^{\alpha} \partial^{\beta} \partial_{\sigma} P ) \big\}_{+} \big] =	\vphantom{\bigg(\bigg)}	\nonumber
			\\
			= \, & \frac{g_{w_3 v_1 p}}{6} \, \tfrac{e}{g_\rho} \, \varepsilon^{\mu \nu \rho \sigma} \, ( \partial_\nu a_\rho ) \, \bigg(	\vphantom{\bigg(\bigg)}	\nonumber
			\\
			& + \rho_{3,\mu \alpha \beta}^{0} \, \bigg\{ ( \partial^\alpha \partial^\beta \partial_\sigma \pi^0 ) +	\vphantom{\bigg(\bigg)}	\nonumber
			\\
			& \quad + 3 \, \big[ ( \partial^\alpha \partial^\beta \partial_\sigma \eta ) \, \cos ( \beta_p ) - ( \partial^\alpha \partial^\beta \partial_\sigma \eta^\prime ) \, \sin ( \beta_p ) \big] \bigg\} +	\vphantom{\bigg(\bigg)}	\nonumber
			\\
			& + \rho_{3,\mu \alpha \beta}^{+} \, ( \partial^\alpha \partial^\beta \partial_\sigma \pi^{-} ) + \rho_{3,\mu \alpha \beta}^{-} \, ( \partial^\alpha \partial^\beta \partial_\sigma \pi^{+} ) -	\vphantom{\bigg(\bigg)}	\nonumber
			\\
			& - 2 \, \bar{K}_{3 , \mu \alpha \beta}^{\ast 0} \, ( \partial^\alpha \partial^\beta \partial_\sigma K^0 ) - 2 \, K_{3 , \mu \alpha \beta}^{\ast 0} \, ( \partial^\alpha \partial^\beta \partial_\sigma \bar{K}^0 ) +	\vphantom{\bigg(\bigg)}	\nonumber
			\\
			& + \bar{K}_{3 , \mu \alpha \beta}^{\ast +} \, ( \partial^\alpha \partial^\beta \partial_\sigma K^- ) + K_{3 , \mu \alpha \beta}^{\ast -} \, ( \partial^\alpha \partial^\beta \partial_\sigma K^+ ) +	\vphantom{\bigg(\bigg)}	\nonumber
			\\
			& + \omega_{3 , \mu \alpha \beta} \, \bigg\{ 3 \, \cos ( \beta_{w_3} ) \, ( \partial^\alpha \partial^\beta \partial_\sigma \pi^0 ) +	\vphantom{\bigg(\bigg)}	\nonumber
			\\
			& \quad + \big[ \cos ( \beta_{w_3} ) \, \cos ( \beta_{p} ) - 2 \, \sin ( \beta_{w_3} ) \, \sin ( \beta_{p} ) \big] \times	\vphantom{\bigg(\bigg)}	\nonumber
			\\
			& \qquad \times ( \partial^\alpha \partial^\beta \partial_\sigma \eta ) +	\vphantom{\bigg(\bigg)}	\nonumber
			\\
			& \quad - \big[ \cos ( \beta_{w_3} ) \, \sin ( \beta_{p} ) + 2 \, \sin ( \beta_{w_3} ) \, \cos ( \beta_{p} ) \big] \times	\vphantom{\bigg(\bigg)}	\nonumber
			\\
			& \qquad \times ( \partial^\alpha \partial^\beta \partial_\sigma \eta^\prime ) \bigg\} -	\vphantom{\bigg(\bigg)}	\nonumber
			\\
			& - \phi_{3 , \mu \alpha \beta} \, \bigg\{ 3 \, \sin ( \beta_{w_3} ) \, ( \partial^\alpha \partial^\beta \partial_\sigma \pi^0 ) +	\vphantom{\bigg(\bigg)}	\nonumber
			\\
			& \quad + \big[ \sin ( \beta_{w_3} ) \, \cos ( \beta_{p} ) + 2 \, \cos ( \beta_{w_3} ) \, \sin ( \beta_{p} ) \big] \times	\vphantom{\bigg(\bigg)}	\nonumber
			\\
			& \qquad \times ( \partial^\alpha \partial^\beta \partial_\sigma \eta ) +	\vphantom{\bigg(\bigg)}	\nonumber
			\\
			& \quad - \big[ \sin ( \beta_{w_3} ) \, \sin ( \beta_{p} ) - 2 \, \cos ( \beta_{w_3} ) \, \cos ( \beta_{p} ) \big] \times	\vphantom{\bigg(\bigg)}	\nonumber
			\\
			& \qquad \times ( \partial^\alpha \partial^\beta \partial_\sigma \eta^\prime ) \bigg\} \, .	\vphantom{\bigg(\bigg)}	\nonumber
		\end{align}
	
		\begin{align}
			& \mathcal{L}_{w_3 a_2 p} =	\vphantom{\bigg(\bigg)}	\label{eq:extended_lagrangian_w3_a2_p}
			\\
			= \, & g_{w_3 a_2 p} \, \varepsilon_{\mu\nu\rho\sigma} \, \mathrm{tr} \big( \tensor{W}{_{3 , }^\mu_\alpha_\beta} \big[ ( \partial^\nu A_2^{\rho\alpha} ) , \, (\partial^\sigma \partial^\beta P) \big]_{-} \big) =	\vphantom{\bigg(\bigg)}	\nonumber
			\\
			= \, & \frac{g_{w_3 a_2 p}}{4}\, \varepsilon_{\mu\nu\rho\sigma} \, \Big(	\vphantom{\bigg(\bigg)}	\nonumber
			\\
			& + \rho^{0,\mu}_{3\;\;\;\alpha\beta} \Big\{ + ( \partial^\nu \bar{K}^{\ast 0,\rho\alpha}_2 ) (\partial^\sigma \partial^\beta K^0) -	\vphantom{\bigg(\bigg)}	\nonumber
			\\
			& \quad - ( \partial^\nu K^{\ast 0,\rho\alpha}_2 ) (\partial^\sigma \partial^\beta \bar{K}^0) + ( \partial^\nu K^{\ast +,\rho\alpha}_2 ) (\partial^\sigma \partial^\beta K^-) -	\vphantom{\bigg(\bigg)}	\nonumber
			\\
			& \quad - ( \partial^\nu K^{\ast -,\rho\alpha}_2 ) (\partial^\sigma \partial^\beta K^+) +	\vphantom{\bigg(\bigg)}	\nonumber
			\\
			& \quad + 2 \, \big[ ( \partial^\nu a^{+,\rho\alpha}_2 ) ( \partial^\sigma \partial^\beta \pi^- ) - ( \partial^\nu a^{-,\rho\alpha}_2 ) ( \partial^\sigma \partial^\beta \pi^+ ) \big] \Big\} +	\vphantom{\bigg(\bigg)}	\nonumber
			\\
			& + \rho^{+,\mu}_{3\;\;\;\;\alpha\beta} \Big\{ \sqrt{2} \, \big[ - ( \partial^\nu K^{\ast -,\rho\alpha}_2 ) (\partial^\sigma \partial^\beta K^0) +	\vphantom{\bigg(\bigg)}	\nonumber
			\\
			& \quad + ( \partial^\nu K^{\ast 0,\rho\alpha}_2 ) (\partial^\sigma \partial^\beta K^-) \big] +	\vphantom{\bigg(\bigg)}	\nonumber
			\\
			& \quad + 2 \, \big[ ( \partial^\nu a^{-,\rho\alpha}_2 ) ( \partial^\sigma \partial^\beta \pi^0 ) - ( \partial^\nu a^{0,\rho\alpha}_2 ) ( \partial^\sigma \partial^\beta \pi^- ) \big] \Big\} +	\vphantom{\bigg(\bigg)}	\nonumber
			\\
			& + \rho^{-,\mu}_{3\;\;\;\;\alpha\beta} \Big\{ \sqrt{2} \, \big[ ( \partial^\nu K^{\ast +,\rho\alpha}_2 ) ( \partial^\sigma \partial^\beta \bar{K}^0 ) -	\vphantom{\bigg(\bigg)}	\nonumber
			\\
			& \quad - ( \partial^\nu \bar{K}^{\ast 0,\rho\alpha}_2 ) ( \partial^\sigma \partial^\beta K^+ ) \big] +	\vphantom{\bigg(\bigg)}	\nonumber
			\\
			& \quad + 2 \, \big[ - ( \partial^\nu a^{+,\rho\alpha}_2 ) ( \partial^\sigma \partial^\beta \pi^0 ) + ( \partial^\nu a^{0,\rho\alpha}_2 ) ( \partial^\sigma \partial^\beta \pi^+ ) \big] \Big\} +	\vphantom{\bigg(\bigg)}	\nonumber
			\\
			& + \bar{K}^{\ast 0,\mu}_{3\;\;\;\;\;\alpha\beta} \Big\{ + ( \partial^\nu K^{\ast 0,\rho\alpha}_2 ) (\partial^\sigma \partial^\beta \pi^0 ) -	\vphantom{\bigg(\bigg)}	\nonumber
			\\
			& \quad - \sqrt{2} \, ( \partial^\nu K^{\ast +,\rho\alpha}_2 ) (\partial^\sigma \partial^\beta \pi^- ) -	\vphantom{\bigg(\bigg)}	\nonumber
			\\
			& \quad - ( \partial^\nu a^{0,\rho\alpha}_2 ) ( \partial^\sigma \partial^\beta K^0 ) + \sqrt{2} \, ( \partial^\nu a^{-,\rho\alpha}_2 ) ( \partial^\sigma \partial^\beta K^+ ) +	\vphantom{\bigg(\bigg)}	\nonumber
			\\
			& \quad + ( \partial^\nu K^{\ast 0,\rho\alpha}_2 ) \big[ ( \partial^\sigma \partial^\beta \eta) ( - \cos \beta_p + \sqrt{2} \sin \beta_p ) +	\vphantom{\bigg(\bigg)}	\nonumber
			\\
			& \qquad + ( \partial^\sigma \partial^\beta \eta^\prime) ( \sin \beta_p + \sqrt{2} \cos \beta_p ) \big] +	\vphantom{\bigg(\bigg)}	\nonumber
			\\
			& \quad + \big[ ( \partial^\mu f_2^{\rho\alpha} ) ( \cos \beta_{a_2} - \sqrt{2} \sin \beta_{a_2} ) +	\vphantom{\bigg(\bigg)}	\nonumber
			\\
			& \qquad + ( \partial^\mu f_2^{\prime \rho\alpha} ) ( - \sin \beta_{a_2} - \sqrt{2} \cos \beta_{a_2} ) \big] ( \partial^\sigma \partial^\beta K^0 ) \Big\} +	\vphantom{\bigg(\bigg)}	\nonumber
			\\
			& + K^{\ast 0,\mu}_{3\;\;\;\;\;\alpha\beta} \Big\{ - ( \partial^\nu \bar{K}^{\ast 0,\rho\alpha}_2 ) (\partial^\sigma \partial^\beta \pi^0 ) +	\vphantom{\bigg(\bigg)}	\nonumber
			\\
			& \quad + \sqrt{2} \, ( \partial^\nu K^{\ast -,\rho\alpha}_2 ) (\partial^\sigma \partial^\beta \pi^+ ) +	\vphantom{\bigg(\bigg)}	\nonumber
			\\
			& \quad + ( \partial^\nu a^{0,\rho\alpha}_2 ) ( \partial^\sigma \partial^\beta \bar{K}^0 ) - \sqrt{2} \, ( \partial^\nu a^{+,\rho\alpha}_2 ) ( \partial^\sigma \partial^\beta K^- ) +	\vphantom{\bigg(\bigg)}	\nonumber
			\\
			& \quad + ( \partial^\nu \bar{K}^{\ast 0,\rho\alpha}_2 ) \big[ ( \partial^\sigma \partial^\beta \eta) ( \cos \beta_p - \sqrt{2} \sin \beta_p ) +	\vphantom{\bigg(\bigg)}	\nonumber
			\\
			& \qquad + ( \partial^\sigma \partial^\beta \eta^\prime) ( - \sin \beta_p - \sqrt{2} \cos \beta_p ) \big] +	\vphantom{\bigg(\bigg)}	\nonumber
			\\
			& \quad + \big[ ( \partial^\mu f_2^{\rho\alpha} ) ( - \cos \beta_{a_2} + \sqrt{2} \sin \beta_{a_2} ) +	\vphantom{\bigg(\bigg)}	\nonumber
			\\
			& \qquad + ( \partial^\mu f_2^{\prime \rho\alpha} ) ( \sin \beta_{a_2} + \sqrt{2} \cos \beta_{a_2} ) \big] ( \partial^\sigma \partial^\beta \bar{K}^0 ) \Big\} +	\vphantom{\bigg(\bigg)}	\nonumber
			\\
			& + K^{\ast +,\mu}_{3\;\;\;\;\;\;\alpha\beta} \Big\{ + ( \partial^\nu K^{\ast -,\rho\alpha}_2 ) (\partial^\sigma \partial^\beta \pi^0 ) +	\vphantom{\bigg(\bigg)}	\nonumber
			\\
			& \quad + \sqrt{2} \, ( \partial^\nu \bar{K}^{\ast 0,\rho\alpha}_2 ) (\partial^\sigma \partial^\beta \pi^- ) -	\vphantom{\bigg(\bigg)}	\nonumber
			\\
			& \quad - ( \partial^\nu a^{0,\rho\alpha}_2 ) ( \partial^\sigma \partial^\beta K^- ) - \sqrt{2} \, ( \partial^\nu a^{-,\rho\alpha}_2 ) ( \partial^\sigma \partial^\beta \bar{K}^0 ) +	\vphantom{\bigg(\bigg)}	\nonumber
			\\
			& \quad + ( \partial^\nu K^{\ast -,\rho\alpha}_2 ) \big[ ( \partial^\sigma \partial^\beta \eta) ( \cos \beta_p - \sqrt{2} \sin \beta_p ) +	\vphantom{\bigg(\bigg)}	\nonumber
			\\
			& \qquad + ( \partial^\sigma \partial^\beta \eta^\prime) ( - \sin \beta_p - \sqrt{2} \cos \beta_p ) \big] +	\vphantom{\bigg(\bigg)}	\nonumber
			\\
			& \quad + \big[ ( \partial^\mu f_2^{\rho\alpha} ) ( - \cos \beta_{a_2} + \sqrt{2} \sin \beta_{a_2} ) +	\vphantom{\bigg(\bigg)}	\nonumber
			\\
			& \qquad + ( \partial^\mu f_2^{\prime \rho\alpha} ) ( \sin \beta_{a_2} + \sqrt{2} \cos \beta_{a_2} ) \big] ( \partial^\sigma \partial^\beta K^- ) \Big\} +	\vphantom{\bigg(\bigg)}	\nonumber
			\\
			& + K^{\ast -,\mu}_{3\;\;\;\;\;\;\alpha\beta} \Big\{ - ( \partial^\nu K^{\ast +,\rho\alpha}_2 ) (\partial^\sigma \partial^\beta \pi^0 ) -	\vphantom{\bigg(\bigg)}	\nonumber
			\\
			& \quad - \sqrt{2}\, ( \partial^\nu K^{\ast 0,\rho\alpha}_2 )\, (\partial^\sigma \partial^\beta \pi^+ ) +	\vphantom{\bigg(\bigg)}	\nonumber
			\\
			& \quad + ( \partial^\nu a^{0,\rho\alpha}_2 ) ( \partial^\sigma \partial^\beta K^+ ) + \sqrt{2} \, ( \partial^\nu a^{+,\rho\alpha}_2 ) ( \partial^\sigma \partial^\beta K^0 ) +	\vphantom{\bigg(\bigg)}	\nonumber
			\\
			& \quad + ( \partial^\nu K^{\ast +,\rho\alpha}_2 ) \big[ ( \partial^\sigma \partial^\beta \eta) ( - \cos \beta_p + \sqrt{2} \sin \beta_p ) +	\vphantom{\bigg(\bigg)}	\nonumber
			\\
			& \qquad + ( \partial^\sigma \partial^\beta \eta^\prime) ( \sin \beta_p + \sqrt{2} \cos \beta_p ) \big] +	\vphantom{\bigg(\bigg)}	\nonumber
			\\
			& \quad + \big[ ( \partial^\mu f_2^{\rho\alpha} ) ( \cos \beta_{a_2} - \sqrt{2} \sin \beta_{a_2} ) +	\vphantom{\bigg(\bigg)}	\nonumber
			\\
			& \qquad + ( \partial^\mu f_2^{\prime \rho\alpha} ) ( - \sin \beta_{a_2} - \sqrt{2} \cos \beta_{a_2} ) \big] ( \partial^\sigma \partial^\beta K^- ) \Big\} +	\vphantom{\bigg(\bigg)}	\nonumber
			\\
			& + \omega^{\;\;\mu}_{3 , \;\alpha\beta} \Big\{ ( - \cos \beta_{w_3} + \sqrt{2} \sin \beta_{w_3} ) \times	\vphantom{\bigg(\bigg)}	\nonumber
			\\
			& \quad \times \big[ ( \partial^\nu \bar{K}^{\ast 0,\rho\alpha}_2 ) ( \partial^\sigma \partial^\beta  K^0 ) - ( \partial^\nu K^{\ast 0,\rho\alpha}_2 ) ( \partial^\sigma \partial^\beta \bar{K}^0 ) +	\vphantom{\bigg(\bigg)}	\nonumber
			\\
			& \qquad - ( \partial^\nu K^{\ast +,\rho\alpha}_2 ) ( \partial^\sigma \partial^\beta K^- ) +	\vphantom{\bigg(\bigg)}	\nonumber
			\\
			& \qquad + ( \partial^\nu K^{\ast -,\rho\alpha}_2 ) ( \partial^\sigma \partial^\beta K^+ ) \big] \Big\} +	\vphantom{\bigg(\bigg)}	\nonumber
			\\
			& + \phi^{\;\;\mu}_{3 , \;\alpha\beta} \Big\{ ( \sin \beta_{w_3} + \sqrt{2} \cos \beta_{w_3} ) \times	\vphantom{\bigg(\bigg)}	\nonumber
			\\
			& \quad \times \big[ ( \partial^\nu \bar{K}^{\ast 0,\rho\alpha}_2 ) ( \partial^\sigma \partial^\beta  K^0 ) - ( \partial^\nu K^{\ast 0,\rho\alpha}_2 ) ( \partial^\sigma \partial^\beta \bar{K}^0 ) +	\vphantom{\bigg(\bigg)}	\nonumber
			\\
			& \qquad - ( \partial^\nu K^{\ast +,\rho\alpha}_2 ) ( \partial^\sigma \partial^\beta K^- ) +	\vphantom{\bigg(\bigg)}	\nonumber
			\\
			& \qquad + ( \partial^\nu K^{\ast -,\rho\alpha}_2 ) ( \partial^\sigma \partial^\beta K^+ ) \big] \Big\} \Big) \, . 	\vphantom{\bigg(\bigg)}	\nonumber
		\end{align}

		\begin{align}
			& \mathcal{L}_{w_3 b_1 p} =	\vphantom{\bigg(\bigg)}	\label{eq:extended_lagrangian_w3_b1_p}
			\\
			= \, & g_{w_3 b_1 p} \, \mathrm{tr} \big[ W_3^{\mu\nu\rho} \big\{ B_{1 , \mu} , \, (\partial_\nu \partial_\rho P) \big\}_{+} \big] =	\vphantom{\bigg(\bigg)}	\nonumber
			\\
			= \, & \frac{g_{w_3 b_1 p}}{4} \, \Big(	\vphantom{\bigg(\bigg)}	\nonumber
			\\
			& + \rho^{0,\mu\nu\rho}_3 \Big\{ - \bar{K}^0_{1 , B,\mu} ( \partial_\nu \partial_\rho K^0 ) - K^0_{1 , B,\mu} ( \partial_\nu \partial_\rho \bar{K}^0 ) +	\vphantom{\bigg(\bigg)}	\nonumber
			\\
			& \quad + K^+_{1 , B,\mu} (\partial_\nu \partial_\rho K^-) + K^-_{1 , B,\mu} (\partial_\nu \partial_\rho K^+) +	\vphantom{\bigg(\bigg)}	\nonumber
			\\
			& \quad + 2 \, \big[ h_{1 , \mu} \cos \beta_{b_1} - h^\prime_{1 , \mu} \sin \beta_{b_1} \big]\, ( \partial_\nu \partial_\rho \pi^0 ) +	\vphantom{\bigg(\bigg)}	\nonumber
			\\
			& \quad + 2 \, b^0_{1 , \mu} \big[ ( \partial_\nu \partial_\rho \eta ) \cos \beta_p - ( \partial_\nu \partial_\rho  \eta^\prime ) \sin \beta_p \big] \Big\} +	\vphantom{\bigg(\bigg)}	\nonumber
			\\
			& + \rho^{+,\mu\nu\rho}_3 \Big\{ \sqrt{2} \, \big[ K^-_{1 , B , \mu} (\partial_\nu \partial_\rho K^0) + K^0_{1 , B,\mu} (\partial_\nu \partial_\rho K^-) \big] +	\vphantom{\bigg(\bigg)}	\nonumber
			\\
			& \quad + 2 \, \big[ h_{1 , \mu} \cos \beta_{b_1} - h^\prime_{1 , \mu} \sin \beta_{b_1} \big] ( \partial_\nu \partial_\rho \pi^- ) +	\vphantom{\bigg(\bigg)}	\nonumber
			\\
			& \quad + 2 \, b^-_{1 , \mu} \big[ ( \partial_\nu \partial_\rho \eta ) \cos \beta_p - ( \partial_\nu \partial_\rho  \eta^\prime ) \sin \beta_p \big] \Big\} +	\vphantom{\bigg(\bigg)}	\nonumber
			\\
			& + \rho^{-,\mu\nu\rho}_3 \Big\{ \sqrt{2} \, \big[ K^+_{1 , B , \mu} (\partial_\nu \partial_\rho \bar{K}^0) + \bar{K}^0_{1 , B,\mu} (\partial_\nu \partial_\rho K^+) \big] +	\vphantom{\bigg(\bigg)}	\nonumber
			\\
			& \quad + 2 \, \big[ h_{1 , \mu} \cos \beta_{b_1} - h^\prime_{1 , \mu} \sin \beta_{b_1} \big] ( \partial_\nu \partial_\rho \pi^+ ) +	\vphantom{\bigg(\bigg)}	\nonumber
			\\
			& \quad + 2 \, b^+_{1 , \mu} \big[ ( \partial_\nu \partial_\rho \eta ) \cos \beta_p - ( \partial_\nu \partial_\rho  \eta^\prime ) \sin \beta_p \big] \Big\} +	\vphantom{\bigg(\bigg)}	\nonumber
			\\
			& + \bar{K}^{\ast 0, \mu\nu\rho}_3 \Big\{ - K^0_{1 , B,\mu} ( \partial_\nu \partial_\rho \pi^0 ) + \sqrt{2} \, K^+_{1 , B,\mu} ( \partial_\nu \partial_\rho \pi^- ) -	\vphantom{\bigg(\bigg)}	\nonumber
			\\
			& \quad - b^0_{1 , \mu} ( \partial_\nu \partial_\rho K^0 ) + \sqrt{2} \, b^-_{1 , \mu} ( \partial_\nu \partial_\rho K^+ ) +	\vphantom{\bigg(\bigg)}	\nonumber
			\\
			& \quad + K^0_{1 , B,\mu} \big[ ( \partial_\nu \partial_\rho \eta) ( \cos \beta_p + \sqrt{2} \sin \beta_p ) +	\vphantom{\bigg(\bigg)}	\nonumber
			\\
			& \qquad + ( \partial_\nu \partial_\rho \eta^\prime) ( - \sin \beta_p + \sqrt{2} \cos \beta_p ) \big] +	\vphantom{\bigg(\bigg)}	\nonumber
			\\
			& \quad + \big[ h_{1 , \mu} ( \cos \beta_{b_1} + \sqrt{2} \sin \beta_{b_1} ) +	\vphantom{\bigg(\bigg)}	\nonumber
			\\
			& \qquad + h^\prime_{1 , \mu} ( - \sin \beta_{b_1} + \sqrt{2} \cos \beta_{b_1} ) \big] ( \partial_\nu \partial_\rho K^0 ) \Big\} +	\vphantom{\bigg(\bigg)}	\nonumber
			\\
			& + K^{\ast 0, \mu\nu\rho}_3 \Big\{ - \bar{K}^0_{1 , B,\mu} ( \partial_\nu \partial_\rho \pi^0 ) + \sqrt{2} \, K^-_{1 , B,\mu} ( \partial_\nu \partial_\rho \pi^+ ) -	\vphantom{\bigg(\bigg)}	\nonumber
			\\
			& \quad - b^0_{1 , \mu} ( \partial_\nu \partial_\rho \bar{K}^0 ) + \sqrt{2} \, b^+_{1 , \mu} ( \partial_\nu \partial_\rho K^- ) +	\vphantom{\bigg(\bigg)}	\nonumber
			\\
			& \quad + \bar{K}^0_{1 , B,\mu} \big[ ( \partial_\nu \partial_\rho \eta) ( \cos \beta_p + \sqrt{2} \sin \beta_p ) +	\vphantom{\bigg(\bigg)}	\nonumber
			\\
			& \qquad + ( \partial_\nu \partial_\rho \eta^\prime) ( - \sin \beta_p + \sqrt{2} \cos \beta_p ) \big] +	\vphantom{\bigg(\bigg)}	\nonumber
			\\
			& \quad + \big[ h_{1 , \mu} ( \cos \beta_{b_1} + \sqrt{2} \sin \beta_{b_1} ) +	\vphantom{\bigg(\bigg)}	\nonumber
			\\
			& \qquad + h^\prime_{1 , \mu} ( - \sin \beta_{b_1} + \sqrt{2} \cos \beta_{b_1} ) \big] ( \partial_\nu \partial_\rho \bar{K}^0 ) \Big\} +	\vphantom{\bigg(\bigg)}	\nonumber
			\\
			& + K^{\ast +, \mu\nu\rho}_3 \Big\{ + K^-_{1 , B,\mu} ( \partial_\nu \partial_\rho \pi^0 ) + \sqrt{2} \, \bar{K}^0_{1 , B,\mu} ( \partial_\nu \partial_\rho \pi^- ) +	\vphantom{\bigg(\bigg)}	\nonumber
			\\
			& \quad + b^0_{1 , \mu} ( \partial_\nu \partial_\rho K^- ) + \sqrt{2} \, b^-_{1 , \mu} ( \partial_\nu \partial_\rho \bar{K}^0 ) +	\vphantom{\bigg(\bigg)}	\nonumber
			\\
			& \quad + K^-_{1 , B,\mu} \big[ ( \partial_\nu \partial_\rho \eta) ( \cos \beta_p + \sqrt{2} \sin \beta_p ) +	\vphantom{\bigg(\bigg)}	\nonumber
			\\
			& \qquad + ( \partial_\nu \partial_\rho \eta^\prime) ( - \sin \beta_p + \sqrt{2} \cos \beta_p ) \big] +	\vphantom{\bigg(\bigg)}	\nonumber
			\\
			& \quad + \big[ h_{1 , \mu} ( \cos \beta_{b_1} + \sqrt{2} \sin \beta_{b_1} ) +	\vphantom{\bigg(\bigg)}	\nonumber
			\\
			& \qquad + h^\prime_{1 , \mu} ( - \sin \beta_{b_1} + \sqrt{2} \cos \beta_{b_1} ) \big] ( \partial_\nu \partial_\rho K^- ) \Big\} +	\vphantom{\bigg(\bigg)}	\nonumber
			\\
			& + K^{\ast -, \mu\nu\rho}_3 \Big\{ + K^+_{1 , B,\mu} ( \partial_\nu \partial_\rho \pi^0 ) + \sqrt{2} \, K^0_{1 , B,\mu} ( \partial_\nu \partial_\rho \pi^+ ) +	\vphantom{\bigg(\bigg)}	\nonumber
			\\
			& \quad + b^0_{1 , \mu} ( \partial_\nu \partial_\rho K^+ ) + \sqrt{2} \, b^+_{1 , \mu} ( \partial_\nu \partial_\rho K^0 ) +	\vphantom{\bigg(\bigg)}	\nonumber
			\\
			& \quad + K^+_{1 , B,\mu} \big[ ( \partial_\nu \partial_\rho \eta) ( \cos \beta_p + \sqrt{2} \sin \beta_p ) +	\vphantom{\bigg(\bigg)}	\nonumber
			\\
			& \qquad + ( \partial_\nu \partial_\rho \eta^\prime) ( - \sin \beta_p + \sqrt{2} \cos \beta_p ) \big] +	\vphantom{\bigg(\bigg)}	\nonumber
			\\
			& \quad + \big[ h_{1 , \mu} ( \cos \beta_{b_1} + \sqrt{2} \sin \beta_{b_1} ) +	\vphantom{\bigg(\bigg)}	\nonumber
			\\
			& \qquad + h^\prime_{1 , \mu} ( - \sin \beta_{b_1} + \sqrt{2} \cos \beta_{b_1} ) \big] ( \partial_\nu \partial_\rho K^+ ) \Big\} +	\vphantom{\bigg(\bigg)}	\nonumber
			\\
			& + \omega^{\mu\nu\rho}_3 \Big\{ ( \cos \beta_{w_3} + \sqrt{2} \sin \beta_{w_3} ) \times	\vphantom{\bigg(\bigg)}	\nonumber
			\\
			& \qquad \times \big[ \bar{K}^0_{1 , B,\mu} ( \partial_\nu \partial_\rho  K^0 ) + K^0_{1 , B,\mu} ( \partial_\nu \partial_\rho \bar{K}^0 ) +	\vphantom{\bigg(\bigg)}	\nonumber
			\\
			& \qquad \quad + K^+_{1 , B,\mu} ( \partial_\nu \partial_\rho K^- ) + K^-_{1 , B,\mu} ( \partial_\nu \partial_\rho K^+ ) \big] +	\vphantom{\bigg(\bigg)}	\nonumber
			\\
			& \quad + 2 \cos \beta_{w_3} \big[ b^0_{1 , \mu} ( \partial_\nu \partial_\rho \pi^0 ) + b^+_{1 , \mu} ( \partial_\nu \partial_\rho \pi^- ) +	\vphantom{\bigg(\bigg)}	\nonumber
			\\
			& \qquad + b^-_{1 , \mu} ( \partial_\nu \partial_\rho \pi^+ ) \big] +	\vphantom{\bigg(\bigg)}	\nonumber
			\\
			& \quad + 2 \cos \beta_{w_3} \big[ h_{1 , \mu} \cos \beta_{b_1} - h^\prime_{1 , \mu} \sin \beta_{b_1} \big] \times	\vphantom{\bigg(\bigg)}	\nonumber
			\\
			& \qquad \times \big[ ( \partial_\nu \partial_\rho \eta ) \cos \beta_p -  ( \partial_\nu \partial_\rho \eta^\prime ) \sin \beta_p  \big] +	\vphantom{\bigg(\bigg)}	\nonumber
			\\
			& \quad + 2 \sqrt{2} \sin \beta_{w_3} \big[ h_{1 , \mu} \sin \beta_{b_1} + h^\prime_{1 , \mu} \cos \beta_{b_1} \big] \times	\vphantom{\bigg(\bigg)}	\nonumber
			\\
			& \qquad \times \big[ ( \partial_\nu \partial_\rho \eta ) \sin \beta_p +  ( \partial_\nu \partial_\rho \eta^\prime ) \cos \beta_p  \big] \Big\} +	\vphantom{\bigg(\bigg)}	\nonumber
			\\
			& + \phi^{\mu\nu\rho}_3 \Big\{ ( -\sin \beta_{w_3} + \sqrt{2} \cos \beta_{w_3} ) \times	\vphantom{\bigg(\bigg)}	\nonumber
			\\
			& \times \big[ \bar{K}^0_{1 , B,\mu} ( \partial_\nu \partial_\rho  K^0 ) + K^0_{1 , B,\mu} ( \partial_\nu \partial_\rho \bar{K}^0 ) +	\vphantom{\bigg(\bigg)}	\nonumber
			\\
			& \qquad + K^+_{1 , B,\mu} ( \partial_\nu \partial_\rho K^- ) + K^-_{1 , B,\mu} ( \partial_\nu \partial_\rho K^+ ) \big] +	\vphantom{\bigg(\bigg)}	\nonumber
			\\
			& \quad - 2 \sin \beta_{w_3} \big[ b^0_{1 , \mu} ( \partial_\nu \partial_\rho \pi^0 ) + b^+_{1 , \mu} ( \partial_\nu \partial_\rho \pi^- ) +	\vphantom{\bigg(\bigg)}	\nonumber
			\\
			& \qquad + b^-_{1 , \mu} ( \partial_\nu \partial_\rho \pi^+ ) \big] +	\vphantom{\bigg(\bigg)}	\nonumber
			\\
			& \quad - 2 \sin \beta_{w_3} \big[ h_{1 , \mu} \cos \beta_{b_1} - h^\prime_{1 , \mu} \sin \beta_{b_1} \big] \times	\vphantom{\bigg(\bigg)}	\nonumber
			\\
			& \qquad \times \big[ ( \partial_\nu \partial_\rho \eta ) \cos \beta_p -  ( \partial_\nu \partial_\rho \eta^\prime ) \sin \beta_p  \big] +	\vphantom{\bigg(\bigg)}	\nonumber
			\\
			& \quad + 2 \sqrt{2} \cos \beta_{w_3} \big[ h_{1 , \mu} \sin \beta_{b_1} + h^\prime_{1 , \mu} \cos \beta_{b_1} \big] \times	\vphantom{\bigg(\bigg)}	\nonumber
			\\
			& \qquad \times \big[ ( \partial_\nu \partial_\rho \eta ) \sin \beta_p +  ( \partial_\nu \partial_\rho \eta^\prime ) \cos \beta_p  \big] \Big\} \Big) \, .	\vphantom{\bigg(\bigg)}	\nonumber
		\end{align}

		\begin{align}
			& \mathcal{L}_{w_3 a_1 p} =	\vphantom{\bigg(\bigg)}	\label{eq:extended_lagrangian_w3_a1_p}
			\\
			= \, & g_{w_3 a_1 p} \, \mathrm{tr} \big( W_3^{\mu\nu\rho} \big[ A_{1 , \mu} , \, ( \partial_\nu \partial_\rho P ) \big]_{-} \big) =	\vphantom{\bigg(\bigg)}	\nonumber
			\\
			= \, & \frac{g_{w_3 a_1 p}}{4}\, \Big(	\vphantom{\bigg(\bigg)}	\nonumber
			\\
			& + \rho^{0,\mu\nu\rho}_3 \Big\{ + \bar{K}^0_{1 , A,\mu} ( \partial_\nu \partial_\rho K^0 ) - K^0_{1 , A,\mu} ( \partial_\nu \partial_\rho \bar{K}^0 ) +	\vphantom{\bigg(\bigg)}	\nonumber
			\\
			& \quad + K^+_{1 , A,\mu} ( \partial_\nu \partial_\rho K^- ) - K^-_{1 , A,\mu} ( \partial_\nu \partial_\rho K^+ ) +	\vphantom{\bigg(\bigg)}	\nonumber
			\\
			& \quad + 2 \, \big[ a^+_{1 , \mu} ( \partial_\nu \partial_\rho \pi^- ) - a^-_{1 , \mu} ( \partial_\nu \partial_\rho \pi^+ ) \big] \Big\} +	\vphantom{\bigg(\bigg)}	\nonumber
			\\
			& + \rho^{+,\mu\nu\rho}_3 \Big\{ \sqrt{2} \, \big[ - K^-_{1 , A,\mu} ( \partial_\nu \partial_\rho K^0 ) + K^0_{1 , A,\mu} ( \partial_\nu \partial_\rho K^- ) \big] +	\vphantom{\bigg(\bigg)}	\nonumber
			\\
			& \quad + 2 \, \big[ a^-_{1 , \mu} ( \partial_\nu \partial_\rho \pi^0 ) - a^0_{1 , \mu} ( \partial_\nu \partial_\rho \pi^- ) \big] \Big\} +	\vphantom{\bigg(\bigg)}	\nonumber
			\\
			& + \rho^{-,\mu\nu\rho}_3 \Big\{ \sqrt{2} \, \big[ K^+_{1 , A,\mu} ( \partial_\nu \partial_\rho \bar{K}^0 ) - \bar{K}^0_{1 , A,\mu} ( \partial_\nu \partial_\rho K^+ ) \big] +	\vphantom{\bigg(\bigg)}	\nonumber
			\\
			& \quad + 2 \, \big[ - a^+_{1 , \mu} ( \partial_\nu \partial_\rho \pi^0 ) + a^0_{1 , \mu} ( \partial_\nu \partial_\rho \pi^+ ) \big] \Big\} +	\vphantom{\bigg(\bigg)}	\nonumber
			\\
			& + \bar{K}^{\ast 0, \mu\nu\rho}_3 \Big\{ + K^0_{1 , A,\mu} ( \partial_\nu \partial_\rho \pi^0 ) - \sqrt{2} \, K^+_{1 , A,\mu} ( \partial_\nu \partial_\rho \pi^- ) -	\vphantom{\bigg(\bigg)}	\nonumber
			\\
			& \quad - a^0_{1 , \mu} ( \partial_\nu \partial_\rho K^0 ) + \sqrt{2} \, a^-_{1 , \mu} ( \partial_\nu \partial_\rho K^+ ) +	\vphantom{\bigg(\bigg)}	\nonumber
			\\
			& \quad + K^0_{1 , A,\mu} \big[ ( \partial_\nu \partial_\rho \eta) ( - \cos \beta_p + \sqrt{2} \sin \beta_p ) +	\vphantom{\bigg(\bigg)}	\nonumber
			\\
			& \qquad + ( \partial_\nu \partial_\rho \eta^\prime) ( \sin \beta_p + \sqrt{2} \cos \beta_p ) \big] +	\vphantom{\bigg(\bigg)}	\nonumber
			\\
			& \quad + \big[ f_{1 , \mu} ( \cos \beta_{a_1} - \sqrt{2} \sin \beta_{a_1} ) +	\vphantom{\bigg(\bigg)}	\nonumber
			\\
			& \qquad + f^\prime_{1 , \mu} ( - \sin \beta_{a_1} - \sqrt{2} \cos \beta_{a_1} ) \big] ( \partial_\nu \partial_\rho K^0 ) \Big\} +	\vphantom{\bigg(\bigg)}	\nonumber
			\\
			& + K^{\ast 0, \mu\nu\rho}_3 \Big\{ - \bar{K}^0_{1 , A,\mu} ( \partial_\nu \partial_\rho \pi^0 ) + \sqrt{2} \, K^-_{1 , A,\mu} ( \partial_\nu \partial_\rho \pi^+ ) +	\vphantom{\bigg(\bigg)}	\nonumber
			\\
			& \quad + a^0_{1 , \mu} ( \partial_\nu \partial_\rho \bar{K}^0 ) - \sqrt{2} \, a^+_{1 , \mu} ( \partial_\nu \partial_\rho K^- ) +	\vphantom{\bigg(\bigg)}	\nonumber
			\\
			& \quad + \bar{K}^0_{1 , A,\mu} \big[ ( \partial_\nu \partial_\rho \eta) ( \cos \beta_p - \sqrt{2} \sin \beta_p ) +	\vphantom{\bigg(\bigg)}	\nonumber
			\\
			& \qquad + ( \partial_\nu \partial_\rho \eta^\prime) ( - \sin \beta_p - \sqrt{2} \cos \beta_p ) \big] +	\vphantom{\bigg(\bigg)}	\nonumber
			\\
			& \quad + \big[ f_{1 , \mu} ( - \cos \beta_{a_1} + \sqrt{2} \sin \beta_{a_1} ) +	\vphantom{\bigg(\bigg)}	\nonumber
			\\
			& \qquad + f^\prime_{1 , \mu} ( \sin \beta_{a_1} + \sqrt{2} \cos \beta_{a_1} ) \big] ( \partial_\nu \partial_\rho \bar{K}^0 ) \Big\} +	\vphantom{\bigg(\bigg)}	\nonumber
			\\
			& + K^{\ast +, \mu\nu\rho}_3 \Big\{ + K^-_{1 , A,\mu} ( \partial_\nu \partial_\rho \pi^0 ) + \sqrt{2} \, \bar{K}^0_{1 , A,\mu} ( \partial_\nu \partial_\rho \pi^- ) -	\vphantom{\bigg(\bigg)}	\nonumber
			\\
			& \quad - a^0_{1 , \mu} ( \partial_\nu \partial_\rho K^- ) - \sqrt{2} \, a^-_{1 , \mu} ( \partial_\nu \partial_\rho \bar{K}^0 ) +	\vphantom{\bigg(\bigg)}	\nonumber
			\\
			& \quad + K^-_{1 , A,\mu} \big[ ( \partial_\nu \partial_\rho \eta) ( \cos \beta_p - \sqrt{2} \sin \beta_p ) +	\vphantom{\bigg(\bigg)}	\nonumber
			\\
			& \qquad + ( \partial_\nu \partial_\rho \eta^\prime) ( - \sin \beta_p - \sqrt{2} \cos \beta_p ) \big] +	\vphantom{\bigg(\bigg)}	\nonumber
			\\
			& \quad + \big[ f_{1 , \mu} ( - \cos \beta_{a_1} + \sqrt{2} \sin \beta_{a_1} ) +	\vphantom{\bigg(\bigg)}	\nonumber
			\\
			& \qquad + f^\prime_{1 , \mu} ( \sin \beta_{a_1} + \sqrt{2} \cos \beta_{a_1} ) \big] ( \partial_\nu \partial_\rho K^- ) \Big\} +	\vphantom{\bigg(\bigg)}	\nonumber
			\\
			& + K^{\ast -, \mu\nu\rho}_3 \Big\{ - K^+_{1 , A,\mu} ( \partial_\nu \partial_\rho \pi^0 ) - \sqrt{2} \, K^0_{1 , A,\mu} ( \partial_\nu \partial_\rho \pi^+ ) +	\vphantom{\bigg(\bigg)}	\nonumber
			\\
			& \quad + a^0_{1 , \mu} ( \partial_\nu \partial_\rho K^+ ) + \sqrt{2} \, a^+_{1 , \mu} ( \partial_\nu \partial_\rho K^0 ) +	\vphantom{\bigg(\bigg)}	\nonumber
			\\
			& \quad + K^+_{1 , A,\mu} \big[ ( \partial_\nu \partial_\rho \eta) ( - \cos \beta_p + \sqrt{2} \sin \beta_p ) +	\vphantom{\bigg(\bigg)}	\nonumber
			\\
			& \qquad + ( \partial_\nu \partial_\rho \eta^\prime) ( \sin \beta_p + \sqrt{2} \cos \beta_p ) \big] +	\vphantom{\bigg(\bigg)}	\nonumber
			\\
			& \quad + \big[ f_{1 , \mu} ( \cos \beta_{a_1} - \sqrt{2} \sin \beta_{a_1} ) +	\vphantom{\bigg(\bigg)}	\nonumber
			\\
			& \qquad + f^\prime_{1 , \mu} ( - \sin \beta_{a_1} - \sqrt{2} \cos \beta_{a_1} ) \big] ( \partial_\nu \partial_\rho K^+ ) \Big\} +	\vphantom{\bigg(\bigg)}	\nonumber
			\\
			& + \omega^{\mu\nu\rho}_3 \Big\{ ( - \cos \beta_{w_3} + \sqrt{2} \sin \beta_{w_3} ) \times	\vphantom{\bigg(\bigg)}	\nonumber
			\\
			& \quad \times \big[ \bar{K}^0_{1 , A,\mu} ( \partial_\nu \partial_\rho  K^0 ) - K^0_{1 , A,\mu} ( \partial_\nu \partial_\rho \bar{K}^0 ) -	\vphantom{\bigg(\bigg)}	\nonumber
			\\
			& \qquad - K^+_{1 , A,\mu} ( \partial_\nu \partial_\rho K^- ) + K^-_{1 , A,\mu} ( \partial_\nu \partial_\rho K^+ ) \big] \Big\} +	\vphantom{\bigg(\bigg)}	\nonumber
			\\
			& + \phi^{\mu\nu\rho}_3 \Big\{ ( \sin \beta_{w_3} + \sqrt{2} \cos \beta_{w_3} ) \times	\vphantom{\bigg(\bigg)}	\nonumber
			\\
			& \quad \times \big[ \bar{K}^0_{1 , A,\mu} ( \partial_\nu \partial_\rho  K^0 ) - K^0_{1 , A,\mu} ( \partial_\nu \partial_\rho \bar{K}^0 ) -	\vphantom{\bigg(\bigg)}	\nonumber
			\\
			& \qquad - K^+_{1 , A,\mu} ( \partial_\nu \partial_\rho K^- ) + K^-_{1 , A,\mu} ( \partial_\nu \partial_\rho K^+ ) \big] \Big\} \Big) \, .	\vphantom{\bigg(\bigg)}	\nonumber
		\end{align}

		\begin{align}
			& \mathcal{L}_{w_3 v_1 v_1} =	\vphantom{\bigg(\bigg)}	\label{eq:extended_lagrangian_w3_v1_v1}
			\\
			= \, & g_{w_3 v_1 v_1} \, \mathrm{tr} \big( W_3^{\mu\nu\rho} \big[ ( \partial_\mu V_{1 , \nu} ) , \, V_{1 , \rho} \big]_{-} \big) =	\vphantom{\bigg(\bigg)}	\nonumber
			\\
			= \, & \frac{g_{w_3 v_1 v_1}}{4} \, \Big(	\vphantom{\bigg(\bigg)}	\nonumber
			\\
			& + \rho^{0,\mu\nu\rho}_3 \Big\{ + ( \partial_\mu \bar{K}^{\ast 0}_\nu ) K^{\ast 0}_\rho - ( \partial_\mu K^{\ast 0}_\nu ) \bar{K}^{\ast 0}_\rho +	\vphantom{\bigg(\bigg)}	\nonumber
			\\
			& \quad + ( \partial_\mu K^{\ast +}_\nu ) K^{\ast -}_\rho - ( \partial_\mu K^{\ast -}_\nu ) K^{\ast +}_\rho +	\vphantom{\bigg(\bigg)}	\nonumber
			\\
			& \quad + 2 \, \big[ (\partial_\mu \rho^+_\nu) \rho^-_\rho - ( \partial_\mu \rho^-_\nu ) \rho^+_\rho \big] \Big\} + \vphantom{\frac{1}{4}} \nonumber
			\\
			& + \rho^{+,\mu\nu\rho}_3 \Big\{ \sqrt{2} \, \big[ - (\partial_\mu K^{\ast -}_\nu) K^{\ast 0}_\rho + (\partial_\mu K^{\ast 0}_\nu) K^{\ast -}_\rho \big] +	\vphantom{\bigg(\bigg)}	\nonumber
			\\
			& \quad + 2 \, \big[ (\partial_\mu \rho^-_\nu) \rho^0_\rho - (\partial_\mu \rho^0_\nu) \rho^-_\rho \big] \Big\} +	\vphantom{\bigg(\bigg)}	\nonumber
			\\
			& + \rho^{-,\mu\nu\rho}_3 \Big\{ \sqrt{2} \, \big[ ( \partial_\mu K^{\ast +}_\nu ) \bar{K}^{\ast 0}_\rho - ( \partial_\mu \bar{K}^{\ast 0}_\nu ) K^{\ast +}_\rho \big] +	\vphantom{\bigg(\bigg)}	\nonumber
			\\
			& \quad + 2\, \big[ - (\partial_\mu \rho^+_\nu) \rho^0_\rho + (\partial_\mu \rho^0_\nu) \rho^+_\rho \big] \Big\} +	\vphantom{\bigg(\bigg)}	\nonumber
			\\
			& + \bar{K}^{\ast 0, \mu\nu\rho}_3 \Big\{ + ( \partial_\mu K^{\ast 0}_\nu ) \rho^0_\rho - ( \partial_\mu \rho^0_\nu ) K^{\ast 0}_\rho +	\vphantom{\bigg(\bigg)}	\nonumber
			\\
			& \quad + \sqrt{2} \, \big[ - ( \partial_\mu K^{\ast +}_\nu ) \rho^-_\rho + ( \partial_\mu \rho^-_\nu ) K^{\ast +}_\rho \big] +	\vphantom{\bigg(\bigg)}	\nonumber
			\\
			& \quad + (\partial_\mu K^{\ast 0}_\nu) \big[ \omega_\rho ( - \cos \beta_{v_1} + \sqrt{2} \sin \beta_{v_1} ) +	\vphantom{\bigg(\bigg)}	\nonumber
			\\
			& \qquad + \phi_\rho ( \sin \beta_{v_1} + \sqrt{2} \cos \beta_{v_1} ) \big] -	\vphantom{\bigg(\bigg)}	\nonumber
			\\
			& \quad - \big[ ( \partial_\mu \omega_\nu ) ( - \cos \beta_{v_1} + \sqrt{2} \sin \beta_{v_1} ) +	\vphantom{\bigg(\bigg)}	\nonumber
			\\
			& \qquad + (\partial_\mu \phi_\nu) ( \sin \beta_{v_1} + \sqrt{2} \cos \beta_{v_1} ) \big] K^{\ast 0}_\rho \Big\} +	\vphantom{\bigg(\bigg)}	\nonumber
			\\
			& + K^{\ast 0, \mu\nu\rho}_3 \Big\{ - (\partial_\mu \bar{K}^{\ast 0}_\nu) \rho^0_\rho + (\partial_\mu \rho^0_\nu) \bar{K}^{\ast 0}_\rho +	\vphantom{\bigg(\bigg)}	\nonumber
			\\
			& \quad + \sqrt{2} \, \big[ (\partial_\mu K^{\ast -}_\nu) \rho^+_\rho - (\partial_\mu \rho^+_\nu) K^{\ast -}_\rho \big] -	\vphantom{\bigg(\bigg)}	\nonumber
			\\
			& \quad - (\partial_\mu \bar{K}^{\ast 0}_\nu) \big[ \omega_\rho ( - \cos \beta_{v_1} + \sqrt{2} \sin \beta_{v_1} ) +	\vphantom{\bigg(\bigg)}	\nonumber
			\\
			& \qquad + \phi_\rho ( \sin \beta_{v_1} + \sqrt{2} \cos \beta_{v_1} ) \big] +	\vphantom{\bigg(\bigg)}	\nonumber
			\\
			& \quad + \big[ (\partial_\mu \omega_\nu) ( - \cos \beta_{v_1} + \sqrt{2} \sin \beta_{v_1} ) +	\vphantom{\bigg(\bigg)}	\nonumber
			\\
			& \qquad + (\partial_\mu \phi_\nu) ( \sin \beta_{v_1} + \sqrt{2} \cos \beta_{v_1} ) \big] \bar{K}^{\ast 0}_\rho \Big\} +	\vphantom{\bigg(\bigg)}	\nonumber
			\\
			& + K^{\ast +, \mu\nu\rho}_3 \Big\{ + (\partial_\mu K^{\ast -}_\nu) \rho^0_\rho - (\partial_\mu \rho^0_\nu) K^{\ast -}_\rho +	\vphantom{\bigg(\bigg)}	\nonumber
			\\
			& \quad + \sqrt{2} \, \big[ (\partial_\mu \bar{K}^{\ast 0}_\nu) \rho^-_\rho - (\partial_\mu \rho^-_\nu) \bar{K}^{\ast 0}_\rho \big] -	\vphantom{\bigg(\bigg)}	\nonumber
			\\
			& \quad - (\partial_\mu K^{\ast -}_\nu) \big[ \omega_\rho ( - \cos \beta_{v_1} + \sqrt{2} \sin \beta_{v_1} ) +	\vphantom{\bigg(\bigg)}	\nonumber
			\\
			& \qquad + \phi_\rho ( \sin \beta_{v_1} + \sqrt{2} \cos \beta_{v_1} ) \big] +	\vphantom{\bigg(\bigg)}	\nonumber
			\\
			& \quad + \big[ (\partial_\mu \omega_\nu) ( - \cos \beta_{v_1} + \sqrt{2} \sin \beta_{v_1} ) +	\vphantom{\bigg(\bigg)}	\nonumber
			\\
			& \qquad + (\partial_\mu \phi_\nu) ( \sin \beta_{v_1} + \sqrt{2} \cos \beta_{v_1} ) \big] K^{\ast -}_\rho \Big\} +	\vphantom{\bigg(\bigg)}	\nonumber
			\\
			& + K^{\ast -, \mu\nu\rho}_3 \Big\{ - (\partial_\mu K^{\ast +}_\nu) \rho^0_\rho + (\partial_\mu \rho^0_\nu) K^{\ast +}_\rho +	\vphantom{\bigg(\bigg)}	\nonumber
			\\
			& \quad + \sqrt{2} \, \big[ - ( \partial_\mu K^{\ast 0}_\nu ) \rho^+_\rho + ( \partial_\mu \rho^+_\nu ) K^{\ast 0}_\rho \big] +	\vphantom{\bigg(\bigg)}	\nonumber
			\\
			& \quad + ( \partial_\mu K^{\ast +}_\nu ) \big[ \omega_\rho ( - \cos \beta_{v_1} + \sqrt{2} \sin \beta_{v_1} ) +	\vphantom{\bigg(\bigg)}	\nonumber
			\\
			& \qquad + \phi_\rho ( \sin \beta_{v_1} + \sqrt{2} \cos \beta_{v_1} ) \big] -	\vphantom{\bigg(\bigg)}	\nonumber
			\\
			& \quad - \big[ (\partial_\mu \omega_\nu) ( - \cos \beta_{v_1} + \sqrt{2} \sin \beta_{v_1} ) +	\vphantom{\bigg(\bigg)}	\nonumber
			\\
			& \qquad + (\partial_\mu \phi_\nu) ( \sin \beta_{v_1} + \sqrt{2} \cos \beta_{v_1} ) \big] K^{\ast +}_\rho \Big\} +	\vphantom{\bigg(\bigg)}	\nonumber
			\\
			& + \omega^{\mu\nu\rho}_3 \Big\{ ( - \cos \beta_{w_3} + \sqrt{2} \sin \beta_{w_3} ) \times	\vphantom{\bigg(\bigg)}	\nonumber
			\\
			& \quad \times \big[ ( \partial_\mu \bar{K}^{\ast 0}_\nu ) K^{\ast 0}_\rho - ( \partial_\mu K^{\ast 0}_\nu ) \bar{K}^{\ast 0}_\rho -	\vphantom{\bigg(\bigg)}	\nonumber
			\\
			& \qquad - ( \partial_\mu K^{\ast +}_\nu ) K^{\ast -}_\rho + ( \partial_\mu K^{\ast -}_\nu ) K^{\ast +}_\rho \big] \Big\} +	\vphantom{\bigg(\bigg)}	\nonumber
			\\
			& + \phi^{\mu\nu\rho}_3 \Big\{ ( \sin \beta_{w_3} + \sqrt{2} \cos \beta_{w_3} ) \times	\vphantom{\bigg(\bigg)}	\nonumber
			\\
			& \quad \times \big[ (\partial_\mu \bar{K}^{\ast 0}_\nu) K^{\ast 0}_\rho - (\partial_\mu K^{\ast 0}_\nu) \bar{K}^{\ast 0}_\rho -	\vphantom{\bigg(\bigg)}	\nonumber
			\\
			& \qquad - (\partial_\mu K^{\ast +}_\nu) K^{\ast -}_\rho + (\partial_\mu K^{\ast -}_\nu) K^{\ast +}_\rho \big] \Big\} \Big) \, .	\vphantom{\bigg(\bigg)}	\nonumber
		\end{align}
	
		\begin{align}
			& \mathcal{L}_{g_3 v_1 p} =	\vphantom{\bigg(\bigg)}	\label{eq:extended_lagrangian_g3_v1_p}
			\\
			= \, & c_{g_3 v_1 p} \, G_{3 , \mu\alpha\beta} \, \varepsilon^{\mu\nu\rho\sigma} \, \mathrm{tr} \big[ \big\{ ( \partial_\nu V_{1 , \rho} ) , \, ( \partial^{\alpha} \partial^{\beta} \partial_{\sigma} P ) \big\}_{+} \big] =	\vphantom{\bigg(\bigg)}	\nonumber
			\\
			= \, & c_{g_3 v_1 p} \, G_{3 , \mu\alpha\beta} \, \varepsilon^{\mu\nu\rho\sigma} \big\{	\vphantom{\bigg(\bigg)}	\nonumber
			\\
			& + ( \partial_{\nu} \bar{K}^{0}_{1 , \mu} ) ( \partial^{\alpha} \partial^{\beta} \partial_{\sigma} K^{0} ) + ( \partial_{\nu} K_{1 , \mu}^{0} ) ( \partial^{\alpha} \partial^{\beta} \partial_{\sigma} \bar{K}^{0} ) +	\vphantom{\bigg(\bigg)}	\nonumber
			\\
			& + ( \partial_{\nu} K_{1 , \mu}^{+} ) ( \partial^{\alpha} \partial^{\beta} \partial_{\sigma} K^{-} ) + ( \partial_{\nu} K_{1 , \mu}^{-} ) ( \partial^{\alpha} \partial^{\beta} \partial_{\sigma} K^{+} ) +	\vphantom{\bigg(\bigg)}	\nonumber
			\\
			& + ( \partial_{\nu} \rho_{1 , \mu}^{0} ) ( \partial^{\alpha} \partial^{\beta} \partial_{\sigma} \pi^{0} ) + ( \partial_{\nu} \rho_{1 , \mu}^{+} ) ( \partial^{\alpha} \partial^{\beta} \partial_{\sigma} \pi^{-} ) +	\vphantom{\bigg(\bigg)}	\nonumber
			\\
			& + ( \partial_{\nu} \rho_{1 , \mu}^{-} ) ( \partial^{\alpha} \partial^{\beta} \partial_{\sigma} \pi^{+} ) +	\vphantom{\bigg(\bigg)}	\nonumber
			\\
			& + \big[ ( \partial_{\nu} \omega_{1 , \mu} ) ( \partial^{\alpha} \partial^{\beta} \partial_{\sigma} \eta ) + ( \partial_{\nu} \phi_{1 , \mu} ) ( \partial^{\alpha} \partial^{\beta} \partial_{\sigma} \eta^{\prime} ) \big] \times	\vphantom{\bigg(\bigg)}	\nonumber
			\\
			& \quad \times \cos ( \beta_{p} - \beta_{v_1} )	\vphantom{\bigg(\bigg)}	\nonumber
			\\
			& + \big[ ( \partial_{\nu} \phi_{1 , \mu} ) ( \partial^{\alpha} \partial^{\beta} \partial_{\sigma} \eta ) - ( \partial_{\nu} \omega_{1 , \mu} ) ( \partial^{\alpha} \partial^{\beta} \partial_{\sigma} \eta^{\prime} ) \big] \times	\vphantom{\bigg(\bigg)}	\nonumber
			\\
			& \quad \times \sin ( \beta_{p} - \beta_{v_1} ) \big\} \, .	\vphantom{\bigg(\bigg)}	\nonumber
		\end{align}

		\begin{align}
			& \mathcal{L}_{g_3 \gamma p} =	\vphantom{\bigg(\bigg)}	\label{eq:extended_lagrangian_g3_gamma_p}
			\\
			= \, & c_{g_3 v_1 p} \, \tfrac{e}{g_\rho} \, G_{3 , \mu \alpha \beta} \, ( \partial_\nu a_\rho ) \, \varepsilon^{\mu\nu\rho\sigma} \, \mathrm{tr} \big[ \big\{ Q , \, ( \partial^{\alpha} \partial^{\beta} \partial_{\sigma} P ) \big\}_{+} \big] =	\vphantom{\bigg(\bigg)}	\nonumber
			\\
			= \, & \frac{c_{g_3 v_1 p}}{3} \, \tfrac{e}{g_\rho} \, G_{3 , \mu \alpha \beta} \, ( \partial_\nu a_\rho ) \, \varepsilon^{\mu\nu\rho\sigma} \big\{	\vphantom{\bigg(\bigg)}	\nonumber
			\\
			& + 3 \, \pi^0 + \big[ \cos ( \beta_{p} ) - \sqrt{2} \, \sin ( \beta_{p} ) \big] \, \eta -	\vphantom{\bigg(\bigg)}	\nonumber
			\\
			& - \big[ \sin ( \beta_{p} ) + \sqrt{2} \, \cos ( \beta_{p} ) \big] \, \eta^\prime \big\} \, .	\vphantom{\bigg(\bigg)}	\nonumber
		\end{align}
	
		\begin{align}
			& \mathcal{L}_{g_3 v_1 a_1} =	\vphantom{\bigg(\bigg)}	\label{eq:gva1lagexp}
			\\
			= \, & c_{g_3 v_1 a_1} \, G_{3 , \mu\alpha\beta} \, \varepsilon^{\mu\nu\rho\sigma} \, \mathrm{tr} \big[ \big\{ ( \partial_\nu V_{1 , \rho} ) , \, ( \partial^{\alpha} \partial^{\beta} A_{1 , \sigma} ) \big\}_{+} \big] =	\vphantom{\bigg(\bigg)}	\nonumber
			\\
			= \, & c_{g_3 v_1 a_1} \, G_{3 , \mu\alpha\beta} \, \varepsilon^{\mu\nu\rho\sigma} \big\{	\vphantom{\bigg(\bigg)}	\nonumber
			\\
			& + ( \partial_{\nu} \bar{K}^{0}_{1 , \mu} ) ( \partial^{\alpha} \partial^{\beta} K_{1 , A , \sigma}^{0} ) + ( \partial_{\nu} K_{1 , \mu}^{0} ) ( \partial^{\alpha} \partial^{\beta} \bar{K}_{1 , A , \sigma}^{0} ) +	\vphantom{\bigg(\bigg)}	\nonumber
			\\
			& + ( \partial_{\nu} K_{1 , \mu}^{+} ) ( \partial^{\alpha} \partial^{\beta} K_{1 , A , \sigma}^{-} ) + ( \partial_{\nu} K_{1 , \mu}^{-} ) ( \partial^{\alpha} \partial^{\beta} K_{1 , A , \sigma}^{+} ) +	\vphantom{\bigg(\bigg)}	\nonumber
			\\
			& + ( \partial_{\nu} \rho_{1 , \mu}^{0} ) ( \partial^{\alpha} \partial^{\beta} a_{1 , \sigma}^{0} ) + ( \partial_{\nu} \rho_{1 , \mu}^{+} ) ( \partial^{\alpha} \partial^{\beta} a_{1 , \sigma}^{-} ) +	\vphantom{\bigg(\bigg)}	\nonumber
			\\
			& + ( \partial_{\nu} \rho_{1 , \mu}^{-} ) ( \partial^{\alpha} \partial^{\beta} a_{1 , \sigma}^{+} ) +	\vphantom{\bigg(\bigg)}	\nonumber
			\\
			& + \big[ ( \partial_{\nu} \omega_{1 , \mu} ) ( \partial^{\alpha} \partial^{\beta} f_{1 , \sigma} ) + ( \partial_{\nu} \phi_{1 , \mu} ) ( \partial^{\alpha} \partial^{\beta} f_{1 , \sigma}^{\prime} ) \big] \times	\vphantom{\bigg(\bigg)}	\nonumber
			\\
			& \quad \times \cos ( \beta_{a_1} - \beta_{v_1} )	\vphantom{\bigg(\bigg)}	\nonumber
			\\
			& + \big[ ( \partial_{\nu} \phi_{1 , \mu} ) ( \partial^{\alpha} \partial^{\beta} f_{1 , \sigma} ) - ( \partial_{\nu} \omega_{1 , \mu} ) ( \partial^{\alpha} \partial^{\beta} f_{1 , \sigma}^{\prime} ) \big] \times	\vphantom{\bigg(\bigg)}	\nonumber
			\\
			& \quad \times \sin ( \beta_{a_1} - \beta_{v_1} ) \big\} \, .	\vphantom{\bigg(\bigg)}	\nonumber
		\end{align}
	
		\begin{align}
			& \mathcal{L}_{g_3 b_1 p} =	\vphantom{\bigg(\bigg)}	\label{eq:extended_lagrangian_g3_b1_p}
			\\
			= \, & c_{g_3 b_1 p} \, G_3^{\mu\nu\rho} \, \mathrm{tr} \big[ \big\{ B_{1 , \mu} , \, ( \partial_{\nu} \partial_{\rho} P ) \big\}_{+} \big] =	\vphantom{\bigg(\bigg)}	\nonumber
			\\
			= \, & c_{g_3 b_1 p} \, G_3^{\mu\nu\rho} \big\{	\vphantom{\bigg(\bigg)}	\nonumber
			\\
			& + \bar{K}^{0}_{1 , B , \mu} ( \partial_{\nu} \partial_{\rho} K^{0} ) + K_{1 , B , \mu}^{0} ( \partial_{\nu} \partial_{\rho} \bar{K}^{0}) +	\vphantom{\bigg(\bigg)}	\nonumber
			\\
			& + K_{1 , B , \mu}^{+} ( \partial_{\nu} \partial_{\rho} K^{-} ) + K_{1 , B , \mu}^{-} ( \partial_{\nu} \partial_{\rho} K^{+} ) +	\vphantom{\bigg(\bigg)}	\nonumber
			\\
			& + b_{1 , \mu}^{0} ( \partial_{\nu} \partial_{\rho} \pi^{0} ) + b_{1 , \mu}^{+} ( \partial_{\nu} \partial_{\rho} \pi^{-} ) + b_{1 , \mu}^{-} ( \partial_{\nu} \partial_{\rho} \pi^{+} ) + \vphantom{\bigg(\bigg)}	\nonumber
			\\
			& + \big[ h_{1 , \mu} ( \partial_{\nu} \partial_{\rho} \eta ) + h_{1 , \mu}^{\prime} ( \partial_{\nu} \partial_{\rho} \eta^{\prime} ) \big] \, \cos ( \beta_{p} - \beta_{b_1} ) +	\vphantom{\bigg(\bigg)}	\nonumber
			\\
			& + \big[ h_{1 , \mu}^{\prime} ( \partial_{\nu} \partial_{\rho} \eta ) - h_{1 , \mu} ( \partial_{\nu} \partial_{\rho} \eta^{\prime} ) \big] \, \sin ( \beta_{p} - \beta_{b_1} ) \big\} \, .	\vphantom{\bigg(\bigg)}	\nonumber
		\end{align}

		\begin{align}
			& \mathcal{L}_{g_3 b_1 a_1} =	\vphantom{\bigg(\bigg)}	\label{eq:extended_lagrangian_g3_b1_a1}
			\\
			= \, & c_{g_3 b_1 a_1} \, G_3^{\mu\nu\rho} \, \mathrm{tr} \big[ \big\{ B_{1 , \mu} , \, ( \partial_{\nu} A_{1 , \rho} ) \big\}_{+} \big] =	\vphantom{\bigg(\bigg)}	\nonumber
			\\
			= \, & c_{g_3 b_1 a_1} \, G_3^{\mu\nu\rho} \big\{	\vphantom{\bigg(\bigg)}	\nonumber
			\\
			& + \bar{K}^{0}_{1 , B , \mu} ( \partial_{\nu} K_{1 , A , \rho}^{0} ) + K_{1 , B , \mu}^{0} ( \partial_{\nu} \bar{K}_{1 , A , \rho}^{0}) +	\vphantom{\bigg(\bigg)}	\nonumber
			\\
			& + K_{1 , B , \mu}^{+} ( \partial_{\nu} K_{1 , A , \rho}^{-} ) + K_{1 , B , \mu}^{-} ( \partial_{\nu} K_{1 , A , \rho}^{+} ) +	\vphantom{\bigg(\bigg)}	\nonumber
			\\
			& + b_{1 , \mu}^{0} ( \partial_{\nu} a_{1 , \rho}^{0} ) + b_{1 , \mu}^{+} ( \partial_{\nu} a_{1 , \rho}^{-} ) + b_{1 , \mu}^{-} ( \partial_{\nu} a_{1 , \rho}^{+} ) + \vphantom{\bigg(\bigg)}	\nonumber
			\\
			& + \big[ h_{1 , \mu} ( \partial_{\nu} f_{1 , \rho} ) + h_{1 , \mu}^{\prime} ( \partial_{\nu} f_{1 , \rho}^{\prime} ) \big] \, \cos ( \beta_{a_1} - \beta_{b_1} ) +	\vphantom{\bigg(\bigg)}	\nonumber
			\\
			& + \big[ h_{1 , \mu}^{\prime} ( \partial_{\nu} f_{1 , \rho} ) - h_{1 , \mu} ( \partial_{\nu} f_{1 , \rho}^{\prime} ) \big] \, \sin ( \beta_{a_1} - \beta_{b_1} ) \big\} \, .	\vphantom{\bigg(\bigg)}	\nonumber
		\end{align}

\section{Coefficients for the decay channels}
\label{app:tables}

	In this appendix we present the explicit forms of the coefficients $\kappa_i$ and $\kappa_i^\gamma$ in Eq.\ \eqref{eq:decay} based on the extended form of Lagrangians presented in the previous section in Tables \ref{tab:kappa_w3_p_p} - \ref{tab:kappa_w3_v1_p}.

	\begin{table}
		\centering
		\renewcommand{\arraystretch}{1.5}
		\caption{\label{tab:kappa_w3_p_p} %
			Coefficients $\kappa_i$ for the decay channels of $W_3 \rightarrow P + P$ decays, that are not kinematically forbidden.
		}
		\begin{ruledtabular}
			\begin{tabular}[c]{ l c }
				\multicolumn{1}{c}{decay process}									&	$\kappa_i$
				\\
				\colrule
				$\rho_{3} (1690) \rightarrow  \pi \, \pi$							&	$1$
				\\
				$\rho_{3} (1690) \rightarrow  \bar{K} \, K$							&	$2 \, \big( \frac{1}{2} \big)^{2}$
				\\
				\colrule
				$K_{3}^{\ast} (1780) \rightarrow  \pi \, \bar{K}$					&	$(\frac{1}{2})^{2}+(\frac{\sqrt{2}}{2})^{2}$
				\\
				$K_{3}^{\ast} (1780) \rightarrow  \bar{K} \, \eta$					&	$\big[ \frac{1}{2} \, ( - \cos \beta_{p} + \sqrt{2} \sin \beta_{p} ) \big]^{2}$
				\\
				$K_{3}^{\ast} (1780) \rightarrow  \bar{K} \, \eta^{\prime} (958)$	&	$\big[ \frac{1}{2} \, ( \sqrt{2} \cos \beta_{p} + \sin \beta_{p} ) \big]^{2}$
				\\
				\colrule
				$\omega_{3} (1670) \rightarrow  \bar{K} \, K$						&	$2 \, \big[ \frac{1}{2} \, ( - \cos \beta_{w_3} + \sqrt{2} \sin \beta_{w_3} ) \big]^{2}$
				\\
				\colrule
				$\phi_{3} (1850) \rightarrow  \bar{K} \, K$							&	$2 \, \big[ \frac{1}{2} \, ( \sqrt{2} \cos \beta_{w_3} + \sin \beta_{w_3} ) \big]^{2}$
			\end{tabular}
		\end{ruledtabular}
	\end{table}
	
	\begin{table}
		\centering
		\renewcommand{\arraystretch}{1.5}
		\caption{\label{tab:kappa_w3_gamma_p} %
			Coefficients $\kappa_i^\gamma$ for the decay channels of radioactive $W_3 \rightarrow \gamma \, P$ decays.
		}
		\begin{ruledtabular}
			\begin{tabular}[c]{ l c }
				\multicolumn{1}{c}{decay process}								&	$\kappa_i^{\gamma}$
				\\
				\colrule
				$\rho_{3}^{\pm/0} (1690) \rightarrow \gamma \, \pi^{\pm/0}$		&	$\big( \frac{1}{6} \big)^{2}$
				\\
				$\rho_{3}^{0} (1690) \rightarrow \gamma \,\eta$					&	$\big( \frac{1}{2} \cos \beta_{p} \big)^{2}$
				\\
				$\rho_{3}^{0} (1690) \rightarrow \gamma \, \eta^{\prime} (958)$	&	$\big( \frac{1}{2} \sin \beta_{p} \big)^{2}$
				\\
				\colrule
				$K_{3}^{\pm} (1780) \rightarrow \gamma \, K^{\pm}$				&	$\big( \frac{1}{6} \big)^{2}$
				\\
				$K_{3}^{0} (1780) \rightarrow \gamma \, K^{0}$					&	$\big( \frac{1}{3} \big)^{2}$
				\\
				\colrule
				$\omega_{3} (1670) \rightarrow \gamma \, \pi^0$					&	$\big( \frac{1}{2} \cos \beta_{w_3} \big)^{2}$
				\\
				$\omega_{3} (1670) \rightarrow \gamma \, \eta$					&	$\big[ \frac{1}{6} \big( \cos \beta_{w_3} \cos \beta_{p} - 2 \sin \beta_{w_3} \sin \beta_{p} \big) \big]^{2}$
				\\
				$\omega_{3} (1670) \rightarrow \gamma \, \eta^{\prime} (958)$	&	$\big[ \frac{1}{6} \big( \cos \beta_{w_3} \sin \beta_{p} + 2 \sin \beta_{w_3} \cos \beta_{p} \big) \big]^{2}$
				\\
				\colrule
				$\phi_{3} (1850) \rightarrow \gamma \, \pi^0$					&	$\big( \frac{1}{2} \sin \beta_{w_3} \big)^{2}$
				\\
				$\phi_{3} (1850) \rightarrow \gamma \, \eta$					&	$\big[ \frac{1}{6} \big( \sin \beta_{w_3} \cos \beta_{p} + 2 \cos \beta_{w_3} \sin \beta_{p} \big) \big]^{2}$
				\\
				$\phi_{3} (1850) \rightarrow \gamma \, \eta^{\prime} (958)$		&	$\big[ \frac{1}{6} \big( \sin \beta_{w_3} \sin \beta_{p} - 2 \cos \beta_{w_3} \cos \beta_{p} \big) \big]^{2}$
			\end{tabular}
		\end{ruledtabular}
	\end{table}

	\begin{table}
		\centering
		\renewcommand{\arraystretch}{1.5}
		\caption{\label{tab:kappa_w3_a2_p} %
			Coefficients $\kappa_i$ for the decay channels of $W_3 \rightarrow A_2 + P$ decays, that are not kinematically forbidden.
		}
		\begin{ruledtabular}
			\begin{tabular}[c]{ l c }
				\multicolumn{1}{c}{decay process}									&	$\kappa_i$
				\\
				\colrule
				$\rho_{3} (1690) \rightarrow  a_{2} (1320) \, \pi$					&	$2 \, \big( \frac{1}{2} \big)^{2}$
				\\
				\colrule
				$K_{3}^{\ast} (1780) \rightarrow  \bar{K}^{\ast}_{2} (1430) \, \pi$	&	$\big( \frac{1}{4} \big)^{2} + \big( \frac{\sqrt{2}}{4} \big)^{2}$
				\\
				$K_{3}^{\ast} (1780) \rightarrow  f_{2} (1270) \, \bar{K}$			&	$\big[ \frac{1}{4} \, \big( \cos \beta_{a_2} - \sqrt{2} \sin \beta_{a_2} \big) \big]^{2}$
			\end{tabular}
		\end{ruledtabular}
	\end{table}

	\begin{table}
		\centering
		\renewcommand{\arraystretch}{1.5}
		\caption{\label{tab:kappa_w3_b1_p} %
			Coefficients $\kappa_i$ for the decay channels of $W_3 \rightarrow B_1 + P$ decays, that are not kinematically forbidden.
		}
		\begin{ruledtabular}
			\begin{tabular}[c]{ l c }
				\multicolumn{1}{c}{decay process}									&	$\kappa_i$
				\\
				\colrule
				$\rho_{3} (1690) \rightarrow  h_1 (1170) \, \pi$					&	$\big( \frac{1}{2} \cos\beta_{b_1} \big)^{2}$
				\\
				$\rho_{3} (1690) \rightarrow  h_1 (1415) \, \pi$					&	$\big( \frac{1}{2} \sin \beta_{b_1} \big)^{2}$
				\\
				\colrule
				$K_{3}^{\ast} (1780) \rightarrow  b_1 (1235) \, K$					&	$\big( \frac{1}{4} \big)^{2} + \big( \frac{\sqrt{2}}{4} \big)^{2}$
				\\
				$K_{3}^{\ast} (1780) \rightarrow \bar{K}_{1 , B} \, \pi$			&	$\big( \frac{1}{4} \big)^{2} + \big( \frac{\sqrt{2}}{4} \big)^{2}$
				\\
				$K_{3}^{\ast} (1780) \rightarrow  h_1 (1170) \, \bar{K}$			&	$\big[ \frac{1}{4} \, \big( \cos \beta_{b_1} + \sqrt{2} \sin \beta_{b_1} \big) \big]^{2}$
				\\
				\colrule
				$\omega_{3} (1670) \rightarrow  b_1 (1235) \, \pi$					&	$3 \, \big( \frac{1}{2} \cos \beta_{w_3} \big)^{2}$
				\\
				\colrule
				$\phi_{3} (1850) \rightarrow  b_1 (1235) \, \pi$					&	$3 \, \big( \frac{1}{2} \sin \beta_{w_3} \big)^{2}$
				\\
				$\phi_{3} (1850) \rightarrow  \bar{K}_{1 , B} \, K$					& $4 \, \big[ \frac{1}{4} \, \big( \sqrt{2} \cos \beta_{w_3} - \sin \beta_{w_3} \big) \big]^{2}$
				\\
				\multirow{2}{*}{$\phi_{3} (1850) \rightarrow  h_1 (1170) \, \eta$}	&	$\big[ \frac{1}{2} \, \big( - \sin \beta_{w_3} \cos \beta_{b_1} \cos \beta_{p} +$
																					\\
																					&	$+ \sqrt{2} \cos \beta_{w_3} \sin \beta_{b_1} \sin \beta_{p} \big) \big]^{2}$
			\end{tabular}
		\end{ruledtabular}
	\end{table}

	\begin{table}
		\centering
		\renewcommand{\arraystretch}{1.5}
		\caption{\label{tab:kappa_w3_a1_p} %
			Coefficients $\kappa_i$ for the decay channels of $W_3 \rightarrow A_1 + P$ decays, that are not kinematically forbidden.
		}
		\begin{ruledtabular}
			\begin{tabular}[c]{ l c }
				\multicolumn{1}{c}{decay process}					&	$\kappa_i$
				\\
				\colrule
				$\rho_{3} (1690) \rightarrow a_1 (1260) \, \pi$		&	$2 \, \big( \frac{1}{2} \big)^{2}$
				\\
				\colrule
				$K_{3}^{\ast} (1780) \rightarrow K_{1 , A} \, \pi$	&	$\big( \frac{1}{4} \big)^{2} + \big( \frac{\sqrt{2}}{4} \big)^{2}$
				\\
				$K_{3}^{\ast} (1780) \rightarrow a_1 (1260) \, K$	&	$\big( \frac{1}{4} \big)^{2} + \big( \frac{\sqrt{2}}{4} \big)^{2}$
			\end{tabular}
		\end{ruledtabular}
	\end{table}

	\begin{table}
		\centering
		\renewcommand{\arraystretch}{1.5}
		\caption{\label{tab:kappa_w3_v1_v1} %
			Coefficients $\kappa_i$ for the decay channels of $W_3 \rightarrow V_1 + V_1$ decays, that are not kinematically forbidden.
		}
		\begin{ruledtabular}
			\begin{tabular}[c]{ l c }
				\multicolumn{1}{c}{decay process}										&	$\kappa_i$
				\\
				\colrule
				$\rho_{3} (1690) \rightarrow \rho (770) \, \rho (770)$					&	$1$
				\\
				\colrule
				$K_{3}^{\ast} (1780) \rightarrow \rho (770) \, \bar{K}^{\ast} (892)$	&	$\big( \frac{1}{2} \big)^{2} + \big( \frac{\sqrt{2}}{2} \big)^{2}$
				\\
				$K_{3}^{\ast} (1780) \rightarrow \bar{K}^{\ast} (892) \, \omega (782)$	&	$\big[ \frac{1}{2} \, \big( \cos \beta_{v_1} - \sqrt{2} \sin \beta_{v_1} \big) \big]^{2}$
				\\
				\colrule
				$\phi_{3} (1850) \rightarrow \bar{K}^{\ast} (892) \, K^{\ast} (892)$	&	$2 \, \big[ \frac{1}{2} \, \big( \sqrt{2} \cos \beta_{w_3} + \sin \beta_{w_3} \big) \big]^{2}$
			\end{tabular}
		\end{ruledtabular}
	\end{table}

	\begin{table}
		\centering
		\renewcommand{\arraystretch}{1.5}
		\caption{\label{tab:kappa_g3_v1_p} %
			Coefficients $\kappa_i$ for the decay channels of $G_3 (4200) \rightarrow V_1 + P$ decays, that are not kinematically forbidden.
		}
		\begin{ruledtabular}
			\begin{tabular}[c]{ l c }
				\multicolumn{1}{c}{decay process}								&	$\kappa_i$
				\\
				\colrule
				$G_3 (4200) \rightarrow \rho (770) \, \pi$						&	$3$
				\\
				$G_3 (4200) \rightarrow \bar{K}^{\ast} (892) \, K$				&	$4$
				\\
				$G_3 (4200) \rightarrow \omega (782) \, \eta$					&	$[ \cos( \beta_{p} - \beta_{v_1} ) ]^{2}$
				\\
				$G_3 (4200) \rightarrow \omega (782) \, \eta^{\prime} (958)$	&	$[ \sin( \beta_{p} - \beta_{v_1} ) ]^{2}$
				\\
				$G_3 (4200) \rightarrow \phi (1020) \, \eta$					&	$[ \sin( \beta_{p} - \beta_{v_1} ) ]^{2}$
				\\
				$G_3 (4200) \rightarrow \phi (1020) \, \eta^{\prime} (958)$		&	$[ \cos( \beta_{p} - \beta_{v_1} ) ]^{2}$
				\\
			\end{tabular}
		\end{ruledtabular}
	\end{table}

	\begin{table}
		\centering
		\renewcommand{\arraystretch}{1.8}
		\caption{\label{tab:kappa_g3_gamma_p} %
			Coefficients $\kappa_i^\gamma$ for the decay channels of radioactive $G_3 (4200) \rightarrow \gamma + P$ decays.
		}
		\begin{ruledtabular}
			\begin{tabular}[c]{ l c }
				\multicolumn{1}{c}{decay process}						&	$\kappa_i^\gamma$
				\\
				\colrule
				$G_3 (4200) \rightarrow \gamma \, \pi^0$				&	$1$
				\\
				$G_3 (4200) \rightarrow \gamma \, \eta$					&	$\big[ \frac{1}{3} \, \big( \cos \beta_{p} - \sqrt{2} \sin \beta_{p} \big) \big]^{2}$
				\\
				$G_3 (4200) \rightarrow \gamma \, \eta^{\prime} (958)$	&	$\big[ \frac{1}{3} \, \big( \sin \beta_{p} + \sqrt{2} \cos \beta_{p} \big) \big]^{2}$
			\end{tabular}
		\end{ruledtabular}
	\end{table}

	\begin{table}
		\centering
		\renewcommand{\arraystretch}{1.5}
		\caption{\label{tab:kappa_g3_v1_a1} %
			Coefficients $\kappa_i$ for the decay channels of $G_3 (4200) \rightarrow V_1 + A_1$ decays, that are not kinematically forbidden.
		}
		\begin{ruledtabular}
			\begin{tabular}[c]{ l c }
				\multicolumn{1}{c}{decay process}								&	$\kappa_i$
				\\
				\colrule
				$G_3 (4200) \rightarrow \rho (770) \, a_1 (1260)$				&	$3$
				\\
				$G_3 (4200) \rightarrow \bar{K}^{\ast} (892) \, K_{1 , A}$		&	$4$
				\\
				$G_3 (4200) \rightarrow \omega (782) \, f_1 (1285)$				&	$[ \cos( \beta_{a_1} - \beta_{v_1} ) ]^{2}$
				\\
				$G_3 (4200) \rightarrow \omega (782) \, f_1^{\prime} (1420)$	&	$[ \sin( \beta_{a_1} - \beta_{v_1} ) ]^{2}$
				\\
				$G_3 (4200) \rightarrow \phi (1020) \, f_1 (1285)$				&	$[ \sin( \beta_{a_1} - \beta_{v_1} ) ]^{2}$
				\\
				$G_3 (4200) \rightarrow \phi (1020) \, f_1^{\prime} (1420)$		&	$[ \cos( \beta_{a_1} - \beta_{v_1} ) ]^{2}$
				\\
			\end{tabular}
		\end{ruledtabular}
	\end{table}

	\begin{table}
		\centering
		\renewcommand{\arraystretch}{1.5}
		\caption{\label{tab:kappa_g3_b1_p} %
			Coefficients $\kappa_i$ for the decay channels of $G_3 (4200) \rightarrow B_1 + P$ decays, that are not kinematically forbidden.
		}
		\begin{ruledtabular}
			\begin{tabular}[c]{ l c }
				\multicolumn{1}{c}{decay process}						&	$\kappa_i$
				\\
				\colrule
				$G_3 (4200) \rightarrow b_1 (1235) \, \pi$				&	$3$
				\\
				$G_3(4200)\rightarrow K_{1 , B} \,K$					&	$4$
				\\
				$G_3(4200)\rightarrow h_1 (1170) \, \eta$				&	$[ \cos ( \beta_{p} - \beta_{b_1} ) ]^{2}$
				\\
				$G_3(4200)\rightarrow h_1 (1170) \,\eta^{\prime}(958)$	&	$[ \sin( \beta_{p} - \beta_{b_1} ) ]^{2}$
				\\
				$G_3(4200)\rightarrow h_1 (1415) \, \eta$				&	$[ \sin ( \beta_{p} - \beta_{b_1} ) ]^{2}$
				\\
				$G_3(4200)\rightarrow h_1 (1415) \, \eta^{\prime}(958)$	&	$[ \cos ( \beta_{p} - \beta_{b_1} ) ]^{2}$
			\end{tabular}
		\end{ruledtabular}
	\end{table}

	\begin{table}
		\centering
		\renewcommand{\arraystretch}{1.5}
		\caption{\label{tab:kappa_g3_b1_a1} %
			Coefficients $\kappa_i$ for the decay channels of $G_3 (4200) \rightarrow B_1 + A_1$ decays, that are not kinematically forbidden.
		}
		\begin{ruledtabular}
			\begin{tabular}[c]{ l c }
				\multicolumn{1}{c}{decay process}							&	$\kappa_i$
				\\
				\colrule
				$G_3 (4200) \rightarrow b_1 (1235) \, a_1 (1260)$			&	$3$
				\\
				$G_3(4200)\rightarrow K_{1 , B} \, K_{1 , A}$				&	$4$
				\\
				$G_3(4200)\rightarrow h_1 (1170) \, f_1 (1285)$				&	$[ \cos ( \beta_{a_1} - \beta_{b_1} ) ]^{2}$
				\\
				$G_3(4200)\rightarrow h_1 (1170) \, f_1^{\prime} (1420)$	&	$[ \sin( \beta_{a_1} - \beta_{b_1} ) ]^{2}$
				\\
				$G_3(4200)\rightarrow h_1 (1415) \, f_1 (1285)$				&	$[ \sin ( \beta_{a_1} - \beta_{b_1} ) ]^{2}$
				\\
				$G_3(4200)\rightarrow h_1 (1415)\, f_1^{\prime} (1420)$		&	$[ \cos ( \beta_{a_1} - \beta_{b_1} ) ]^{2}$
			\end{tabular}
		\end{ruledtabular}
	\end{table}

	\begin{table*}
		\centering
		\renewcommand{\arraystretch}{1.5}
		\caption{\label{tab:kappa_w3_v1_p} %
			Coefficients $\kappa_i$ for the decay channels of $W_3 \rightarrow V_1 + P$ decays, that are not kinematically forbidden.
		}
		\begin{tabular}[c]{ l @{\qquad} c }
			\toprule
			\multicolumn{1}{c}{decay process}												&	$\kappa_i$
			\\
			\colrule
			$\rho_{3} (1690) \rightarrow  \rho (770) \, \eta$								&	$\big( \frac{1}{2} \cos \beta_{p} \big)^{2}$
			\\
			$\rho_{3} (1690) \rightarrow  \bar{K}^{\ast} (892) \, K$						&	$4 \, \big( \frac{1}{4} \big)^{2}$
			\\
			$\rho_{3} (1690) \rightarrow  \omega (782) \, \pi$								&	$\big( \frac{1}{2} \cos \beta_{v_1} \big)^{2}$
			\\
			$\rho_{3} (1690) \rightarrow  \phi (1020) \, \pi$								&	$\big( \frac{1}{2} \sin \beta_{v_1} \big)^{2}$
			\\
			\colrule
			$K_{3}^{\ast} (1780) \rightarrow  \rho (770) \, K$								&	$\big( \frac{1}{4} \big)^{2} + \big( \frac{\sqrt{2}}{4} \big)^{2}$
			\\
			$K_{3}^{\ast} (1780) \rightarrow  \bar{K}^{\ast} (892) \, \pi$					&	$\big( \frac{1}{4} \big)^{2} + \big( \frac{\sqrt{2}}{4} \big)^{2}$
			\\
			$K_{3}^{\ast} (1780) \rightarrow  \bar{K}^{\ast} (892) \, \eta$					&	$\big[ \frac{1}{4} \, \big( \cos \beta_{p} + \sqrt{2} \sin \beta_{p} \big) \big]^{2}$
			\\
			$K_{3}^{\ast} (1780) \rightarrow  \bar{K}^{\ast} (892) \, \eta^{\prime} (958)$	&	$\big[ \frac{1}{4} \, \big( \sqrt{2} \cos \beta_{p} - \sin \beta_{p} \big) \big]^{2}$
			\\
			$K_{3}^{\ast} (1780) \rightarrow  \omega(782) \, \bar{K}$						&	$\big[ \frac{1}{4} \, \big( \cos \beta_{v_1} + \sqrt{2} \sin \beta_{v_1} \big) \big]^{2}$
			\\
			$K_{3}^{\ast} (1780) \rightarrow  \phi (1020) \, \bar{K}$						&	$\big[ \frac{1}{4} \, \big( \sqrt{2} \cos \beta_{v_1} - \sin \beta_{v_1} \big) \big]^{2}$
			\\
			\colrule
			$\omega_{3} (1670) \rightarrow  \rho (770) \, \pi$								&	$3 \, \big( \frac{1}{2} \cos \beta_{w_3} \big)^{2}$
			\\
			$\omega_{3} (1670) \rightarrow  \bar{K}^{\ast} (892) \, K$						&	$4 \, \big[ \frac{1}{4} \, \big( \cos \beta_{w_3} + \sqrt{2} \sin \beta_{w_3} \big) \big]^{2} $
			\\
			$\omega_{3} (1670) \rightarrow  \omega (782) \, \eta$							&	$\big[ \frac{1}{2} \, \big( \cos \beta_{w_3} \cos \beta_{v_1} \cos \beta_{p} + \sqrt{2} \sin \beta_{w_3} \sin \beta_{v_1} \sin \beta_{p} \big) \big]^{2}$
			\\
			$\omega_{3} (1670) \rightarrow  \phi (1020) \, \eta$							&	$\big[ \frac{1}{2} \, \big( - \cos \beta_{w_3} \sin \beta_{v_1} \cos \beta_{p} + \sqrt{2} \sin \beta_{w_3} \cos \beta_{v_1} \sin \beta_{p} \big) \big]^{2}$
			\\
			\colrule
			$\phi_{3} (1850) \rightarrow  \rho (770) \, \pi$								&	$3 \, \big( \frac{1}{2} \sin \beta_{w_3} \big)^{2}$
			\\
			$\phi_{3} (1850) \rightarrow  \bar{K}^{\ast} (892) \, K$						&	$4 \, \big[ \frac{1}{4} \, \big( \sqrt{2} \cos \beta_{w_3} - \sin \beta_{w_3} \big) \big]^{2}$
			\\
			$\phi_{3} (1850) \rightarrow  \omega (782) \, \eta$								&	$\big[ \frac{1}{2} \, \big( - \sin \beta_{w_3} \cos \beta_{v_1} \cos \beta_{p} + \sqrt{2} \cos \beta_{w_3} \sin \beta_{v_1} \sin \beta_{p} \big) \big]^{2}$
			\\
			$\phi_{3} (1850) \rightarrow  \omega (782) \, \eta^{\prime} (958)$				&	$\big[ \frac{1}{2} \, \big( \sin \beta_{w_3} \cos \beta_{v_1} \sin \beta_{p} + \sqrt{2} \cos \beta_{w_3} \sin \beta_{v_1} \cos \beta_{p} \big) \big]^{2}$
			\\
			$\phi_{3} (1850) \rightarrow  \phi (1020) \, \eta$								&	$\big[ \frac{1}{2} \, \big( \sin \beta_{w_3} \sin \beta_{v_1} \cos \beta_{p} + \sqrt{2} \cos \beta_{w_3} \cos \beta_{v_1} \sin \beta_{p} \big) \big]^{2}$
			\\
			\botrule
		\end{tabular}
	\end{table*}

\section{Unpolarized invariant decay amplitudes}
\label{app:deacyamplitudes}

	This appendix is devoted to the calculation of the unpolarized invariant decay amplitudes. These form the normal decay amplitudes by multiplication with factors $\kappa_i$ depending on the specific decay channels. We apply the following Feynman rules for each interaction vertex in order to derive the expressions for the decay amplitudes,
		\begin{align}
			\partial^{\mu} & \mapsto \mathrm{i} k_{\circ}^{\mu} \, ,	\vphantom{\bigg(\bigg)}
			\\
			V_1^\mu & \mapsto \epsilon^{\mu} ( \lambda_{v_1} , \vec{k}_{v_1} ) \, ,	\vphantom{\bigg(\bigg)}
			\\
			B_1^\mu & \mapsto \epsilon^{\mu} ( \lambda_{b_1} , \vec{k}_{b_1} ) \, ,	\vphantom{\bigg(\bigg)}
			\\
			A_1^\mu & \mapsto \epsilon^{\mu} ( \lambda_{a_1} , \vec{k}_{a_1} ) \, ,	\vphantom{\bigg(\bigg)}
			\\
			A_2^{\mu\nu} & \mapsto \epsilon^{\mu\nu} ( \lambda_{a_2} , \vec{k}_{a_2} ) \, ,	\vphantom{\bigg(\bigg)}
			\\
			W_3^{\mu\nu\rho} & \mapsto \epsilon^{\mu\nu\rho} ( \lambda_{w_3} , \vec{k}_{w_3} ) \, ,	\vphantom{\bigg(\bigg)}
			\\
			G_3^{\mu\nu\rho} & \mapsto \epsilon^{\mu\nu\rho} ( \lambda_{g_3} , \vec{k}_{g_3} ) \, ,	\vphantom{\bigg(\bigg)}
			\\
			a^\mu & \mapsto \epsilon^{\mu} ( \lambda_\gamma , \vec{k}_\gamma ) \, ,	\vphantom{\bigg(\bigg)}
		\end{align}
	where $\epsilon^{\mu(\nu)(\rho)}$ are the polarization vectors (tensors), $\lambda$ is the specific polarization\footnote{For massive spin-$3$ tensor fields $\lambda = - 3 , \,  - 2 , \, - 1 , \, 0 , \, + 1 , \, + 2 , \, + 3$, for massive spin-2 tensor fields $\lambda = - 2 , \, - 1 , \, 0 , \, + 1 , \, + 2$, whereas for massive vector fields $\lambda = - 1 , \, 0 , \, + 1$, representing the spin orientations in a magnetic field, or respective degrees of freedom of the massive fields. For massless vector fields (photons) there are only two polarizations $\lambda = 1, \, 2$.}, $k^{\mu}$ is the four momentum, and $\vec{k}$ is the three momentum.	
	\begin{enumerate}
		\item	For the interaction Lagrangian
			\begin{align}
				\mathcal{L}_{w_3 p p} = g_{w_3 p p} \, W_{3 , \mu\nu\rho} \, P^{(1)} \, \big( \partial^{\mu} \partial^{\nu} \partial^{\rho} P^{(2)} \big) \, ,
			\end{align}
		which describes decays of massive spin-$3$ fields into two massive pseudoscalar fields, the unpolarized invariant decay amplitude has the following form
			\begin{align}
				& \mathrm{i} \mathcal{M}_{w_3 p p} =	\vphantom{\bigg(\bigg)}
				\\
				= \, & \mathrm{i} g_{w_3 p p} \, \epsilon_{\mu\nu\rho} ( \lambda_{w_3} , \vec{k}_{w_3} ) \, k_{p^{(2)}}^\mu \, k_{p^{(2)}}^\nu \, k_{p^{(2)}}^\rho \, .	\vphantom{\bigg(\bigg)}	\nonumber
			\end{align}
		The square of the amplitude reads
			\begin{align}
				& \tfrac{1}{7} \, | \mathcal{M}_{w_3 p p} |^{2} =	\vphantom{\sum_{\lambda_{w_3} = -3}^{+3}}	\label{eq:amplitude_square_w3_p_p}
				\\
				= \, & g_{w_3 p p}^{2} \, \tfrac{1}{7} \sum_{\lambda_{w_3} = -3}^{+3} \epsilon_{\mu \nu \rho} ( \lambda_{w_3} , \vec{k}_{w_3} ) \, \epsilon_{\bar{\mu} \bar{\nu} \bar{\rho}} ( \lambda_{w_3} , \vec{k}_{w_3} ) \times	\vphantom{\sum_{\lambda_{w_3} = -3}^{+3}}	\nonumber
				\\
				& \times k_{p^{(2)}}^\mu \, k_{p^{(2)}}^\nu \, k_{p^{(2)}}^\rho \, k_{p^{(2)}}^{\bar{\mu}} \, k_{p^{(2)}}^{\bar{\nu}} \, k_{p^{(2)}}^{\bar{\rho}} =	\vphantom{\sum_{\lambda_{w_3} = -3}^{+3}}	\nonumber
				\\
				= \, & g_{w_3 p p}^{2} \, \tfrac{2}{35} \, \big| \vec{k}_{p^{(1)},p^{(2)}} \big|^{6} \, .	\vphantom{\sum_{\lambda_{w_3} = -3}^{+3}}	\nonumber
			\end{align}
			
		\item The unpolarized invariant decay amplitude for the interaction of massive spin-$3$ fields, massive vector fields and massive pseudoscalar fields 
			\begin{align}
				& \mathcal{L}_{w_3 v_1 p} =	\vphantom{\bigg(text\bigg)}
				\\
				= \, & g_{w_3 v_1 p} \, \varepsilon^{\mu\nu\rho\sigma} \, W_{3 , \mu\alpha\beta} \, ( \partial_{\nu} V_{1 , \rho} ) \, ( \partial^{\alpha} \partial^{\beta} \partial_{\sigma} P )	\vphantom{\bigg(text\bigg)}	\nonumber
			\end{align}
		reads
			\begin{align}
				& \mathrm{i} \mathcal{M}_{w_3 v_1 p} =	\vphantom{\bigg(text\bigg)}
				\\
				= \, & g_{w_3 v_1 p} \, \varepsilon^{\mu \nu \rho \sigma} \, \epsilon_{\mu \alpha \beta} ( \lambda_{w_3} , \vec{k}_{w_3} ) \times	\vphantom{\bigg(text\bigg)}	\nonumber
				\\
				& \times k_{v_1 \nu} \, \epsilon_{\rho} ( \lambda_{v_1} , \vec{k}_{v_1} ) \, k_p^\alpha \, k_p^\beta \, k_{p \sigma} \, .	\vphantom{\bigg(text\bigg)}	\nonumber
			\end{align}
		The square of the amplitude is
			\begin{align}
				& \tfrac{1}{7} \, | \mathcal{M}_{w_3 v_1 p} |^{2} =	\vphantom{\sum_{\lambda_{w_3} = -3}^{+3}}	\label{eq:amplitude_square_w3_v1_p}
				\\
				= \, & g_{w_3 v_1 p}^{2} \, \tfrac{1}{7} \, \varepsilon^{\mu \nu \rho \sigma} \, \varepsilon^{\bar{\mu} \bar{\nu} \bar{\rho} \bar{\sigma}} \times	\vphantom{\sum_{\lambda_{w_3} = -3}^{+3}}	\nonumber
				\\
				& \times \sum_{\lambda_{w_3} = -3}^{+3} \epsilon_{\mu \alpha \beta} ( \lambda_{w_3} , \vec{k}_{w_3} ) \, \epsilon_{\bar{\mu} \bar{\alpha} \bar{\beta}} ( \lambda_{w_3} , \vec{k}_{w_3} ) \times	\vphantom{\sum_{\lambda_{w_3} = -3}^{+3}}	\nonumber
				\\
				& \times k_{v_1 \nu} \, k_{v_1 \bar{\nu}} \sum_{\lambda_{v_1} = -1}^{+1} \epsilon_\rho ( \lambda_{v_1} , \vec{k}_{v_1} ) \, \epsilon_{\bar{\rho}} ( \lambda_{v_1} , \vec{k}_{v_1} ) \times	\vphantom{\sum_{\lambda_{w_3} = -3}^{+3}}	\nonumber
				\\
				& \times k_p^\alpha \, k_p^\beta \,k_{p \sigma} \, k_p^{\bar{\alpha}} \, k_p^{\bar{\beta}} \,k_{p \bar{\sigma}} =	\vphantom{\sum_{\lambda_{w_3} = -3}^{+3}}	\nonumber
				\\
				= \, & g_{w_3 v_1 p}^{2} \, \tfrac{8}{105} \, \big| \vec{k}_{v_1 , p} \big|^{6} \, m_{w_3}^{2} \, .	\vphantom{\sum_{\lambda_{w_3} = -3}^{+3}}	\nonumber
			\end{align}
	
		\item	For the decays of massive spin-$3$ fields into massive pseudoscalar fields and photons, the generic interaction Lagrangian is of the general type
			\begin{align}
				& \mathcal{L}_{w_3 \gamma p} =	\vphantom{\bigg(\bigg)}
				\\
				= \, & g_{w_3 v_1 p} \, \tfrac{e}{g_\rho} \, \varepsilon^{\mu\nu\rho\sigma} \, W_{3 , \mu\alpha\beta} \, ( \partial_\nu a_\rho ) \, ( \partial^{\alpha} \partial^{\beta} \partial_{\sigma} P ) \, .	\vphantom{\bigg(\bigg)}	\nonumber
			\end{align}
			The corresponding decay amplitude reads
			\begin{align}
				& \mathrm{i} \mathcal{M}_{w_3 \gamma p} =	\vphantom{\bigg(\bigg)}
				\\
				= \, & g_{w_3 v_1 p} \, \tfrac{e}{g_\rho} \, \varepsilon^{\mu \nu \rho \sigma} \, \epsilon_{\mu \alpha \beta} ( \lambda_{w_3} , \vec{k}_{w_3} ) \times	\vphantom{\bigg(\bigg)}	\nonumber
				\\
				& \times k_{\gamma , \nu} \, \epsilon_{\rho} ( \lambda_{\gamma} , \vec{k}_{\gamma} ) \, k_{p}^\alpha \, k_{p}^\beta \, k_{p , \sigma} \, .	\vphantom{\bigg(\bigg)}	\nonumber
			\end{align}
			The averaged square of the amplitude is
			\begin{align}
				& \tfrac{1}{7} \, | \mathcal{M}_{w_3 \gamma p} |^2 =	\vphantom{\Bigg(\Bigg)}	\label{eq:amplitude_square_w3_gamma_p}
				\\
				= \, & g^2_{w_3 v_1 p} \, \big( \tfrac{e}{g_\rho} \big)^2 \, \varepsilon^{\mu \nu \rho \sigma} \, \varepsilon^{\bar{\mu} \bar{\nu} \bar{\rho} \bar{\sigma}} \, k_{p}^\alpha \, k_{p}^\beta \, k_{p , \sigma} \, k_{p}^{\bar{\alpha}} \, k_{p}^{\bar{\beta}} \, k_{p , \bar{\sigma}} \times	\vphantom{\Bigg(\Bigg)}	\nonumber
				\\
				& \times \sum_{\lambda_{w_3} = -3}^{+3} \epsilon_{\mu \alpha \beta} ( \lambda_{w_3} , \vec{k}_{w_3} ) \, \epsilon_{\bar{\mu} \bar{\alpha} \bar{\beta}} ( \lambda_{w_3} , \vec{k}_{w_3} ) \times	\vphantom{\Bigg(\Bigg)}	\nonumber
				\\
				&  \times k_{\gamma , \nu} \, k_{\gamma , \bar{\nu}} \, \sum_{\lambda_{\gamma} = 1}^{2} \epsilon_{\rho} ( \lambda_{\gamma} , \vec{k}_{\gamma} ) \, \epsilon_{\bar{\rho}} ( \lambda_{\gamma} , \vec{k}_{\gamma} ) =	\vphantom{\Bigg(\Bigg)}	\nonumber
				\\
				= \, & g_{w_3 v_1 p}^{2} \, \big( \tfrac{e}{g_\rho} \big)^2 \, \tfrac{8}{105} \, \big| \vec{k}_{\gamma , p} \big|^{6} \, m_{w_3}^{2} \, .	\vphantom{\Bigg(\Bigg)}	\nonumber
			\end{align}
		
		\item	Decays of massive spin-$3$ fields into massive spin-2 tensor and massive pseudoscalar fields is described by the following interaction
			\begin{align}
				& \mathcal{L}_{w_3 a_2 p} =	\vphantom{\bigg(\bigg)}
				\\
				= \, & g_{w_3 a_2 p} \, \varepsilon_{\mu\nu\rho\sigma} \, \tensor{W}{_{3 , }^\mu_\alpha_\beta} \, ( \partial^{\nu} A_2^{\rho \alpha} ) \, ( \partial^{\sigma} \partial^{\beta} P )	\vphantom{\bigg(\bigg)}	\nonumber
			\end{align}
		which leads to the amplitude
			\begin{align}
				& \mathrm{i} \mathcal{M}_{w_3 a_2 p} =	\vphantom{\bigg(\bigg)}
				\\
				= \, & - \mathrm{i} g_{w_3 a_2 p} \, \varepsilon_{\mu \nu \rho \sigma} \, \eta_{\alpha \gamma} \, \epsilon^{\mu \alpha \beta} ( \lambda_{w_3} , \vec{k}_{w_3} ) \times	\vphantom{\bigg(\bigg)}	\nonumber
				\\
				& \times k_{a_2}^\nu \, \epsilon^{\rho \gamma} ( \lambda_{a_2} , \vec{k}_{a_2} ) \, k_p^\sigma \, k_{p \beta} \, .	\vphantom{\bigg(\bigg)}	\nonumber
			\end{align}
		Squaring the amplitude, we find,
			\begin{align}
				& \tfrac{1}{7} \,| \mathcal{M}_{w_3 a_2 p} |^{2} =	\vphantom{\Bigg(\Bigg)}	\label{eq:amplitude_square_w3_a2_p}
				\\
				= \, & g_{w_3 a_2 p}^{2} \, \tfrac{1}{7} \, \varepsilon_{\mu \nu \rho \sigma} \, \varepsilon_{\bar{\mu} \bar{\nu} \bar{\rho} \bar{\sigma}} \, \eta_{\alpha \gamma} \, \eta_{\bar{\alpha} \bar{\gamma}} \times	\vphantom{\Bigg(\Bigg)}	\nonumber
				\\
				& \times \sum_{\lambda_{w_3} = -3}^{+3} \, \epsilon^{\mu \alpha \beta} ( \lambda_{w_3} , \vec{k}_{w_3} ) \, \epsilon^{\bar{\mu} \bar{\alpha} \bar{\beta}} ( \lambda_{w_3} , \vec{k}_{w_3} ) \times \nonumber
				\\
				& \times \sum_{\lambda_{a_2} = -2}^{2} \epsilon^{\rho \gamma} ( \lambda_{a_2} , \vec{k}_{a_2} ) \, \epsilon^{\bar{\rho} \bar{\gamma}} ( \lambda_{a_2} , \vec{k}_{a_2} ) \, k_{a_2}^{\nu} \, k_{a_2}^{\bar{\nu}} \times	\vphantom{\Bigg(\Bigg)}	\nonumber
				\\
				& \times k_p^{\sigma} \, k_{p \beta} \, k_p^{\bar{\sigma}} \, k_{p \bar{\beta}} =	\vphantom{\Bigg(\Bigg)}	\nonumber
				\\
				= \, & g_{w_3 a_2 p}^{2} \, \tfrac{2}{105} \, \big| \vec{k}_{a_2 , p} \big|^4 \, \frac{m_{w_3}^2}{m_{a_2}^2} \, \big( 2 \, \big| \vec{k}_{a_2 , p} \big|^2 + 7 \, m_{a_2}^2 \big) \, .	\vphantom{\Bigg(\Bigg)}	\nonumber
			\end{align}
		
		\item	The amplitudes for the decays of massive spin-$3$ fields into massive pseudoscalar and massive axial-/pseudovector fields, which are described via the interaction Lagrangians 
			\begin{align}
				\mathcal{L}_{w_3 b_1 p} = g_{w_3 b_1 p} \, W_3^{\mu\nu\rho} \, B_{1 , \mu} \, ( \partial_{\nu}	\partial_{\rho} P )
			\end{align}
		and
			\begin{align}
				\mathcal{L}_{w_3 a_1 p} = g_{w_3 a_1 p} \, W_3^{\mu\nu\rho} \, A_{1 , \mu} \, ( \partial_{\nu} \partial_{\rho} P ) \, ,
			\end{align}
		have the same shape,
			\begin{align}
				& \mathrm{i} \mathcal{M}_{w_3 b_1 p} =	\vphantom{\bigg(\bigg)}
				\\
				= \, & - g_{w_3 b_1 p} \, \epsilon^{\mu \nu \rho} ( \lambda_{w_3} , \vec{k}_{w_3} ) \, \epsilon_{\mu} ( \lambda_{b_1} , \vec{k}_{b_1} ) \, k_{p \nu} \, k_{p \rho} \, ,	\vphantom{\bigg(\bigg)}	\nonumber
			\end{align}
			\begin{align}
				& \mathrm{i} \mathcal{M}_{w_3 a_1 p} =	\vphantom{\bigg(\bigg)}
				\\
				= \, & - g_{w_3 a_1 p} \, \epsilon^{\mu\nu\rho} ( \lambda_{w_3} , \vec{k}_{w_3} ) \, \epsilon_{\mu} ( \lambda_{a_1} , \vec{k}_{a_1} ) \, k_{p \nu} \, k_{p \rho} \, .	\vphantom{\bigg(\bigg)}	\nonumber
			\end{align}
		Their squares have the following form,
			\begin{align}
				& \tfrac{1}{7} \, | \mathcal{M}_{w_3 b_1 p} |^{2} =	\vphantom{\Bigg(\Bigg)}	\label{eq:amplitude_square_w3_b1_p}
				\\
				= \, & g_{w_3 b_1 p}^{2} \, \tfrac{1}{7} \, k_{p \nu} \, k_{p \rho} \, k_{p \bar{\nu}} \, k_{p \bar{\rho}} \times	\vphantom{\Bigg(\Bigg)}	\nonumber
				\\
				& \times \sum_{\lambda_{w_3} = -3}^{+3} \epsilon^{\mu \nu \rho} ( \lambda_{w_3} , \vec{k}_{w_3} ) \, \epsilon^{\bar{\mu} \bar{\nu} \bar{\rho}} ( \lambda_{w_3} , \vec{k}_{w_3} ) \times	\vphantom{\Bigg(\Bigg)}	\nonumber
				\\
				&  \times \sum_{\lambda_{b_1} = - 1}^{+ 1} \epsilon_{\mu} ( \lambda_{b_1} , \vec{k}_{b_1} ) \, \epsilon_{\bar{\mu}} ( \lambda_{b_1} , \vec{k}_{b_1} ) =	\vphantom{\Bigg(\Bigg)}	\nonumber
				\\
				= \, & g_{w_3 b_1 p}^{2} \, \tfrac{2}{105} \, \big| \vec{k}_{b_1 , p} \big|^4 \, \Bigg( 7 + 3 \, \frac{\big| \vec{k}_{b_1 , p} \big|^2}{m_{b_1}^2} \Bigg) \, ,	\vphantom{\Bigg(\Bigg)}	\nonumber
			\end{align}
			\begin{align}
				& \tfrac{1}{7} \, | \mathcal{M}_{w_3 a_1 p} |^{2} =	\vphantom{\bigg(\bigg)}	\label{eq:amplitude_square_w3_a1_p}
				\\
				= \, & g_{w_3 a_1 p}^{2} \, \tfrac{1}{7} \, k_{p \nu} \, k_{p \rho} \, k_{p \bar{\nu}} \, k_{p \bar{\rho}} \times	\vphantom{\Bigg(\Bigg)}	\nonumber
				\\
				& \times \sum_{\lambda_{w_3} = -3}^{+3} \epsilon^{\mu \nu \rho} ( \lambda_{w_3} , \vec{k}_{w_3} ) \, \epsilon^{\bar{\mu} \bar{\nu} \bar{\rho}} ( \lambda_{w_3} , \vec{k}_{w_3} ) \times	\vphantom{\Bigg(\Bigg)}	\nonumber
				\\
				&  \times \sum_{\lambda_{a_1} = - 1}^{+ 1} \epsilon_{\mu} ( \lambda_{a_1} , \vec{k}_{a_1} ) \, \epsilon_{\bar{\mu}} ( \lambda_{a_1} , \vec{k}_{a_1} ) =	\vphantom{\Bigg(\Bigg)}	\nonumber
				\\
				= \, & g_{w_3 a_1 p}^{2} \, \tfrac{2}{105} \, \big| \vec{k}_{a_1 , p} \big|^4 \, \Bigg( 7 + 3 \, \frac{\big| \vec{k}_{a_1 , p} \big|^2}{m_{a_1}^2} \Bigg) \, .	\vphantom{\Bigg(\Bigg)}	\nonumber
			\end{align}
		
		\item	The interaction Lagrangian which describes the decay of massive spin-$3$ fields into two massive vector fields
			\begin{align}
				\mathcal{L}_{w_3 v_1 v_1} = g_{w_3 v_1 v_1} \, W_3^{\mu\nu\rho} \, ( \partial_{\mu} V_{1 , \nu}^{(1)} ) \, V_{1 , \rho}^{(2)}
			\end{align}
		leads to the following amplitude
			\begin{align}
				& \mathrm{i} \mathcal{M}_{w_3 v_1 v_1} =	\vphantom{\bigg(\bigg)}
				\\
				= \, & \mathrm{i} g_{w_3 v_1 v_1} \, \epsilon^{\mu\nu\rho} ( \lambda_{w_3} , \vec{k}_{w_3} ) \, k_{v_1 \mu}^{(1)} \, \epsilon_{\nu} ( \lambda_{v_1}^{(1)} , \vec{k}_{v_1}^{(1)} ) \times	\vphantom{\bigg(\bigg)}	\nonumber
				\\
				& \times \epsilon_{\rho} ( \lambda_{v_1}^{(2)} , k_{v_2}^{(2)} ) \, .	\vphantom{\bigg(\bigg)}	\nonumber
			\end{align}
		The square of this amplitude is given by
			\begin{align}
				& \tfrac{1}{7} \, | \mathcal{M}_{w_3 v_1 v_1} |^{2} =	\vphantom{\Bigg(\Bigg)}	\label{eq:amplitude_square_w3_v1_v1}
				\\
				= \, & g_{w_3 v_1 v_1}^{2} \, \tfrac{1}{7} \sum_{\lambda_{w_3} = -3}^{+3} \, \epsilon^{\mu \nu \rho} ( \lambda_{w_3} , \vec{k}_{w_3} ) \, \epsilon^{\bar{\mu} \bar{\nu} \bar{\rho}} ( \lambda_{w_3} , \vec{k}_{w_3} ) =	\vphantom{\Bigg(\Bigg)}	\nonumber
				\\
				& \times k_{v_1 \mu}^{(1)} \, k_{v_1 \bar{\mu}}^{(1)} \sum_{\lambda_{v_1}^{(1)} = - 1}^{+ 1} \epsilon_{\nu} ( \lambda_{v_1}^{(1)} , \vec{k}_{v_1}^{(1)} ) \, \epsilon_{\bar{\nu}} ( \lambda_{v_1}^{(1)} , \vec{k}_{v_1}^{(1)} ) \times	\vphantom{\Bigg(\Bigg)}	\nonumber
				\\
				& \times \sum_{\lambda_{v_1}^{(2)} = - 1}^{+ 1} \epsilon_{\rho} ( \lambda_{v_1}^{(2)} , \vec{k}_{v_1}^{(2)} ) \, \epsilon_{\bar{\rho}} ( \lambda_{v_1}^{(2)} , \vec{k}_{v_1}^{(2)} ) =	\vphantom{\Bigg(\Bigg)}	\nonumber
				\\
				= \, & g_{w_3 v_1 v_1}^{2} \, \tfrac{1}{105} \, \big( m_{v_1^{(1)}}^2 \, m_{v_1^{(2)}}^2 \big)^{-1} \, \big| \vec{k}_{v_1^{(1)} , v_1^{(2)}} \big|^2 \times	\vphantom{\Bigg(\Bigg)}	\nonumber
				\\
				& \times \big[ 6 \, \big| \vec{k}_{v_1^{(1)} , v_1^{(2)}} \big|^4 + 35 \, m_{v_1^{(1)}}^2 \,  m_{v_1^{(2)}}^2 +	\vphantom{\Bigg(\Bigg)}	\nonumber
				\\
				& \quad + 14 \, \big| \vec{k}_{v_1^{(1)} , v_1^{(2)}} \big|^2 \, \big( m_{v_1^{(1)}}^2 + m_{v_1^{(2)}}^2 \big) \big] \, .	\vphantom{\Bigg(\Bigg)}	\nonumber
			\end{align}		
			
		\item	For the decays of massive spin-$3$ fields into massive vector and axial-vector fields, we use the following Lagrangian
			\begin{align}
				& \mathcal{L}_{w_3 v_1 a_1} =	\vphantom{\bigg(\bigg)}	\label{eq:amplitude_square_w3_v1_a1}
				\\
				= \, & g_{w_3 v_1 a_1} \, \varepsilon^{\mu \nu \rho \sigma} \, W_{3 , \mu \alpha \beta} \, ( \partial_\nu V_{1 , \rho} ) \, ( \partial^\alpha \partial^\beta A_{1 , \sigma} ) \, ,	\vphantom{\bigg(\bigg)}	\nonumber
			\end{align}
		to obtain the amplitude
			\begin{align}
				&	\mathrm{i} \mathcal{M}_{w_3 v_1 a_1} =	\vphantom{\bigg(\bigg)}
				\\
				= \, & - \mathrm{i} g_{w_3 v_1 a_1} \, \varepsilon^{\mu \nu \rho \sigma} \, \epsilon_{\mu \alpha \beta} ( \lambda_{w_3} , \vec{k}_{w_3} ) \, k_{v_1 \nu} \, \epsilon_{\rho} ( \lambda_{v_1} , \vec{k}_{v_1} ) \times	\vphantom{\bigg(\bigg)}	\nonumber
				\\
				& \times k_{a_1}^\alpha \, k_{a_1}^\beta \, \epsilon_{\sigma} ( \lambda_{a_1} , \vec{k}_{a_1} ) \, .	\vphantom{\bigg(\bigg)}	\nonumber
			\end{align}
		The square of the amplitude is
			\begin{align}
				& \tfrac{1}{7} \, | \mathcal{M}_{w_3 v_1 a_1} |^2 =	\vphantom{\Bigg(\Bigg)}
				\\
				= \, & g_{w_3 v_1 a_1}^2 \, \tfrac{1}{7} \, \varepsilon^{\mu \nu \rho \sigma} \, \varepsilon^{\bar{\mu} \bar{\nu} \bar{\rho} \bar{\sigma}} \times	\vphantom{\Bigg(\Bigg)}	\nonumber
				\\
				& \times \sum_{\lambda_{w_3} = -3}^{+3} \, \epsilon_{\mu \alpha \beta} ( \lambda_{w_3} , \vec{k}_{w_3} ) \, \epsilon_{\bar{\mu} \bar{\alpha} \bar{\beta}} ( \lambda_{w_3} , \vec{k}_{w_3} ) \times	\vphantom{\Bigg(\Bigg)}	\nonumber
				\\
				& \times k_{v_1 \nu} \,  k_{v_1 \bar{\nu}} \,  \sum_{\lambda_{v_1} = - 1}^{+ 1} \epsilon_{\rho} ( \lambda_{v_1} , \vec{k}_{v_1} ) \, \epsilon_{\bar{\rho}} ( \lambda_{v_1} , \vec{k}_{v_1} )	\vphantom{\Bigg(\Bigg)}	\nonumber
				\\
				& \times k_{a_1}^\alpha \, k_{a_1}^\beta \, k_{a_1}^{\bar{\alpha}} \, k_{a_1}^{\bar{\beta}} \, \sum_{\lambda_{a_1} = - 1}^{+ 1} \epsilon_{\sigma} ( \lambda_{a_1} , \vec{k}_{a_1} ) \, \epsilon_{\bar{\sigma}} ( \lambda_{a_1} , \vec{k}_{a_1} ) =	\vphantom{\Bigg(\Bigg)}	\nonumber
				\\
				= \, & g_{w_3 v_1 a_1}^2 \, \tfrac{4}{105} \, \big| \vec{k}_{v_1 , a_1} \big|^4 \, \bigg[ \big| \vec{k}_{v_1 , a_1} \big|^2 \, \bigg( 3 + 2 \, \frac{m_{w_2}^2}{m_{a_1}^2} \bigg) + 7 \, m_{v_1}^2 \bigg] \, .	\vphantom{\Bigg(\Bigg)}	\nonumber
			\end{align}
	\end{enumerate}

	We remark that the form of the amplitudes $\mathcal{M}$ of most of the decay channels was already derived in Ref.\ \cite{Wang:2016enc}. In contrast to Ref.\ \cite{Wang:2016enc} we neglect higher order derivative couplings and purely rely on the lowest order contribution in derivatives of the fields. Furthermore, we do not include form factors.\\

	In order to obtain the results of this appendix one has to average over all incoming spin polarizations, sum up all possible outgoing polarizations and consider the following completeness relations for massive particles, which are discussed in the subsequent appendix,
		\begin{align}
			& \sum_{\lambda = -1}^{+1} \epsilon_{\mu} ( \lambda , \vec{k} ) \, \epsilon_{\nu} ( \lambda , \vec{k} ) = - G_{\mu\nu} \, ,	\vphantom{\sum_{\lambda = -3}^{+3}}	\label{eq:completenessvector}
		\end{align}
		\begin{align}
			& \sum_{\lambda = -2}^{+2} \epsilon_{\mu\nu} ( \lambda , \vec{k} ) \, \epsilon_{\alpha\beta} ( \lambda , \vec{k} ) =	\vphantom{\sum_{\lambda = -3}^{+3}}	\label{eq:completenesstensor}
			\\
			= \, & - \tfrac{1}{3} \, G_{\mu\nu} \, G_{\alpha\beta} + \tfrac{1}{2} \, ( G_{\mu\alpha} \, G_{\nu\beta} + G_{\mu\beta} \, G_{\nu\alpha} ) \, ,	\vphantom{\sum_{\lambda = -3}^{+3}}	\nonumber
		\end{align}
		\begin{align}
			& \sum_{\lambda = -3}^{+3} \epsilon_{\mu\nu\rho} ( \lambda , \vec{k} ) \, \epsilon_{\alpha\beta\gamma} ( \lambda , \vec{k} ) =	\vphantom{\sum_{\lambda = -3}^{+3}}	\label{eq:completeness3tensor}
			\\
			= \, & \tfrac{1}{15} \, \big[ G_{\mu\nu} \, ( G_{\rho\alpha} \, G_{\beta\gamma} + G_{\rho\beta} \, G_{\alpha\gamma} + G_{\rho\gamma} \, G_{\alpha\beta} ) +	\vphantom{\sum_{\lambda = -3}^{+3}}	\nonumber
			\\
			& + G_{\mu\rho} \, (  G_{\nu\alpha} \, G_{\beta\gamma} + G_{\nu\beta} \, G_{\alpha\gamma} + G_{\nu\gamma} \, G_{\alpha\beta} ) +	\vphantom{\sum_{\lambda = -3}^{+3}}	\nonumber
			\\
			& + G_{\nu\rho} \, ( G_{\mu\alpha}\, G_{\beta\gamma} + G_{\mu\beta} \, G_{\alpha\gamma} + G_{\mu\gamma} \, G_{\alpha\beta} ) \big] -	\vphantom{\sum_{\lambda = -3}^{+3}}	\nonumber
			\\
			& - \tfrac{1}{6} \, \big[ G_{\mu\alpha} \, ( G_{\nu\beta} \, G_{\rho\gamma} + G_{\nu\gamma} \, G_{\rho\beta} ) +	\vphantom{\sum_{\lambda = -3}^{+3}}	\nonumber
			\\
			& + G_{\mu\beta} \, ( G_{\nu\alpha} \, G_{\rho\gamma} + G_{\nu\gamma} \, G_{\rho\alpha} ) +	\vphantom{\sum_{\lambda = -3}^{+3}}	\nonumber
			\\
			& + G_{\mu\gamma} \, ( G_{\nu\alpha} \, G_{\rho\beta} + G_{\nu\beta} \, G_{\rho\alpha} ) \big] \, ,	\vphantom{\sum_{\lambda = -3}^{+3}}	\nonumber
		\end{align}
	where
		\begin{align}
			&	G_{\mu\nu} \equiv \eta_{\mu\nu} - \frac{k_{\mu} k_{\nu}}{k^{2}} \, ,	&&	k^2 = m^2 \, ,
		\end{align}
	and $(\eta_{\mu\nu}) = \mathrm{diag}( +1 , -1 , -1 , -1 )$.
	
	For the square of the amplitudes of the radioactive decays in Eq.\ \eqref{eq:amplitude_square_w3_gamma_p} the following completeness relation for photons is used \cite{Greiner:1996zu},
		\begin{align}
			& \sum_{\lambda = 1}^{2} \epsilon_{\mu} ( \lambda, \vec{k} ) \, \epsilon_{\nu} ( \lambda, \vec{k} ) =	\vphantom{\sum_{\lambda = 1}^{2}}
			\\
			= \, & - \eta_{\mu \nu} - \frac{k_\mu k_\nu}{( k \cdot n )^2} + \frac{k_\mu n_\nu + k_\nu n_\mu}{( k \cdot n )} \, .	\vphantom{\sum_{\lambda = 1}^{2}}	\nonumber
		\end{align}
	Here, $n = ( 1, 0, 0, 0 )^T$ is the unit vector in temporal direction and $k$ is the four momentum of the outgoing photon.\\
	
	The summations over Minkwoski spacetime indices in the squares of the amplitudes were performed on a symbolic level with \textit{Wolfram Mathematica 12.1} \cite{Mathematica} using \textit{Sum} and \textit{ParallelSum}. We work in the center of mass frame of the decaying particle. In order to reduce the number of addends in the sums drastically, \textit{w.l.o.g.}\ we choose the three momenta of the outgoing particles in $z$ direction.

\section{Completeness relations}
\label{app:polarization}

	In this appendix, we present how the completeness relations for the polarization vectors and tensors of massive higher-spin fields, like Eqs.\ \eqref{eq:completenessvector}, \eqref{eq:completenesstensor} and \eqref{eq:completeness3tensor}, are constructed. These completeness relations are used for the calculation of the unpolarized decay amplitudes in the previous App.\ \ref{app:deacyamplitudes}.
	
	We begin by deriving the completeness relation for massive spin-1 fields explicitly. This derivation is based on the discussion in Ref.\ \cite{Greiner:1996zu} and presented here for reasons of consistency. Starting from this result for spin-1 fields, we generalize the procedure and show, how the completeness relations for massive higher-spin fields can be constructed. We then provide an explicit derivation of the completeness relations for massive spin-$3$ tensor fields. We close the discussion, by elaborating on the degrees of freedom of massive fields of arbitrary integer spin-$J$ and their description via rank-$J$ tensor fields.\\
	
	None of the results in this appendix is original to our work. However, we think that the discussion might still be useful for a nonexpert reader.

\subsection{Spin-1 fields}

	The construction of the completeness relation for the polarization vectors $\epsilon^{\mu} ( \lambda , \vec{k} )$ of the massive (real) vector fields $V^{\mu} (x)$ is based on the field equations that describe free massive vector particles.\footnote{For free massive axial- and pseudovector fields the 	construction is identical.} These field equations are given by the Proca equations, which are the Klein-Gordon (KG) equation and a transversality condition (which is explicitly not a gauge),
		\begin{align}
			&	0 = ( \Box + m^{2} ) \, V^{\mu} (x) \, ,	&&	0 = \partial_{\mu} V^{\mu} (x) \, .	\label{eq:proca_equations}
		\end{align}
	For neutral massive vector mesons, the solutions to the KG equation are plane waves,
\begin{widetext}
		\begin{align}
			V^{\mu} (x) = \int \frac{\mathrm{d}^{3} k}{(2 \uppi)^{3}} \frac{1}{2 E(\vec{k})} \, \sum_{\lambda = -1}^{+1} \epsilon^{\mu} ( \lambda , \vec{k} ) \, \Big\{ a ( \lambda , \vec{k} ) \, \mathrm{e}^{+ \mathrm{i} [ E (\vec{k}) \, t - \vec{k} \cdot \vec{x} ]} + a^{\ast} ( \lambda , \vec{k} ) \, \mathrm{e}^{- \mathrm{i} [ E (\vec{k}) \, t + \vec{k} \cdot \vec{x} ]} \Big\} \, ,
		\end{align}
\end{widetext}
	where $a ( \lambda , \vec{k} ) = a^{\ast} ( \lambda , - \vec{k} )$ are the Fourier amplitudes and $E^2 (\vec{k}) = \vec{k}^{\,2} + m^{2}$ is the square of the energy. The transversality condition further implies that the polarization vectors $\epsilon ( \lambda , \vec{k} )$ have to be transversal to the four momentum $k = ( E (\vec{k}) , \vec{k} )^T$ of the the fields,
		\begin{align}
			0 = k_{\mu} \, \epsilon^{\mu} ( \lambda , \vec{k} ) \, .	\label{eq:transversaltiy}
		\end{align}
	From this equation, we can construct the completeness relation for the polarization vectors.\newline We start with the most general ansatz for a dimensionless rank-two tensor that only depends on the four momentum $k = ( E (\vec{k}) , \vec{k})^T$,
		\begin{align}
			\sum_{\lambda = -1}^{+1} \epsilon_{\mu} ( \lambda , \vec{k} ) \, \epsilon_{\nu} ( \lambda , \vec{k} ) = a \, \eta_{\mu\nu} + b \, \frac{k_{\mu} \, k_{\nu}}{k^{2}} \, ,
		\end{align}
	where $a$ and $b$ are dimensionless constants. The transversality condition \eqref{eq:transversaltiy} implies
		\begin{align}
			0 \overset{!}{=} k^{\mu} \sum_{\lambda = -1}^{+1} \epsilon_{\mu} ( \lambda , \vec{k} ) \, \epsilon_{\nu} ( \lambda , \vec{k} ) = k_{\nu} \, ( a + b ) \, ,
		\end{align}
	which simplifies our ansatz to
		\begin{align}
			\sum_{\lambda = -1}^{+1} \epsilon_{\mu} ( \lambda , \vec{k} ) \, \epsilon_{\nu} ( \lambda  , \vec{k} ) = a \, \bigg( \eta_{\mu\nu} - \frac{k_{\mu} \, k_{\nu}}{k^{2}} \bigg) \, .
		\end{align}
	The coefficient $a$ can be fixed by choosing an appropriate normalization condition. This is done by demanding that in the rest frame $k = ( m , \vec{0} )^T$ for the spatial components the completeness relation should reduce to the Kronecker delta, to have a three-dimensional orthonormal basis of polarization vectors.
		\begin{align}
			\delta_{ij} \overset{!}{=} \sum_{\lambda = -1}^{+1} \epsilon_i ( \lambda , \vec{0} ) \, \epsilon_j( \lambda , \vec{0} ) = - a \, \delta_{ij} \, .	\label{eq:restframe_orthonormal}
		\end{align}
	We obtain the completeness relation
		\begin{align}
			\sum_{\lambda = -1}^{+1} \epsilon_{\mu} ( \lambda , \vec{k} ) \, \epsilon_{\nu} ( \lambda , \vec{k} ) = \, & - \bigg( \eta_{\mu\nu} - \frac{k_{\mu} \, k_{\nu}}{k^{2}} \bigg) \equiv
			\\
			\equiv \, &  - G_{\mu\nu} \, ,	\vphantom{\sum_{\lambda = -1}^{+1}}	\nonumber
		\end{align}
	and define the projector $G_{\mu\nu}$, which projects on the subspace of the Minkowski space that is transversal to the four momentum vector $k = ( E (\vec{k}) , \vec{k} )^T$ of the massive spin-1 field.

\subsection{Higher-spin fields}

	For all kind of higher-spin tensor fields, we proceed analogously. The starting point are the so called Fierz-Pauli equations for a massive field with integer spin $J$. The field is described by a rank-$J$ tensor field that fulfills the following equations (see also Refs.\ \cite{Koenigstein:2015asa,Greiner:1996zu,Fierz:1939i,Fierz:1939ix,Huang:2002ex,Bouatta:2004kk,Barut:1986dd} for more details on wave equations and degrees of freedom of higher-spin fields),
		\begin{align}
			0 = \, & ( \Box + m^2 ) \, \varphi^{\alpha_1 \cdots \alpha_{J} } (x) \, ,	\vphantom{\bigg(\bigg)}	\label{eq:fp_1}
		\end{align}
	and $\forall \, 1 \leq m \neq n \leq J$,
		\begin{align}
			0 = \, & \varphi^{\alpha_1 \cdots \alpha_{m} \cdots \alpha_{n} \cdots \alpha_{J}} (x) - \varphi^{\alpha_1 \cdots \alpha_{n} \cdots \alpha_{m} \cdots \alpha_{J}} (x) \, ,	\vphantom{\bigg(\bigg)}	\label{eq:fp_2}
			\\
			0 = \, & \eta_{\alpha_{m} \alpha_{n}} \, \varphi^{\alpha_1 \cdots \alpha_{m} \cdots \alpha_{n} \cdots \alpha_{J}} (x) \, ,	\label{eq:fp_3}	\vphantom{\bigg(\bigg)}
		\end{align}
	and $\forall \, 1 \leq m \leq J$,
		\begin{align}
			0 = \, & \partial_{\alpha_{m}} \varphi^{\alpha_1 \cdots \alpha_{m} \cdots \alpha_{J}} (x) \, .	\vphantom{\bigg(\bigg)}	\label{eq:fp_4}
		\end{align}
	These equations reduce the $4^{J}$ degrees of freedom of a general rank-$J$ tensor field in 4-dimensional Minkowski space to the $(2J+1)$ degrees of freedom that are necessary to describe a massive spin-$J$ field. (We will come back to this point at the end of this appendix.) Note that the Proca equations \eqref{eq:proca_equations} for spin-1 fields are merely a special case of the Fierz-Pauli equations \eqref{eq:fp_1} - \eqref{eq:fp_4}.
	
	Completely analogous to the vector field, we can thus find a solution to the KG equation and derive the following constraints for the corresponding polarization tensors, $\forall\,1\leq m\neq n\leq J \, ,$
		\begin{align}
			0 = \, &  \epsilon^{\alpha_1 \cdots \alpha_{m} \cdots \alpha_{n} \cdots \alpha_{J}} ( \lambda , \vec{k} ) -	\vphantom{\bigg(\bigg)}	\label{eq:symmetry_gen}
			\\
			& - \epsilon^{\alpha_1 \cdots \alpha_{n} \cdots \alpha_{m} \cdots \alpha_{J}} ( \lambda , \vec{k} ) \, ,	\vphantom{\bigg(\bigg)}	\nonumber
			\\
			0 = \, & \eta_{\alpha_m \alpha_n} \, \epsilon^{\alpha_1 \cdots \alpha_{m} \cdots \alpha_{n} \cdots \alpha_{J}} ( \lambda , \vec{k} ) \, ,	\vphantom{\bigg(\bigg)}	\label{eq:trace_gen}
		\end{align}
	$\forall\,1\leq m\leq J \, ,$
		\begin{align}
			0 = \, & k_{\alpha_{m}} \epsilon^{\alpha_1 \cdots \alpha_{m} \cdots \alpha_{J}} ( \lambda , \vec{k} ) \, , \label{eq:lorentz_gen}
		\end{align}
	Next, to find the corresponding completeness relation, one has to start with an appropriate ansatz for
		\begin{align}
			\sum_{\lambda = - J}^{+ J} \epsilon_{\alpha_1 \cdots \alpha_{J}} ( \lambda , \vec{k} ) \, \epsilon_{\beta_1 \cdots \beta_{J}} ( \lambda , \vec{k} ) = \ldots \, .
		\end{align}
	Consequently, one should write down the most general rank-$2J$ tensor, which only depends on the four momentum $k = ( E ( \vec{k} ) , \vec{k} )^T$ and the metric tensor. This means that all possible combinations of $\eta_{\mu\nu}$ and $\frac{k_{\mu} k_{\nu}}{k^{2}}$, with all possible permutations of the indices $\alpha_1 , \ldots , \alpha_{J}$, $\beta_1 , \ldots , \beta_{J}$ of the polarization tensors have to be included,
		\begin{align}
			& \sum_{\lambda = - J}^{J} \epsilon_{\alpha_1 \cdots \alpha_{J}} ( \lambda , \vec{k} ) \, \epsilon_{\beta_1 \cdots \beta_{J}} ( \lambda , \vec{k} ) =
			\\
			= \, & + a_1 \,\eta_{\alpha_1 \alpha_{2}} \cdots \eta_{\beta_{J-1}\beta_{J}} + \text{all permutations} +	\vphantom{\sum_{\lambda = 1}^{2 J + 1}}	\nonumber
			\\
			& + b_1 \, \eta_{\alpha_1\alpha_{2}} \cdots \frac{k_{\beta_{J-1}} k_{\beta_{J}}}{k^{2}} + \text{all permutations} +	\vphantom{\sum_{\lambda = 1}^{2 J + 1}}	\nonumber
			\\
			& + \ldots +	\vphantom{\sum_{\lambda = 1}^{2 J + 1}}	\nonumber
			\\
			& + z_1 \, \frac{k_{\alpha_1} k_{\alpha_{2}}}{k^{2}} \cdots \frac{k_{\beta_{J-1}} k_{\beta_{J}}}{k^{2}} + \text{all permutations} \, .	\vphantom{\sum_{\lambda = 1}^{2 J + 1}}	\nonumber
		\end{align}
	Here, $a_i$, $b_i$, $\ldots$, $z_i$ stand for the various dimensionless coefficients of the ansatz. An alternative approach is to write down the ansatz for the rank-$2J$ tensor on the r.h.s.\ of the above equation in terms of projection operators
		\begin{align}
			&	G_{\mu\nu} = \eta_{\mu\nu} - \frac{k_{\mu} k_{\nu}}{k^{2}} \, ,	&&	P_{\mu\nu} = \frac{k_{\mu} k_{\nu}}{k^{2}} \, ,	\label{eq:projection_operators}
		\end{align}
	which also form a complete basis.\\
	
	What follows is identical for both approaches:\newline To determine the various coefficients, one has to use Eqs.\ \eqref{eq:symmetry_gen}, \eqref{eq:trace_gen} and \eqref{eq:lorentz_gen} (or combinations of these constraints) successively until one reaches the point, where only one overall coefficient/factor is left, see for example Ref.\ \cite{Koenigstein:2015asa} for massive spin-2 fields. The leftover constant can only be fixed by appropriate normalization. For most problems, a normalization, which produces an orthonormal basis of polarization tensors in the rest frame, like for the vectors in Eq.\ \eqref{eq:restframe_orthonormal}, is suitable.
	
	Following these procedures, one eventually arrives at the completeness relations \eqref{eq:completenesstensor} and \eqref{eq:completeness3tensor} for the polarization tensors of spin-2 and spin-$3$ fields.
	
	We remark, that an expression for arbitrary spin fields was already presented in Refs.\ \cite{Harnad:1972fe,Durand:1967zza}. The expressions for spin-2 and spin-$3$ were already presented among others in Refs.\ \cite{Wang:2016enc,Wang:2016hbl,Bergstrom:1990kf,Koenigstein:2015asa,Zee:2003mt}.

\subsection{Spin-3 fields revisited}

	Here, we explicitly present the derivation of the completeness relation for massive spin-$3$ fields.
	
	We start by adopting the constraints \eqref{eq:symmetry_gen}, \eqref{eq:trace_gen} and \eqref{eq:lorentz_gen} to the polarization tensors of rank-3 tensor fields. Thus, the polarization tensors of spin-$3$ fields have to fulfill the following conditions,
		\begin{align}
			0 = \, & \epsilon^{(\mu\nu\rho)} ( \lambda , \vec{k} ) \, ,	\vphantom{\bigg(\bigg)}	\label{eq:symmetry}
			\\
			0 = \, & \tensor{\epsilon}{_\mu^\mu^\nu} ( \lambda , \vec{k} ) \, ,	\vphantom{\bigg(\bigg)}	\label{eq:trace}
			\\
			0 = \, & k_{\mu} \epsilon^{\mu\nu\rho} ( \lambda , \vec{k} ) \, .	\vphantom{\bigg(\bigg)}	\label{eq:lorentz}
		\end{align}
	Next we construct the completeness relation for these polarization tensors. We use the definition
		\begin{align}
			G_{\mu\nu} \equiv \eta_{\mu\nu} - \frac{k_{\mu} k_{\nu}}{k^{2}}
		\end{align}
	and write down the most general ansatz for the completeness relation, which contains all possible combinations of the metric and the four momentum or the projection operators \eqref{eq:projection_operators} respectively,
		\begin{align}
			& \sum_{\lambda = -3}^{+3} \epsilon_{\mu\nu\rho} ( \lambda , \vec{k} ) \, \epsilon_{\alpha\beta\gamma} ( \lambda , \vec{k} ) =	\vphantom{\Bigg(\Bigg)}
			\\
			= \, & + a_1 \, G_{\mu\nu} G_{\rho\alpha} G_{\beta\gamma} + a_2 \, G_{\mu\nu} G_{\rho\beta} G_{\alpha\gamma} +	\vphantom{\Bigg(\Bigg)}	\nonumber
			\\
			& + a_3 \, G_{\mu\nu} G_{\rho\gamma} G_{\alpha\beta} + a_4 \, G_{\mu\rho} G_{\nu\alpha} G_{\beta\gamma} +	\vphantom{\Bigg(\Bigg)}	\nonumber
			\\
			& + a_5 \, G_{\mu\rho} G_{\nu\beta} G_{\alpha\gamma} + a_6 \, G_{\mu\rho} G_{\nu\gamma} G_{\alpha\beta} +	\vphantom{\Bigg(\Bigg)}	\nonumber
			\\
			& + a_7 \, G_{\mu\alpha} G_{\nu\rho} G_{\beta\gamma} + a_8 \, G_{\mu\alpha} G_{\nu\beta} G_{\rho\gamma} +	\vphantom{\Bigg(\Bigg)}	\nonumber
			\\
			& + a_9 \, G_{\mu\alpha} G_{\nu\gamma} G_{\rho\beta} + a_{10} \, G_{\mu\beta} G_{\nu\rho} G_{\alpha\gamma} +	\vphantom{\Bigg(\Bigg)}	\nonumber
			\\
			& + a_{11} \, G_{\mu\beta} G_{\nu\alpha} G_{\rho\gamma} + a_{12} \, G_{\mu\beta} G_{\nu\gamma} G_{\rho\alpha} +	\vphantom{\Bigg(\Bigg)}	\nonumber
			\\
			& + a_{13} \, G_{\mu\gamma} G_{\nu\rho} G_{\alpha\beta} + a_{14} \, G_{\mu\gamma} G_{\nu\alpha} G_{\rho\beta} +	\vphantom{\Bigg(\Bigg)}	\nonumber
			\\
			& + a_{15} \, G_{\mu\gamma} G_{\nu\beta} G_{\rho\alpha} +	\vphantom{\Bigg(\Bigg)}	\nonumber
			\\
			& + b_1 \, G_{\mu\nu} G_{\rho\alpha} k_{\beta} k_{\gamma} + b_2 \, G_{\mu\nu} G_{\rho\beta} k_{\alpha} k_{\gamma} +	\vphantom{\Bigg(\Bigg)}	\nonumber
			\\
			& + b_3 \, G_{\mu\nu} G_{\rho\gamma} k_{\alpha} k_{\beta} + b_4 \, G_{\mu\nu} G_{\alpha\beta} k_{\rho} k_{\gamma} +	\vphantom{\Bigg(\Bigg)}	\nonumber
			\\
			& + b_5 \, G_{\mu\nu} G_{\alpha\gamma} k_{\rho} k_{\beta} + b_6 \, G_{\mu\nu} G_{\beta\gamma} k_{\rho} k_{\alpha} +	\vphantom{\Bigg(\Bigg)}	\nonumber
			\\
			& + b_7 \, G_{\mu\rho} G_{\nu\alpha} k_{\beta} k_{\gamma} + b_8 \, G_{\mu\rho} G_{\nu\beta} k_{\alpha} k_{\gamma} +	\vphantom{\Bigg(\Bigg)}	\nonumber
			\\
			& + b_9 \, G_{\mu\rho} G_{\nu\gamma} k_{\alpha} k_{\beta} + b_{10} \, G_{\mu\rho} G_{\alpha\beta} k_{\nu} k_{\gamma} +	\vphantom{\Bigg(\Bigg)}	\nonumber
			\\
			& + b_{11} \, G_{\mu\rho} G_{\alpha\gamma} k_{\nu} k_{\beta} + b_{12} \, G_{\mu\rho} G_{\beta\gamma} k_{\nu} k_{\alpha} +	\vphantom{\Bigg(\Bigg)}	\nonumber
			\\
			& + b_{13} \, G_{\mu\alpha} G_{\nu\rho} k_{\beta} k_{\gamma} + b_{14} \, G_{\mu\alpha} G_{\nu\beta} k_{\rho} k_{\gamma} +	\vphantom{\Bigg(\Bigg)}	\nonumber
			\\
			& + b_{15} \, G_{\mu\alpha} G_{\nu\gamma} k_{\rho} k_{\beta} + b_{16} \, G_{\mu\alpha} G_{\rho\beta} k_{\nu} k_{\gamma} +	\vphantom{\Bigg(\Bigg)}	\nonumber
			\\
			& + b_{17} \, G_{\mu\alpha} G_{\rho\gamma} k_{\nu} k_{\beta} + b_{18} \, G_{\mu\alpha} G_{\beta\gamma} k_{\nu} k_{\rho} +	\vphantom{\Bigg(\Bigg)}	\nonumber
			\\
			& + b_{19} \, G_{\mu\beta} G_{\nu\rho} k_{\alpha} k_{\gamma} + b_{20} \, G_{\mu\beta} G_{\nu\alpha} k_{\rho} k_{\gamma} +	\vphantom{\Bigg(\Bigg)}	\nonumber
			\\
			& + b_{21} \, G_{\mu\beta} G_{\nu\gamma} k_{\rho} k_{\alpha} + b_{22} \, G_{\mu\beta} G_{\rho\alpha} k_{\nu} k_{\gamma} +	\vphantom{\Bigg(\Bigg)}	\nonumber
			\\
			& + b_{23} \, G_{\mu\beta} G_{\rho\gamma} k_{\nu} k_{\alpha} + b_{24} \, G_{\mu\beta} G_{\alpha\gamma} k_{\nu} k_{\rho} +	\vphantom{\Bigg(\Bigg)}	\nonumber
			\\
			& + b_{25} \, G_{\mu\gamma} G_{\nu\rho} k_{\alpha} k_{\beta} + b_{26} \, G_{\mu\gamma} G_{\nu\alpha} k_{\rho} k_{\beta} +	\vphantom{\Bigg(\Bigg)}	\nonumber
			\\
			& + b_{27} \, G_{\mu\gamma} G_{\nu\beta} k_{\rho} k_{\alpha} + b_{28} \, G_{\mu\gamma} G_{\rho\alpha} k_{\nu} k_{\beta} +	\vphantom{\Bigg(\Bigg)}	\nonumber
			\\
			& + b_{29} \, G_{\mu\gamma} G_{\rho\beta} k_{\nu} k_{\alpha} + b_{30} \, G_{\mu\gamma} G_{\alpha\beta} k_{\nu} k_{\rho} +	\vphantom{\Bigg(\Bigg)}	\nonumber
			\\
			& + b_{31} \, k_{\mu} k_{\nu} G_{\rho\alpha} G_{\beta\gamma} + b_{32} \, k_{\mu} k_{\nu} G_{\rho\beta} G_{\alpha\gamma} +	\vphantom{\Bigg(\Bigg)}	\nonumber
			\\
			& + b_{33} \, k_{\mu} k_{\nu} G_{\rho\gamma} G_{\alpha\beta} + b_{34} \, k_{\mu} k_{\rho} G_{\nu\alpha} G_{\beta\gamma} +	\vphantom{\Bigg(\Bigg)}	\nonumber
			\\
			& + b_{35} \, k_{\mu} k_{\rho} G_{\nu\beta} G_{\alpha\gamma} + b_{36} \, k_{\mu} k_{\rho} G_{\nu\gamma} G_{\alpha\beta} +	\vphantom{\Bigg(\Bigg)}	\nonumber
			\\
			& + b_{37} \, k_{\mu} k_{\alpha} G_{\nu\rho} G_{\beta\gamma} + b_{38} \, k_{\mu} k_{\alpha} G_{\nu\beta} G_{\rho\gamma} +	\vphantom{\Bigg(\Bigg)}	\nonumber
			\\
			& + b_{39} \, k_{\mu} k_{\alpha} G_{\nu\gamma} G_{\rho\beta} + b_{40} \, k_{\mu} k_{\beta} G_{\nu\rho} G_{\alpha\gamma} +	\vphantom{\Bigg(\Bigg)}	\nonumber
			\\
			& + b_{41} \, k_{\mu} k_{\beta} G_{\nu\alpha} G_{\rho\gamma} + b_{42} \, k_{\mu} k_{\beta} G_{\nu\gamma} G_{\rho\alpha} +	\vphantom{\Bigg(\Bigg)}	\nonumber
			\\
			& + b_{43} \, k_{\mu} k_{\gamma} G_{\nu\rho} G_{\alpha\beta} + b_{44} \, k_{\mu} k_{\gamma} G_{\nu\alpha} G_{\rho\beta} +	\vphantom{\Bigg(\Bigg)}	\nonumber
			\\
			& + b_{45} \, k_{\mu} k_{\gamma} G_{\nu\beta} G_{\rho\alpha} +	\vphantom{\Bigg(\Bigg)}	\nonumber
			\\
			& + c_{1} \, G_{\mu\nu} k_{\rho} k_{\alpha} k_{\beta} k_{\gamma} + c_{2} \, G_{\mu\rho} k_{\nu} k_{\alpha} k_{\beta} k_{\gamma} +	\vphantom{\Bigg(\Bigg)}	\nonumber
			\\
			& + c_{3} \, G_{\mu\alpha} k_{\nu} k_{\rho} k_{\beta} k_{\gamma} + c_{4} \, G_{\mu\beta} k_{\nu} k_{\rho} k_{\alpha} k_{\gamma} +	\vphantom{\Bigg(\Bigg)}	\nonumber
			\\
			& + c_{5} \, G_{\mu\gamma} k_{\nu} k_{\rho} k_{\alpha} k_{\beta} + c_{6} \, G_{\nu\rho} k_{\mu} k_{\alpha} k_{\beta} k_{\gamma} +	\vphantom{\Bigg(\Bigg)}	\nonumber
			\\
			& + c_{7} \, G_{\nu\alpha} k_{\mu} k_{\rho} k_{\beta} k_{\gamma} + c_{8} \, G_{\nu\beta} k_{\mu} k_{\rho} k_{\alpha} k_{\gamma} +	\vphantom{\Bigg(\Bigg)}	\nonumber
			\\
			& + c_{9} \, G_{\nu\gamma} k_{\mu} k_{\rho} k_{\alpha} k_{\beta} + c_{10} \, G_{\rho\alpha} k_{\mu} k_{\nu} k_{\beta} k_{\gamma} +	\vphantom{\Bigg(\Bigg)}	\nonumber
			\\
			& + c_{11} \, G_{\rho\beta} k_{\mu} k_{\nu} k_{\alpha} k_{\gamma} + c_{12} \, G_{\rho\gamma} k_{\mu} k_{\nu} k_{\alpha} k_{\beta} +	\vphantom{\Bigg(\Bigg)}	\nonumber
			\\
			& + c_{13} \, G_{\alpha\beta} k_{\mu} k_{\nu} k_{\rho} k_{\gamma} + c_{14} \, G_{\alpha\gamma} k_{\mu} k_{\nu} k_{\rho} k_{\beta} +	\vphantom{\Bigg(\Bigg)}	\nonumber
			\\
			& + c_{15} \, G_{\beta\gamma} k_{\mu} k_{\nu} k_{\rho} k_{\alpha} +	\vphantom{\Bigg(\Bigg)}	\nonumber
			\\
			& + d \, k_{\mu} k_{\nu} k_{\rho} k_{\alpha} k_{\beta} k_{\gamma} \, .	\vphantom{\Bigg(\Bigg)}	\nonumber
		\end{align}
	Next, we have to determine the coefficients $a_i$, $b_i$, $c_i$ and $d$ by applying the constraints \eqref{eq:symmetry}, \eqref{eq:trace} and \eqref{eq:lorentz}. Note that for convenience only, the coefficients $a_i$ are dimensionless, while $b_i$, $c_i$ and $d$ contain factors of $\frac{1}{k^2}$, in order to have the r.h.s.\ of the equation dimensionless.
	
	First, we use Eq.\ \eqref{eq:symmetry}. The symmetry under the exchange of $\mu \leftrightarrow \nu$ yields
		\begin{align}
			&	a_{4} = a_{7} \, ,		&&	a_{5} = a_{10} \, ,		&&	a_{6} = a_{13} \, ,	\vphantom{\frac{1}{2}}
			\\
			&	a_{8} = a_{11} \, ,		&&	a_{9} = a_{14} \, ,		&&	a_{12} = a_{15} \, ,	\vphantom{\frac{1}{2}}	\nonumber
			\\
			&	b_{7} = b_{13} \, ,		&&	b_{8} = b_{19} \, ,		&&	b_{9} = b_{25} \, ,	\vphantom{\frac{1}{2}}	\nonumber
			\\
			&	b_{10} = b_{43} \, ,	&&	b_{11} = b_{40} \, ,	&&	b_{12} = b_{37}\, ,	\vphantom{\frac{1}{2}}	\nonumber
			\\
			&	b_{14} = b_{20} \, ,	&&	b_{15} = b_{26} \, ,	&&	b_{16} = b_{44} \, ,	\vphantom{\frac{1}{2}}	\nonumber
			\\
			&	b_{17} = b_{41} \, ,	&&	b_{18} = b_{34} \, ,	&&	b_{21} = b_{27} \, ,	\vphantom{\frac{1}{2}}	\nonumber
			\\
			&	b_{22} = b_{45} \, ,	&&	b_{23} = b_{38} \, ,	&&	b_{24} = b_{35} \, ,	\vphantom{\frac{1}{2}}	\nonumber
			\\
			&	b_{28} = b_{42} \, ,	&&	b_{29} = b_{39} \, ,	&&	b_{30} = b_{36} \, ,	\vphantom{\frac{1}{2}}	\nonumber
			\\
			&	c_{2} \, = c_{6} \, ,		&&	c_{3} = c_{7} \, ,		&&	c_{4} = c_{8} \, ,	\vphantom{\frac{1}{2}}	\nonumber
			\\
			&	c_{5} \, = c_{9} \, ,		&&							&&	\vphantom{\frac{1}{2}}	\nonumber
		\end{align}
	whereas the for the symmetry of the completeness relation under the exchange $\nu \leftrightarrow \rho$ we obtain
		\begin{align}
			&	a_{1} = a_{4} \, ,		&&	a_{2} = a_{5} \, ,		&&	a_{3} = a_{6} \, ,	\vphantom{\frac{1}{2}}
			\\
			&	a_{8} = a_{9} \, ,		&&	a_{11} = a_{12} \, ,	&&	a_{14} = a_{15} \, ,	\vphantom{\frac{1}{2}}	\nonumber
			\\
			&	b_{1} = b_{7} \, ,		&&	b_{2} = b_{8}	\, ,	&&	b_{3} = b_{9}	\, ,	\vphantom{\frac{1}{2}}	\nonumber
			\\
			&	b_{4} = b_{10} \, ,		&&	b_{5} = b_{11}	\, ,	&&	b_{6} = b_{12} \, ,	\vphantom{\frac{1}{2}}	\nonumber
			\\
			&	b_{14} = b_{16} \, ,	&&	b_{15} = b_{17} \, ,	&&	b_{20} = b_{22} \, ,	\vphantom{\frac{1}{2}}	\nonumber
			\\
			&	b_{21} = b_{23} \, ,	&&	b_{26} = b_{28} \, ,	&&	b_{27} = b_{29} \, ,	\vphantom{\frac{1}{2}}	\nonumber
			\\
			&	b_{31} = b_{34}	\, ,	&&	b_{32} = b_{35} \, ,	&&	b_{33} = b_{36} \, ,	\vphantom{\frac{1}{2}}	\nonumber
			\\
			&	b_{38} = b_{39} \, ,	&&	b_{41} = b_{42} \, ,	&&	b_{44} = b_{45} \, ,	\vphantom{\frac{1}{2}}	\nonumber
			\\
			&	c_{1} = c_{2} \, ,		&&	c_{7} = c_{10} \, ,		&&	c_{8} = c_{11} \, ,	\vphantom{\frac{1}{2}}	\nonumber
			\\
			&	c_{9} = c_{12} \, .		&&							&&	\vphantom{\frac{1}{2}}	\nonumber
		\end{align}
	The last symmetry, which includes all other possible index interchanges, is $\mu \, \nu \, \rho \leftrightarrow \alpha \, \beta \, \gamma$, from which follows that
		\begin{align}
			&	a_{1} = a_{13} \, ,		&&	a_{2} = a_{6} \, ,		&&	a_{4} = a_{10} \, ,	\vphantom{\frac{1}{2}}
			\\
			&	a_{12} = a_{14} \, ,	&&	b_{1} = b_{30} \, ,		&&	b_{2} = b_{36} \, ,	\vphantom{\frac{1}{2}}	\nonumber
			\\
			&	b_{3} = b_{33} \, ,		&&	b_{5} = b_{10} \, ,		&&	b_{6} = b_{43} \, ,	\vphantom{\frac{1}{2}}	\nonumber
			\\
			&	b_{7} = b_{24} \, ,		&&	b_{8} = b_{35} \, ,		&&	b_{9} = b_{32} \, ,	\vphantom{\frac{1}{2}}	\nonumber
			\\
			&	b_{12} = b_{40} \, ,	&&	b_{13} = b_{18} \, ,	&&	b_{15} = b_{16} \, ,	\vphantom{\frac{1}{2}}	\nonumber
			\\
			&	b_{19} = b_{34} \, ,	&&	b_{21} = b_{44} \, ,	&&	b_{22} = b_{26} \, ,	\vphantom{\frac{1}{2}}	\nonumber
			\\
			&	b_{23} = b_{41} \, ,	&&	b_{25} = b_{31} \, ,	&&	b_{27} = b_{45} \, ,	\vphantom{\frac{1}{2}}	\nonumber
			\\
			&	b_{29} = b_{42} \, ,	&&	c_{1} = c_{13} \, ,		&&	c_{2} = c_{14} \, ,	\vphantom{\frac{1}{2}}	\nonumber
			\\
			&	c_{4} = c_{7} \, ,		&&	c_{5} = c_{10} \, ,		&&	c_{6} = c_{15} \, ,	\vphantom{\frac{1}{2}}	\nonumber
			\\
			&	c_{9} = c_{11} \, .		&&							&&	\vphantom{\frac{1}{2}}	\nonumber
		\end{align}
	Therefore, we define new coefficients
		\begin{align}
			a \equiv \, & a_1 = a_2 = a_3 = a_4 = a_5 = a_6 = a_7 = a_{10} = a_{13} \, ,	\vphantom{\frac{1}{2}}	\nonumber
			\\
			a^\prime \equiv \, & a_8 = a_9 = a_{11} = a_{12} = a_{14} = a_{15} \, ,	\vphantom{\frac{1}{2}}	\nonumber
			\\
			b \equiv \, & b_1 = b_2 = b_3 = b_7 = b_8 = b_9 = b_{13} = b_{18} = b_{19} =	\vphantom{\frac{1}{2}}	\nonumber
			\\
			= \, & b_{24} = b_{25} = b_{30} = b_{31} = b_{32} = b_{33} = b_{34} = b_{35} =	\vphantom{\frac{1}{2}}	\nonumber
			\\
			= \, & b_{36} \, ,	\vphantom{\frac{1}{2}}	\nonumber
			\\
			b^{\prime} \equiv \, & b_{14} = b_{15} = b_{16} = b_{17} = b_{20} = b_{21} = b_{22} = b_{23} =	\vphantom{\frac{1}{2}}	\nonumber
			\\
			= \, & b_{26} = b_{27} = b_{28} = b_{29} = b_{38} = b_{39} = b_{41} = b_{42} =	\vphantom{\frac{1}{2}}	\nonumber
			\\
			= \, & b_{44} = b_{45} \, ,	\vphantom{\frac{1}{2}}	\nonumber
			\\
			\tilde{b} \equiv \, & b_4 = b_5 = b_6 = b_{10} = b_{11} = b_{12} = b_{37} = b_{40} = b_{43} \, ,	\vphantom{\frac{1}{2}}	\nonumber
			\\
			c \equiv \, & c_1 = c_2 = c_6 = c_{13} = c_{14} = c_{15} \, ,	\vphantom{\frac{1}{2}}	\nonumber
			\\
			c^{\prime} \equiv \, & c_3 = c_4 = c_5 = c_7 = c_8 = c_9 = c_{10} = c_{11} = c_{12} \, ,	\vphantom{\frac{1}{2}}	\nonumber
		\end{align}
	and replace the coefficients of our ansatz.
		\begin{align}
			& \sum_{\lambda = -3}^{+3} \epsilon_{\mu\nu\rho} ( \lambda , \vec{k} ) \, \epsilon_{\alpha\beta\gamma} ( \lambda , \vec{k} ) =
			\\
			= & + a \,  G_{\mu\nu} G_{\rho\alpha} G_{\beta\gamma} + a \, G_{\mu\nu} G_{\rho\beta} G_{\alpha\gamma} +	\vphantom{\sum_{\lambda = -3}^{+3}}	\nonumber
			\\
			& + a \, G_{\mu\nu} G_{\rho\gamma} G_{\alpha\beta} + a \, G_{\mu\rho} G_{\nu\alpha} G_{\beta\gamma} +	\vphantom{\sum_{\lambda = -3}^{+3}}	\nonumber
			\\
			& + a \, G_{\mu\rho} G_{\nu\beta} G_{\alpha\gamma} + a \, G_{\mu\rho} G_{\nu\gamma} G_{\alpha\beta} +	\vphantom{\sum_{\lambda = -3}^{+3}}	\nonumber
			\\
			& + a \, G_{\mu\alpha} G_{\nu\rho} G_{\beta\gamma} + a^\prime \, G_{\mu\alpha} G_{\nu\beta} G_{\rho\gamma} +	\vphantom{\sum_{\lambda = -3}^{+3}}	\nonumber
			\\
			& + a^\prime \, G_{\mu\alpha} G_{\nu\gamma} G_{\rho\beta} + a \, G_{\mu\beta} G_{\nu\rho} G_{\alpha\gamma} +	\vphantom{\sum_{\lambda = -3}^{+3}}	\nonumber
			\\
			& + a^\prime \, G_{\mu\beta} G_{\nu\alpha} G_{\rho\gamma} + a^\prime \, G_{\mu\beta} G_{\nu\gamma} G_{\rho\alpha} +	\vphantom{\sum_{\lambda = -3}^{+3}}	\nonumber
			\\
			& + a \, G_{\mu\gamma} G_{\nu\rho} G_{\alpha\beta} + a^\prime \, G_{\mu\gamma} G_{\nu\alpha} G_{\rho\beta} +	\vphantom{\sum_{\lambda = -3}^{+3}}	\nonumber
			\\
			& + a^\prime \, G_{\mu\gamma} G_{\nu\beta} G_{\rho\alpha} +	\vphantom{\sum_{\lambda = -3}^{+3}}	\nonumber
			\\
			& + b \, G_{\mu\nu} G_{\rho\alpha} k_{\beta} k_{\gamma} + b \, G_{\mu\nu} G_{\rho\beta} k_{\alpha} k_{\gamma} +	\vphantom{\sum_{\lambda = -3}^{+3}}	\nonumber
			\\
			& + b \, G_{\mu\nu} G_{\rho\gamma} k_{\alpha} k_{\beta} + \tilde{b} \, G_{\mu\nu} G_{\alpha\beta} k_{\rho} k_{\gamma} +	\vphantom{\sum_{\lambda = -3}^{+3}}	\nonumber
			\\
			& + \tilde{b} \, G_{\mu\nu} G_{\alpha\gamma} k_{\rho} k_{\beta} + \tilde{b} \, G_{\mu\nu} G_{\beta\gamma} k_{\rho} k_{\alpha} +	\vphantom{\sum_{\lambda = -3}^{+3}}	\nonumber
			\\
			& + b \, G_{\mu\rho} G_{\nu\alpha} k_{\beta} k_{\gamma} + b \, G_{\mu\rho} G_{\nu\beta} k_{\alpha} k_{\gamma} +	\vphantom{\sum_{\lambda = -3}^{+3}}	\nonumber
			\\
			& + b \, G_{\mu\rho} G_{\nu\gamma} k_{\alpha} k_{\beta} + \tilde{b} \, G_{\mu\rho} G_{\alpha\beta} k_{\nu} k_{\gamma} +	\vphantom{\sum_{\lambda = -3}^{+3}}	\nonumber
			\\
			& + \tilde{b} \, G_{\mu\rho} G_{\alpha\gamma} k_{\nu} k_{\beta} + \tilde{b} \, G_{\mu\rho} G_{\beta\gamma} k_{\nu} k_{\alpha} +	\vphantom{\sum_{\lambda = -3}^{+3}}	\nonumber
			\\
			& + b \, G_{\mu\alpha} G_{\nu\rho} k_{\beta} k_{\gamma} + b^\prime \, G_{\mu\alpha} G_{\nu\beta} k_{\rho} k_{\gamma} +	\vphantom{\sum_{\lambda = -3}^{+3}}	\nonumber
			\\
			& + b^\prime \, G_{\mu\alpha} G_{\nu\gamma} k_{\rho} k_{\beta} + b^\prime \, G_{\mu\alpha} G_{\rho\beta} k_{\nu} k_{\gamma} +	\vphantom{\sum_{\lambda = -3}^{+3}}	\nonumber
			\\
			& + b^{\prime} G_{\mu\alpha} G_{\rho\gamma} k_{\nu} k_{\beta} + b \, G_{\mu\alpha} G_{\beta\gamma} k_{\nu} k_{\rho} +	\vphantom{\sum_{\lambda = -3}^{+3}}	\nonumber
			\\
			& + b \, G_{\mu\beta} G_{\nu\rho} k_{\alpha} k_{\gamma} + b^\prime \, G_{\mu\beta} G_{\nu\alpha} k_{\rho} k_{\gamma} +	\vphantom{\sum_{\lambda = -3}^{+3}}	\nonumber
			\\
			& + b^\prime \, G_{\mu\beta} G_{\nu\gamma} k_{\rho} k_{\alpha} + b^\prime \, G_{\mu\beta} G_{\rho\alpha} k_{\nu} k_{\gamma} +	\vphantom{\sum_{\lambda = -3}^{+3}}	\nonumber
			\\
			& + b^{\prime} G_{\mu\beta} G_{\rho\gamma} k_{\nu} k_{\alpha} + b \, G_{\mu\beta} G_{\alpha\gamma} k_{\nu} k_{\rho} +	\vphantom{\sum_{\lambda = -3}^{+3}}	\nonumber
			\\
			& + b \, G_{\mu\gamma} G_{\nu\rho} k_{\alpha} k_{\beta} + b^\prime \, G_{\mu\gamma} G_{\nu\alpha} k_{\rho} k_{\beta} +	\vphantom{\sum_{\lambda = -3}^{+3}}	\nonumber
			\\
			& + b^\prime \, G_{\mu\gamma} G_{\nu\beta} k_{\rho} k_{\alpha} + b^\prime \, G_{\mu\gamma} G_{\rho\alpha} k_{\nu} k_{\beta} +	\vphantom{\sum_{\lambda = -3}^{+3}}	\nonumber
			\\
			& + b^{\prime} G_{\mu\gamma} G_{\rho\beta} k_{\nu} k_{\alpha} + b \, G_{\mu\gamma} G_{\alpha\beta} k_{\nu} k_{\rho} +	\vphantom{\sum_{\lambda = -3}^{+3}}	\nonumber
			\\
			& + b \, k_{\mu} k_{\nu} G_{\rho\alpha} G_{\beta\gamma} + b \, k_{\mu} k_{\nu} G_{\rho\beta} G_{\alpha\gamma} +	\vphantom{\sum_{\lambda = -3}^{+3}}	\nonumber
			\\
			& + b \, k_{\mu} k_{\nu} G_{\rho\gamma} G_{\alpha\beta} + b \, k_{\mu} k_{\rho} G_{\nu\alpha} G_{\beta\gamma} +	\vphantom{\sum_{\lambda = -3}^{+3}}	\nonumber
			\\
			& + b \, k_{\mu} k_{\rho} G_{\nu\beta} G_{\alpha\gamma} + b \, k_{\mu} k_{\rho} G_{\nu\gamma} G_{\alpha\beta} +	\vphantom{\sum_{\lambda = -3}^{+3}}	\nonumber
			\\
			& + \tilde{b} \, k_{\mu} k_{\alpha} G_{\nu\rho} G_{\beta\gamma} + b^{\prime} k_{\mu} k_{\alpha} G_{\nu\beta} G_{\rho\gamma} +	\vphantom{\sum_{\lambda = -3}^{+3}}	\nonumber
			\\
			& + b^\prime \, k_{\mu} k_{\alpha} G_{\nu\gamma} G_{\rho\beta} + \tilde{b} \, k_{\mu} k_{\beta} G_{\nu\rho} G_{\alpha\gamma} +	\vphantom{\sum_{\lambda = -3}^{+3}}	\nonumber
			\\
			& + b^{\prime} k_{\mu} k_{\beta} G_{\nu\alpha} G_{\rho\gamma} + b^\prime \, k_{\mu} k_{\beta} G_{\nu\gamma} G_{\rho\alpha} +	\vphantom{\sum_{\lambda = -3}^{+3}}	\nonumber
			\\
			& + \tilde{b} \, k_{\mu} k_{\gamma} G_{\nu\rho} G_{\alpha\beta} + b^{\prime} k_{\mu} k_{\gamma} G_{\nu\alpha} G_{\rho\beta} +	\vphantom{\sum_{\lambda = -3}^{+3}}	\nonumber
			\\
			& + b^\prime \, k_{\mu} k_{\gamma} G_{\nu\beta} G_{\rho\alpha} +	\vphantom{\sum_{\lambda = -3}^{+3}}	\nonumber
			\\
			& + c \, G_{\mu\nu} k_{\rho} k_{\alpha} k_{\beta} k_{\gamma} + c \, G_{\mu\rho} k_{\nu} k_{\alpha} k_{\beta} k_{\gamma} +	\vphantom{\sum_{\lambda = -3}^{+3}}	\nonumber
			\\
			& + c^\prime \, G_{\mu\alpha} k_{\nu} k_{\rho} k_{\beta} k_{\gamma} + c^\prime \, G_{\mu\beta} k_{\nu} k_{\rho} k_{\alpha} k_{\gamma} +	\vphantom{\sum_{\lambda = -3}^{+3}}	\nonumber
			\\
			& + c^{\prime} G_{\mu\gamma} k_{\nu} k_{\rho} k_{\alpha} k_{\beta} + c \, G_{\nu\rho} k_{\mu} k_{\alpha} k_{\beta} k_{\gamma} +	\vphantom{\sum_{\lambda = -3}^{+3}}	\nonumber
			\\
			& + c^\prime \, G_{\nu\alpha} k_{\mu} k_{\rho} k_{\beta} k_{\gamma} + c^{\prime} G_{\nu\beta} k_{\mu} k_{\rho} k_{\alpha} k_{\gamma} +	\vphantom{\sum_{\lambda = -3}^{+3}}	\nonumber
			\\
			& + c^\prime \, G_{\nu\gamma} k_{\mu} k_{\rho} k_{\alpha} k_{\beta} + c^\prime \, G_{\rho\alpha} k_{\mu} k_{\nu} k_{\beta} k_{\gamma} +	\vphantom{\sum_{\lambda = -3}^{+3}}	\nonumber
			\\
			& + c^{\prime} G_{\rho\beta} k_{\mu} k_{\nu} k_{\alpha} k_{\gamma} + c^\prime \, G_{\rho\gamma} k_{\mu} k_{\nu} k_{\alpha} k_{\beta} +	\vphantom{\sum_{\lambda = -3}^{+3}}	\nonumber
			\\
			& + c \, G_{\alpha\beta} k_{\mu} k_{\nu} k_{\rho} k_{\gamma} + c \, G_{\alpha\gamma} k_{\mu} k_{\nu} k_{\rho} k_{\beta} +	\vphantom{\sum_{\lambda = -3}^{+3}}	\nonumber
			\\
			& + c \, G_{\beta\gamma} k_{\mu} k_{\nu} k_{\rho} k_{\alpha} +	\vphantom{\sum_{\lambda = -3}^{+3}}	\nonumber
			\\
			& + d \, k_{\mu} k_{\nu} k_{\rho} k_{\alpha} k_{\beta} k_{\gamma} \, .	\vphantom{\sum_{\lambda = -3}^{+3}}	\nonumber
		\end{align}
	To determine the remaining eight constants, we use Eqs.\ \eqref{eq:trace} and \eqref{eq:lorentz}. We note that
		\begin{align}
			k^\mu G_{\mu\nu} = 0
		\end{align}
	and contract the last expression with $k^{\mu} k^{\nu} k^{\rho} k^{\alpha}$,
		\begin{align}
			0 = \, & k^{\mu} k^{\nu} k^{\rho} k^{\alpha} \sum_{\lambda = -3}^{+3} \epsilon_{\mu\nu\rho} ( \lambda , \vec{k} ) \, \epsilon_{\alpha\beta\gamma} ( \lambda , \vec{k} ) =
			\\
			= \, &  (  k^2 )^4 \, ( c \, G_{\beta\gamma} + d \, k_{\beta} k_{\gamma} ) \, .	\vphantom{\sum_{\lambda = -3}^{+3}}	\nonumber
		\end{align}
	The only solution to this equation is $c = d = 0$, which eliminates two of the remaining constants. In complete analogy the contraction with $k^{\mu} k^{\nu} k^{\alpha} k^{\beta}$ leads to
		\begin{align}
			0 = \, & k^{\mu} k^{\nu} k^{\alpha} k^{\beta} \sum_{\lambda = -3}^{+3} \epsilon_{\mu\nu\rho} ( \lambda , \vec{k} ) \, \epsilon_{\alpha\beta\gamma} ( \lambda , \vec{k} ) =
			\\
			= \, & ( k^2 )^4 \, c^\prime \, G_{\rho\gamma} \, ,	\vphantom{\sum_{\lambda = -3}^{+3}}	\nonumber
		\end{align}
	which sets $c^{\prime} = 0$. Next, we contract with $k^{\mu} k^{\nu} \eta^{\rho\alpha} \eta^{\beta\gamma}$ and find
		\begin{align}
			0 = \, & k^{\mu} k^{\nu} \eta^{\rho\alpha} \eta^{\beta\gamma} \sum_{\lambda = -3}^{+3} \epsilon_{\mu\nu\rho} ( \lambda , \vec{k} ) \, \epsilon_{\alpha\beta\gamma} ( \lambda , \vec{k} ) =
			\\
			= \, & ( k^2 )^2 \, b \, ( \tensor{G}{_\mu^\mu} \tensor{G}{_\nu^\nu} + 2 \, G_{\mu\nu} G^{\mu\nu} ) =	\vphantom{\sum_{\lambda = -3}^{+3}}	\nonumber
			\\
			= \, & 15 \, b \, ( k^2)^2 \, ,	\vphantom{\sum_{\lambda = -3}^{+3}}	\nonumber
		\end{align}
	which states that also $b=0$. To eliminate $b^{\prime}$ and $\tilde{b}$, we have to contract the whole expression with $k^{\mu} k^{\alpha} \eta^{\nu\beta} \eta^{\rho\gamma}$,
		\begin{align}
			0 = \, & k^{\mu} k^{\alpha} \eta^{\nu\beta} \eta^{\rho\gamma} \sum_{\lambda = -3}^{+3} \epsilon_{\mu\nu\rho} ( \lambda , \vec{k} ) \, \epsilon_{\alpha\beta\gamma} ( \lambda , \vec{k} ) =
			\\
			= \, & ( k^2 )^2 \, ( 3 \, \tilde{b} + 12 \, b^\prime ) \, ,	\vphantom{\sum_{\lambda = -3}^{+3}}	\nonumber
		\end{align}
	and with $k^{\mu} k^{\alpha} \eta^{\nu\rho} \eta^{\beta\gamma}$,
		\begin{align}
			0 = \, & k^{\mu} k^{\alpha} \eta^{\nu\rho} \eta^{\beta\gamma} \sum_{\lambda=1}^{7} \epsilon_{\mu\nu\rho} ( \lambda , \vec{k} ) \, \epsilon_{\alpha\beta\gamma} ( \lambda , \vec{k} ) =
			\\
			= \, & ( k^2 )^2 \, ( 9 \, \tilde{b} + 6 \, b^\prime ) \, .	\vphantom{\sum_{\lambda=1}^{7}}	\nonumber
		\end{align}
	Thus, we conclude $b^\prime = \tilde{b} = 0$.
	Before we calculate the relation between the coefficients $a$ and $a^\prime$, we provide the remainder expression for the completeness relation.
		\begin{align}
			& \sum_{\lambda=1}^{7} \epsilon_{\mu\nu\rho} ( \lambda , \vec{k} ) \, \epsilon_{\alpha\beta\gamma} ( \lambda , \vec{k} ) =	\vphantom{\Bigg(\Bigg)}
			\\
			= \, & + a \, G_{\mu\nu} G_{\rho\alpha} G_{\beta\gamma} + a \, G_{\mu\nu} G_{\rho\beta} G_{\alpha\gamma} +	\vphantom{\Bigg(\Bigg)}	\nonumber
			\\
			& + a \, G_{\mu\nu} G_{\rho\gamma} G_{\alpha\beta} + a \, G_{\mu\rho} G_{\nu\alpha} G_{\beta\gamma} +	\vphantom{\Bigg(\Bigg)}	\nonumber
			\\
			& + a \, G_{\mu\rho} G_{\nu\beta} G_{\alpha\gamma} + a \, G_{\mu\rho} G_{\nu\gamma} G_{\alpha\beta} +	\vphantom{\Bigg(\Bigg)}	\nonumber
			\\
			& + a \, G_{\mu\alpha} G_{\nu\rho} G_{\beta\gamma} + a^\prime \, G_{\mu\alpha} G_{\nu\beta} G_{\rho\gamma} +	\vphantom{\Bigg(\Bigg)}	\nonumber
			\\
			& + a^\prime \, G_{\mu\alpha} G_{\nu\gamma} G_{\rho\beta} + a \, G_{\mu\beta} G_{\nu\rho} G_{\alpha\gamma} +	\vphantom{\Bigg(\Bigg)}	\nonumber
			\\
			& + a^\prime \, G_{\mu\beta} G_{\nu\alpha} G_{\rho\gamma} + a^\prime \, G_{\mu\beta} G_{\nu\gamma} G_{\rho\alpha} +	\vphantom{\Bigg(\Bigg)}	\nonumber
			\\
			& + a \, G_{\mu\gamma} G_{\nu\rho} G_{\alpha\beta} + a^\prime \, G_{\mu\gamma} G_{\nu\alpha} G_{\rho\beta} +	\vphantom{\Bigg(\Bigg)}	\nonumber
			\\
			& + a^\prime \, G_{\mu\gamma} G_{\nu\beta} G_{\rho\alpha} \, .	\vphantom{\Bigg(\Bigg)}	\nonumber
		\end{align}
	The easiest way to eliminate $a^{\prime}$ in favor of $a$ is, to contract this expression with $\eta^{\mu\nu} \eta^{\rho\alpha} \eta^{\beta\gamma}$.
		\begin{align}
			0 = \, & \eta^{\mu\nu} \eta^{\rho\alpha} \eta^{\beta\gamma} \sum_{\lambda = -3}^{+3} \epsilon_{\mu\nu\rho} ( \lambda , \vec{k} ) \, \epsilon_{\alpha\beta\gamma} ( \lambda , \vec{k} ) =
			\\
			= \, & 75 \, a + 30 \, a^\prime \, .	\vphantom{\sum_{\lambda = -3}^{+3}}	\nonumber
		\end{align}
	Thus, $a^{\prime}=-\frac{5}{2}a$. Reordering the completeness relation gives
		\begin{align}
			& \sum_{\lambda = -3}^{+3} \epsilon_{\mu\nu\rho} ( \lambda , \, \vec{k} ) \, \epsilon_{\alpha\beta\gamma} ( \lambda , \, \vec{k} ) =	\vphantom{\sum_{\lambda = -3}^{+3}}
			\\
			= \, & a \, \big\{ G_{\mu\nu} \, ( G_{\rho\alpha} \, G_{\beta\gamma} + G_{\rho\beta} \, G_{\alpha\gamma} + G_{\rho\gamma} \, G_{\alpha\beta} ) +	\vphantom{\sum_{\lambda = -3}^{+3}}	\nonumber
			\\
			& + G_{\mu\rho} \, (  G_{\nu\alpha} \, G_{\beta\gamma} + G_{\nu\beta} \, G_{\alpha\gamma} + G_{\nu\gamma} \, G_{\alpha\beta} ) +	\vphantom{\sum_{\lambda = -3}^{+3}}	\nonumber
			\\
			& + G_{\nu\rho} \, ( G_{\mu\alpha}\, G_{\beta\gamma} + G_{\mu\beta} \, G_{\alpha\gamma} + G_{\mu\gamma} \, G_{\alpha\beta} ) -	\vphantom{\sum_{\lambda = -3}^{+3}}	\nonumber
			\\
			& - \tfrac{5}{2} \, \big[ G_{\mu\alpha} \, ( G_{\nu\beta} \, G_{\rho\gamma} + G_{\nu\gamma} \, G_{\rho\beta} ) +	\vphantom{\sum_{\lambda = -3}^{+3}}	\nonumber
			\\
			& + G_{\mu\beta} \, ( G_{\nu\alpha} \, G_{\rho\gamma} + G_{\nu\gamma} \, G_{\rho\alpha} ) +	\vphantom{\sum_{\lambda = -3}^{+3}}	\nonumber
			\\
			& + G_{\mu\gamma} \, ( G_{\nu\alpha} \, G_{\rho\beta} + G_{\nu\beta} \, G_{\rho\alpha} ) \big] \big\} \, .	\vphantom{\sum_{\lambda = -3}^{+3}}	\nonumber
		\end{align}
	The last constant $a$ can be chosen arbitrarily. Nevertheless it might be convenient to orthonormalize the polarization tensors,
		\begin{align}
			\epsilon_{\mu\nu\rho} ( \lambda , \vec{k} ) \, \epsilon^{\mu\nu\rho} ( \lambda^\prime , \vec{k} ) = - \delta_{\lambda \lambda^\prime} \, ,
		\end{align}
	which is the higher-spin equivalent to the orthonormality condition Eq.\ \eqref{eq:restframe_orthonormal}.	Then it directly follows that
		\begin{align}
			- 7 = \, & - \sum_{\lambda -3}^{+3} \delta_{\lambda \lambda} =
			\\
			= \, & \sum_{\lambda = -3}^{+3} \epsilon_{\mu\nu\rho} ( \lambda , \vec{k} ) \, \epsilon^{\mu\nu\rho} ( \lambda , \vec{k} ) = - 105 \, a \, .	\nonumber
		\end{align}
	The final result is.
		\begin{align}
			& \sum_{\lambda = -3}^{+3} \epsilon_{\mu\nu\rho} ( \lambda , \vec{k} ) \, \epsilon_{\alpha\beta\gamma} ( \lambda , \vec{k} ) =	\vphantom{\sum_{\lambda = -3}^{+3}}
			\\
			= \, & \tfrac{1}{15} \, \big[ G_{\mu\nu} \, ( G_{\rho\alpha} \, G_{\beta\gamma} + G_{\rho\beta} \, G_{\alpha\gamma} + G_{\rho\gamma} \, G_{\alpha\beta} ) +	\vphantom{\sum_{\lambda = -3}^{+3}}	\nonumber
			\\
			& + G_{\mu\rho} \, (  G_{\nu\alpha} \, G_{\beta\gamma} + G_{\nu\beta} \, G_{\alpha\gamma} + G_{\nu\gamma} \, G_{\alpha\beta} ) +	\vphantom{\sum_{\lambda = -3}^{+3}}	\nonumber
			\\
			& + G_{\nu\rho} \, ( G_{\mu\alpha}\, G_{\beta\gamma} + G_{\mu\beta} \, G_{\alpha\gamma} + G_{\mu\gamma} \, G_{\alpha\beta} ) \big] -	\vphantom{\sum_{\lambda = -3}^{+3}}	\nonumber
			\\
			& - \tfrac{1}{6} \, \big[ G_{\mu\alpha} \, ( G_{\nu\beta} \, G_{\rho\gamma} + G_{\nu\gamma} \, G_{\rho\beta} ) +	\vphantom{\sum_{\lambda = -3}^{+3}}	\nonumber
			\\
			& + G_{\mu\beta} \, ( G_{\nu\alpha} \, G_{\rho\gamma} + G_{\nu\gamma} \, G_{\rho\alpha} ) +	\vphantom{\sum_{\lambda = -3}^{+3}}	\nonumber
			\\
			& + G_{\mu\gamma} \, ( G_{\nu\alpha} \, G_{\rho\beta} + G_{\nu\beta} \, G_{\rho\alpha} ) \big] \, .	\vphantom{\sum_{\lambda = -3}^{+3}}	\nonumber
		\end{align}

\subsection{Spin-\texorpdfstring{$J$}{J} tensor degrees of freedom}

	As a last step, we recapitulate for the sake of completeness, how to determine the number of degrees of freedom for a massive bosonic field (see Ref.\ \cite{Fierz:1939i, Fierz:1939ix} for the original discussion). A Lorentz tensor $\phi^{...\mu\nu...}$ of arbitrary rank $J$ in four-dimensional Minkowski space has $4^{J}$ degrees of freedom. The KG equation \eqref{eq:fp_1} does not reduce the degrees of freedom, whereas the symmetry condition \eqref{eq:fp_2} does. A combinatorial analysis is necessary and we use for simplicity an urn model: for each index of the rank-$J$ tensor components there are four possible entries, which is equivalent to four different balls in the urn. The tensor field is symmetric in all indices, which means that the order of drawings does not matter. Indices can also be ``drawn'' several times. This leads to
		\begin{align}
			\frac{( 4 - 1 + J )!}{J! \, ( 4 - 1 )!} = \binom{4 - 1 + J}{4 - 1}
		\end{align}
	possibilities, or better degrees of freedom. The next reduction is done using the traceless condition \eqref{eq:fp_3} which leads to
		\begin{align}
			\binom{4 - 1 + ( J - 2 )}{4 - 1}
		\end{align}
	additional constraints [this is exactly the number of degrees of freedom of a symmetric tensor of rank $(J - 2)$]. The last constraint, also refereed to as Lorentz constraint or transversality constraint \eqref{eq:fp_4}, leads to restrictions:
		\begin{align}
			0 = \partial_{\mu} \phi^{ \cdots \mu \nu \cdots} \, .
		\end{align}
	Now it is easy to conclude, that the number of these constraints are exactly the degrees of freedom of a rank $(J - 1)$ symmetric traceless tensor which are:
		\begin{align}
			\binom{4 - 1 + ( J - 1 )}{4 - 1} - \binom{4 - 1 + ( J - 3 )}{4 - 1} \, .
		\end{align}
	In total, the Fierz-Pauli equations lead to
\begin{widetext}
		\begin{align}
			\binom{4 - 1 + J}{4 - 1} - \binom{4 - 1 + ( J - 2 )}{4 - 1} - \left[ \binom{4 - 1 + ( J - 1 )}{4 - 1} - \binom{4 - 1 + ( J - 3 )}{4 - 1} \right] = 2 J + 1 \, .
		\end{align}
\end{widetext}
	degrees of freedom for a rank-$J$ massive tensor field.
	
	On the other hand, we know that a massive particle of total spin-$J$ has also $2 J + 1$ degrees of freedom, which correspond to the $2 J + 1$ possible orientations of the particles spin in a magnetic field. Furthermore, one could study the behavior of particles of spin-$J$ under rotations or Lorentz transformations, which would also emphasize the use massive rank-$J$ tensor fields to describe those spin-$J$ particles.

\bibliography{spin-3}

\end{document}